%% file: thesis.tex
\documentclass{scrbook} 
\usepackage{wbh-thesis} 
\pdfoutput=1

\title{A Study about Distribution and Acceptance of Conversational Agents for Mental Health in Germany: Keep the Human in the Loop?}
\Autor[f]{Christina Lukas} 
\Fachbereich{INF} 
\Betreuung[m]{Prof. Dr. Rüdiger Breitschwerdt} 
\Matrikelnummer{853485}
\Abgabetermin{03.12.2024}

\makeglossaries
\newacronym{mHealth}{mHealth}{Mental Health}
\newacronym{ca}{CA}{Conversational agent}
\newacronym{WHO}{WHO}{World Health Organization}
\newacronym{ICD-10}{ICD-10}{International Classiﬁcation of Diseases 10th Revision}
\newacronym{ChatGPT}{ChatGPT}{Chat Generative Pre-trained Transformer}
\newacronym{XAI}{XAI}{Explainable AI}
\newacronym{HITL}{HITL}{Human in the loop}
\newacronym{AuI}{AuI}{Augmented Intelligence}
\newacronym{AI}{AI}{Artificial Intelligence}
\newacronym{DMHI}{DMHI}{Digital mHealth interventions}
\newacronym{DiGA}{DiGA}{Digitale Gesundheitsanwendung}
\newacronym{FtF}{FtF}{Face-to-Face}
\newacronym{CBT}{CBT}{Cognitive Behavioral Therapy}
\newacronym{DVG}{DVG}{Digitale Versorgungs Gesetz}
\newacronym{BfArM}{BfArM}{Bundesinstitut für Arzneimittel und Medizinprodukte}
\newacronym{TK}{TK}{Techniker Krankenkasse}
\newacronym{NLP}{NLP}{Natural Language Processing}
\newacronym{LLM}{LLM}{Large Language Model}
\newacronym{IRB}{IRB}{Institutional Review Board}
\newacronym{HI}{HI}{Human Intelligence}
\newacronym{ML}{ML}{Machine Learning}
\newacronym{TAM}{TAM}{Technology Acceptance Model}
\newacronym{HITAM}{HITAM}{Health Information Technology Acceptance Model}
\newacronym{UTAUT}{UTAUT}{Unified Theory of Acceptance and Use of Technology}
\newacronym{PE}{PE}{Performance Expectancy}
\newacronym{EE}{EE}{Effort Expectancy}
\newacronym{SI}{SI}{Social Influence}
\newacronym{FC}{FC}{Facilitating Conditions}
\newacronym{BI}{BI}{Behavioral Intention}
\newacronym{UB}{UB}{Use Behavior}

\newacronym{PU}{PU}{Perceived Usefulnes}
\newacronym{PEU}{PEU}{Perceived Ease of Use}
\newacronym{SN}{SN}{Subjective Norm}
\newacronym{OQ}{OQ}{Output Quality}
\newacronym{RD}{RD}{Result Demonstrability}
\newacronym{CS}{CS}{Computer Self-efficacy}
\newacronym{PEC}{PEC}{Perceptions of External Control}
\newacronym{CoA}{CoA}{Computer Anxiety}
\newacronym{CP}{CP}{Computer Playfulness}

\newacronym{GAD}{GAD}{Generalized Anxiety Disorder}
\newacronym{K6}{K6}{Kessler Psychological Distress Scale 6}
\newacronym{QOL}{QOL}{Quality of Life}
\newacronym{PWBI}{PWBI}{Physician Well-Being Index}
\newacronym{CHERRIES}{CHERRIES}{Checklist for Reporting Results of Internet E-Surveys}

\begin{document}
  \frontmatter        

  \maketitle          
  \addchap*{Zusammenfassung}%
  \input{zusammenfassung} 
  
  \addchap*{Abstract}%
  \input{abstract} 
  
  \renewcommand*\contentsname{Table of Content}
  \tableofcontents    
  \mainmatter         
\begin{spacing}{1.55}

\include{content/introduction}
\include{content/literatur-review}

\include{content/theoretical-framework}

\include{content/methods}

\include{content/results}

\include{content/discussion}

\end{spacing}
  
  \appendix  
  \listoftables                  
  \listoffigures              
  \printglossary[type=\acronymtype, title={Acronyms}]
 
  \printbibliography[title={References}]      

\end{document}

%% file: zusammenfassung.tex
\begin{otherlanguage}{ngerman}
Mentale Gesundheit (mHealth) ist mehr als die Abwesenheit von psychischen Störungen. Es handelt sich um ein komplexes Kontinuum, das von Mensch zu Mensch unterschiedlich erlebt wird \cite{WHO-MentalHealth2022}. Die Fehlzeiten am Arbeitsplatz, in der Schule, im Studium oder in der Ausbildung aufgrund psychischer oder mentaler Probleme und die Zahl der Hilfesuchenden nehmen zu. Die wachsende Nachfrage nach Beratung oder psychischer Unterstützung erfordert einen genaueren Blick auf bereits bestehende Lösungen wie Telemedizin, digitale Gesundheitsanwendungen und gemischte Therapie, deren Akzeptanz, Vorzüge und Hindernisse. Mehr oder weniger plötzlich kamen neue Technologien wie KI-basierte Konversations Agenten (CAs) z.\,B. ChatGPT und Siri auf den Markt und werden als neue, zusätzliche Lösung bereitgestellt. Nicht nur, um die Patientenversorgung zu verbessern, sondern auch, um das medizinische Personal zu entlasten.

Es stellen sich daher die Fragen, ob KI-basierte CAs bereits im Einsatz sind, ob sie akzeptiert werden und inwiefern unterscheidet sich die Meinung der Öffentlichkeit von der der Fachleute? Zwischen blindem Vertrauen und absoluter Ablehnung dieser Technologie liegt das Potenzial, menschliche und künstliche Intelligenz in Form von erweiterter Intelligenz (AuI) zu kombinieren. Inwieweit ist diese Kombination in der Beratung, Diagnose oder Behandlung akzeptabel?

Um Antworten auf diese Fragen zu erhalten, wurden zwei quantitative Online-Umfragen durchgeführt, bei denen 444 Teilnehmende aus der Bevölkerung und 351 mHealth-Spe-zialisten in Deutschland befragt wurden. Für ein differenziertes Verständnis, werden zusätzlich neben soziodemographischen Angaben, Vorteile und Hindernisse sowie der aktuelle mHealth- oder Wohlfühlstatus abgefragt. Die statistischen Auswertungen zeigen, dass die Bevölkerung bereits KI-basierte CAs zur Kummerbewältigung nutzt und dem gegenüber aufgeschlossen ist, diese auch in Kombination mit Fachkräften einzusetzen. Um den Anschluss nicht zu verlieren, ist es wichtig, über den Einsatz, Nutzen, aber im besonderen auch die Risiken zu informieren. Mit den aus dieser Studie gewonnenen Erkenntnissen können Lösungen erarbeitet werden, die sowohl künstliche als auch menschliche Intelligenz integrieren, um dem wachsenden Bedarf nachzukommen und Fachkräfte entlasten zu können.

\end{otherlanguage}

%% file: abstract.tex
Mental health (mHealth) is more than the absence of mental disorder. It is a complex continuum which is experienced differently from person to person \cite{WHO-MentalHealth2022}. Absenteeism from work, school, university or training due to psychological problems and the number of people seeking help is increasing. The growing demand for counselling or psychical support requires a closer look at existing solutions such as telemedicine, digital health applications and blended therapy, their acceptance, benefits and barriers. More or less suddenly, new technologies such as AI-based conversational agents (CAs) like ChatGPT and Siri came onto the market and are being provided as a new, additional solution. Not only to improve patient care, but also to relieve the burden on medical staff. 

The question therefore arises as to whether AI-based CAs are already in use and whether they are accepted, and how does the public's view differ from that of professionals? Between blind trust and absolute rejection of this technology lies the potential to combine human and artificial intelligence in the form of augmented intelligence (AuI). To what extent is this combination acceptable in counselling, diagnose or treatment?

To obtain answers to these questions, two quantitative online surveys have been conducted in which 444 participants of the population and 351 mHealth specialists in Germany were questioned. In addition to sociodemographic information, advantages and obstacles as well as the current mHealth or well-being status have been asked for a differentiated understanding. The statistical analyses show that the population already uses AI-based CAs to combat grief and is more open to using this technology in combination with specialists. In order not to lose touch, it is important to provide information about the use, benefits and, in particular, the risks. The knowledge gained from this study can be used to develop solutions that utilize both artificial and human intelligence in order to meet the growing demand and at the same time relieve the burden on specialists.

\subsubsection{Keywords}
Artificial Intelligence, Mental Health, Augmented Intelligence, AI Acceptance, Mental Health Professionals and Population

%% file: content/introduction.tex
\chapter{Introduction}
The following introduction aims to provide an overview and motivation for the research topic -- usage and acceptance of \acrlong{AI} (\acrshort{AI}-) based conversational agents (\acrshort{ca}) for people who are living in Germany on the one side and professionals in \acrfull{mHealth} environments on the other side. This part outlines the significance and relevance of the topic, identifies existing research gaps, and presents the objectives of addressing these gaps and gaining further insights together with some hypothesis.

\section{Background}
"There is no health without mental health" \cite[p. 11]{WHO2005}. But what does mHealth exactly mean and include? The absence of a clear and consensual definition and its core concepts have already been discussed in several publications \cite[cf. p. 1]{Manwell2014}, \cite[cf. p. 34]{Fusar-Poli2020}. Manwell et al. ask 50 people with expertise in the field of mHealth from eight countries to rank four of the most known current definitions of mHealth they prefer \cite[cf. p. 4]{Manwell2014}. Furthermore a disagreement was displayed in the answer of the core concepts, which are highly dependent on the used empirical frame. Fusar-Poli et al. state that "good mHealth can be deﬁned as a state of well-being that allows individuals to cope with the normal stresses of life and function productively" \cite[p. 34]{Fusar-Poli2020}, which stays in line with the definition and fact sheet of the \acrfull{WHO} \cite{WHO-MentalHealth2022}. In their review the authors identified 14 core domains -- including besides next mHealth literacy, cognitive, and social skills as well as physical health -- which define good mHealth. It is important to appreciate the affinity between mental illness and other health conditions \cite[cf. p. 359]{Prince2007}. This work refers to the WHO definition and understands mHealth as a complex, multifaceted condition that can be experienced differently by individuals and can change in scope and over time \cite{WHO-MentalHealth2022}.   

In 2022, mental illnesses -- including depression and anxiety -- ranked third among the disease groups that cause the most days of absence from work in Germany \cite[cf. p. 2]{DAK2022}. Mental and behavioural disorders furthermore occupied the third position in terms of the number of deaths by the \acrfull{ICD-10} chapter in Germany in 2022, following diseases of the circulatory system and cancer \cite{DStatist2022}. In the same year, a total of 10,119 people died by suicide, with approximately 75\,\% of them are men \cite{DStatist2022_sucide}. In Germany, around 27.8\,\% of the adult population is affected by a mental illness annually \cite[cf. p. 88]{Jacobi2016}. This corresponds to approximately 17.8 million people, of whom only 18.9\,\% seek help from service providers each year \cite[cf. p. 1]{DGPPN2023}. 

Long waiting lists for professional mHealth care, often lasting several months \cite[cf. p. 3]{DGPPN2023}, increase the pressure for additional services. The demand for free counselling services such as krisenchat gGmbH \cite{krisenchat_gGmbH}, Nummer Gegen Kummer e.\,V. \cite{NummerGegenKummer_eV} and TelefonSeelsorge\textsuperscript{\textregistered} Deutschland e.\,V. \cite{Telefonseelsorge_eV} remains high. Coaches become increasingly prevalent, and large numbers of promising mobile and web apps are entering or even floating the market. The demand for personalized content, even during a depression episode, is increasing \cite[cf. p. 2 -- 3]{Mayer2022}. In today's healthcare landscape, patients desire to be actively involved in their treatment process, acquire health literacy, and participate in shared decision-making \cite[cf. p. 14]{Mayer2022}. Consequently, it is therefore not surprising that after access to the internet and various search engines, AI models such as \acrfull{ChatGPT} are emerging as new innovative solutions to expand the range of psycho-social support on offer. 

In an online survey, conducted in 2021 with a total of 100 mHealth professionals or experts from Northern Ireland, Ireland, Scotland, Sweden and Finland, Sweeney et al. observe that mental healthcare professionals with greater experience tend to have a stronger belief in the use of chatbots to help their clients better manage their own mHealth \cite[cf. p. 11]{Sweeney2021}. 65\,\% of the participants agreed that there are benefits associated with mental healthcare chatbots. The perceived importance of chatbots was quite high, with 74\,\% of respondents indicating this. Additionally, 79\,\% agreed that these chatbots could help their clients to improve the management of their own health, reduce stress, improve diet, medication or treatment adherence \cite[cf. p. 1]{Sweeney2021}. However, these professionals also identified the biggest challenges as the lack of intelligence or knowledge of chatbots to accurately assess clients and concerns regarding data privacy and confidentiality. Additionally, 90\,\% of the professionals expressed concern that their clients may not feel adequately connected to their healthcare providers when using chatbots \cite[cf. p. 11 -- 12]{Sweeney2021}. In a previous study, conducted by Palanica et al. in 2019 with a total of 100 practising physicians across the United States, the results regarding acceptance of using chatbots for mHealth were less strong. Less than half (44\,\%) of the respondents believed that chatbots for mHealth would become very likely or somewhat likely relevant in the next five years \cite[cf. p. 1]{Palanica2019}. Both studies provided lists of commonly known chatbots for mHealth, most, if not all, of which used decision trees and not 'real' AI. The popularity and public access to powerful generative AI started in November 2022 with the release and public availability of ChatGPT, which one year after its release is already being used by one in three Germans \cite{Kero2023, dpa2023}. Nevertheless, 30\,\% of the physicians practising in the US and participating in the survey of 2019 had used chatbots for healthcare previously, and 16\,\% of the participating mHealth professionals in 2021 \cite[cf. p. 3]{Palanica2019} \cite[cf. p. 5]{Sweeney2021}. Comparing these studies, it becomes evident that there are differences in experts' perceptions and views of chatbots for health, based on time and culture. However, both publications suggest that chatbots are best applied to assist physicians rather than replace them.

Still, the integration of AI is a highly contentious topic, as it involves the handling of sensitive health and personal data. Advantages and disadvantages regarding bias, security, availability, cost, ethics, and more are debated and evaluated. 
Additionally the need for explainability through \acrfull{XAI} is a relevant topic that needs to be incorporated into text-based conversations between patients, experts or clinicians and CAs \cite[cf. p. 47]{Nov2021}. Müller et al. write that truly intelligent systems need to be able to understand context, and this understanding is currently something that only human experts can do, sometimes, not always \cite[cf. p. 120]{mueller2021}. This motivates to focus on a \acrfull{HITL} approach, to gain more experience and knowledge about human-AI interaction and resulting behaviour. Therefore it is worth considering combining both intelligence into what is known as \acrfull{AuI}. With a HITL, greater control over potential health harms can be achieved, and perhaps acceptance on both sides, patients and mHealth professionals, can be increased. However, the benefits of constant availability and speed are significantly reduced compared to solely AI interaction. Measuring the acceptance in both groups -- the population and specialists -- is relevant to find out if this is a adequate solution to achieve the quadruple aim.

\section{Relevance}
Different social, relational interactions or events can trigger mHealth problems such as depression, anxiety and psychological disorder. Some examples are loss of a job, partner, friend, or financial problems, poverty, illness as well as social exclusion and dysfunction in the family \cite[cf. p. 2]{Kalam2024}. Additionally to personal conflicts, genetics, political debates, war and crises like pandemics can influence the healthiness. Beyond the factors already mentioned, the increasing temperature can have impact on mHealth. As part of the TempPredict Study Mason et al. figure out that body temperature alterations are potentially relevant factors in depression etiology \cite[cf. p. 1]{Mason2024}. Higher levels of depressive symptoms were associated with higher body temperatures during time awake \cite[cf. p. 9]{Mason2024}. Bundo et al. took a closer look on ambient temperature and their effect on mood. For the general population increasing temperature seems to have a positive effect on their mood. However, individuals with certain psychiatric disorders, such as anxiety, depression, and schizophrenia, may exhibit altered responses to heat, which may explain their increased morbidity when exposed to high temperatures \cite[cf. p. 1 -- 2]{Bundo2023}. With regard to climate change, the researchers emphasize the need to take special care of vulnerable people in order to protect their lives. Similar findings could be identified in the review article of Liu et al. \cite[cf. p. 1 -- 2]{Li2023}. These hypothesis underline the importance of taking care of mHealth, cause not only personal mood and life situation can have an effect on individuals mHealth, also the climate change and the associated increase of temperature.

It is not only the number of people using mHealth services that is increasing, but pressure on healthcare professionals as well. According to current statistics, an increasing number of doctors, including psychologists, are reaching retirement age in Germany \cite[cf. p. 2]{DGPPN2023}. To optimize the performance of the health system, the so called 'quadruple aim' was introduced in 2014 as a compass \cite[cf. p. 573]{Bodenheimer2014}. This framework seeks to enhance patient experience, improve population health, reduce costs, and improve the work-life of healthcare providers.

The well-being of professionals is important for the treatment of their patients. Less job satisfaction and impaired health can have an impact on health care quality \cite[cf. p. 1]{Werdecker2021}. So called 'illegitimate tasks' like administration and documentation, can have an high impact as stressors on job satisfaction. In their online-survey with 548 physicians in Germany participating, Werdecker et al. could illustrate that illegitimate tasks are negatively associated with life and job satisfaction as well as it seems to increase burnout symptoms \cite[cf. p. 9]{Werdecker2021}. In addition they found in their results that burnout was not as severe among German general practitioners as among physicians in total \cite[cf. p. 10]{Werdecker2021}. Organizational characteristics, policies, and work culture influence the physicians' well-being, hence specific strategies may support an environment where physicians thrive. Tawfik et al. recommend that reducing documentation burden and improving practice efficiency may help to balance job demands and resources \cite[cf. p. 11 -- 12]{Tawfik2019}. In their article they summarized four organizational strategies to drive progress, including: developing physician-leaders, cultivating community and organizational culture, improving practice efficiency and optimizing administrative policies \cite[cf. p. 12]{Tawfik2019}. Zumbrunn et al. conduct a survey in Switzerland with 472 general internal medicine residents \cite[cf. p. 1]{Zumbrunn2020}. Approximately 20\,\% of the participants appear to have a reduced well-being and many show signs of distress or state career choice regrets \cite[cf. p. 1]{Zumbrunn2020}. In contrast to the results from Tawfik et al. they did not find an association between well-being and time spent on administrative tasks \cite[cf. p. 6]{Zumbrunn2020}. As literature shows, each physician, counsellor and therapists is a human being and each single person is dealing differently with well-being and workload.

One potential solution to achieve the quadruple aim and address the current dilemma is to utilize AI that is responsible, evidence-based, free from bias, as well as designed and deployed to promote equity \cite[cf. p. 1]{Crigger2022}. However, as such solutions do not yet exist, and AI in healthcare remains highly controversial, a step forward could be the use of AuI. This approach offers more control and safety because humans are still involved as experts in the process. The use of CAs is not intended to replace health professionals but rather to help patients navigate the health system more effectively and empower them to manage their own health in a better way. This approach should save time and money for both patients and health professionals \cite[cf. p. 196]{Dingler2021}. To figure out whether this could be a viable approach, population and professionals have to be asked about their current experience and acceptance to use this technology and intervention in the field of mHealth.

\section{Research Gap}
The use of AI in health related topics was reignited with the introduction of publicly accessible and easy to handle models like ChatGPT. Even though this model is not specifically trained for health related topics, a study from February 2023 shows, that people are inclined to use it for self-diagnosis \cite[cf. p. 1]{Shahsavar2023}. Digital mHealth interventions (\acrshort{DMHI}) are easy and sometimes accessible for free. In Germany some selected health applications (\acrshort{DiGA}) -- evaluated and paid by health insurance -- can be accessed by recipe of the physician. The acceptance to use them continuously is still mixed on both sides, population and professionals. For a couple of years the use of telehealth has been simplified, the acceptance on both sides is high. But this synchronous process -- especially for video conferencing -- still needs solid internet connection, which is sometimes hard to establish in rural areas. In addition, there are still people who do not want to share their worries, feelings or fears with a human professional due to fear of stigmatisation.

In contrast to this, AI-based CAs are easy to access, already used for study, school, work, and private interests. This raises the question of whether AI is already being used to treat psychological conditions, for example to alleviate grief. Some studies mention the lack of sufficient studies assessing medical professionals perspective on the advantages, disadvantages and use of AI in medicine \cite[cf. p. 2]{Pedro2023}. Prevention and healing strategies not solely base on evidence but also require acceptance to use or offer them. To the current knowledge, no study exists that examines the experience and acceptance of AI-based CAs in the field of mHealth, considering both involved roles -- population and professionals in Germany. In addition less is known about the approval involving this technology as blended therapy.

\section{Hypotheses and Objectives}
Based on the available literature, the mentioned relevance and gaps, the already existing, available and with media attention promoted AI-based CAs since November 2022, the following hypotheses are put forward:
\begin{itemize}
    \item H1: Although AI-based CAs are frequently used in general, most of the people and specialists don't use them for mHealth or patient treatment.
    \item H2: The belief that AI can help with mild mental disorders is higher for the population than for specialists.
    \item H3: Drivers for the acceptance of AI-based CAs for mHealth are: male gender, younger age, increased anxiety, increased psychological distress, increased workload, less job experience, frequency of use or offer of telemedicine and experience with AI.
    \item H4: Specialists will moderately accept AI-based CAs as additional expert during counselling, treatment, or diagnosis.
    \item H5: Specialists will not accept AI-based CAs as companion for the patients during counselling, treatment, or diagnosis.
    \item H6: The population will highly accept AI-based CAs as additional experts used by professionals for counselling, treatment or diagnosis.
    \item H7: The degree of acceptance using AI-based CAs during counselling, treatment, or diagnosis will decrease with less specialist involvement, among German population.
\end{itemize}

The aim of this study is to analyse the current frequency of use of AI-based CAs (e.\,g. ChatGPT, Siri, Gemini) in general and in the field of mHealth in Germany, the perceived benefits and obstacles as well as the level of acceptance. In addition, the possible use as AuI in counselling, diagnosis or treatment will be examined for acceptance. To this end, two anonymous, quantitative online surveys will be conducted among people aged 18 and over and professionals from the fields of therapy, psychology or counselling in order to include both roles -- the general public and professionals. Answers will be given to the following questions:

\begin{itemize}
    \item RQ1: How frequently are AI-based CAs already being used for mHealth by the general population compared to mHealth specialists in Germany?
    \item RQ2: How does the degree of acceptance of using AI in mHealth compare between the general population and specialists in Germany?
    \item RQ3: To what extent is the use of AuI in counselling, diagnosis, and treatment acceptable to both -- the general population and specialists in Germany?
\end{itemize}

All questions should provide valuable insights into the current and potential use of CAs in the context of mHealth by understanding the views of the general population and professionals. With these findings, it should be possible to develop CAs for facilitating interactions between patients and healthcare professionals to reach the quadruple aim, to provide help during patient care and relieve professionals.

%% file: content/literatur-review.tex
\chapter{Literature Review}
The number of children, students, young adults and employees absent from school, university or work due to psychological problems or stress is constantly increasing. Although not a new phenomenon, mental disorders are widespread and result in various impairments and participation limitations, as already mentioned in 1998 \cite[cf. p. 77]{Jacobi2016}. Patients often avoid seeking the help of a specialist due to fears of stigma, travel time, or absence from family or work \cite[cf. p. 5]{Gulliver2010}. To address this issue, various technologies and approaches have been introduced. This literature review section presents a comprehensive examination of existing research and scholarly discourse on this topic. It provides an overview of teletherapy, digital health applications, and blended therapy, focusing on their acceptance in Germany. Additionally, it explores the concept of AI as a counsellor or therapist and the potential of the HITL approach

\section{Teletherapy in Germany}
Telepsychiatry and telehealth are defined as the use of communication or information technologies to provide or support healthcare at a distance \cite[cf. p. 230]{Brunt2023}. These can be categorized into synchronous and asynchronous types, such as video calls, phone calls, email, and messaging \cite[cf. p. 230]{Brunt2023}.

\subsection{Historical Context and Growth}
As early as the 1970s, the first remote video calls were provided in the USA to patients in rural areas \cite[cf. p. 230]{Brunt2023}. Specialists were scarce, but patients could access specialists in major cities by visiting a rural hospital. However, due to extremely slow data transfer rates, natural conversation was not possible \cite[cf. p. 230 -- 231]{Brunt2023}. Several decades later, with the rise of the internet and mobile phone availability, telehealth began to gain significant interest.

Since 2018, patients in Germany have been permitted to receive exclusive remote treatment -- by telephone or online -- with their doctors in individual cases \cite[cf. p. 1003]{Klinge2021}. During the COVID-19 pandemic, this offering increased dramatically. Teletherapy is currently a promising and growing way to deliver mHealth services, although it remains underused \cite[cf. p. 1]{Carlo2021}). However, technical issues and low bandwidth, especially in rural areas, still limit access possibilities. In these regions, remote treatment could be a viable alternative for patients due to the greater distances to the nearest doctor \cite[cf. p. 1006]{Klinge2021}. Additionally, patients need at least some knowledge of technology, such as computer and internet access, as well as a low frustration tolerance when connections or technology fail. Private space, where patients can interact freely and undisturbed, is necessary for treatment success.

\subsection{Assessing the Impact and Acceptance}
In a focus group interview, Klinge et al. identify advantages and disadvantages that medical experts encountered with remote treatment \cite[cf. p. 1003]{Klinge2021}. Advantages include avoiding infection risks -- during a pandemic or flu epidemic -- better accessibility for patients with limited mobility, more attractive part-time opportunities for professionals, and reduced strain on reception desks and waiting rooms \cite[cf. p. 1003]{Klinge2021}. Disadvantages include limited diagnostic capabilities and increased organizational and technical effort \cite[cf. p. 1003]{Klinge2021}. Patients furthermore note faster and easier access to specialists for initial therapeutic proposals \cite[cf. p. 1004]{Klinge2021}. However, they highlight the limited possibilities for accurate diagnosis. Overall, acceptance of remote treatment decreased with increasing age and lower digital affinity \cite[cf. p. 1004]{Klinge2021}.

In their research article, Hajesmaeel-Gohari et al. evaluate the most used questionnaires for telemedicine services \cite[cf. p. 1]{Hajesmaeel-Gohari2021}. They found out that attention to user needs, acceptance, satisfaction, usability and implementation processes are necessary to optimize telemedicine efforts \cite[cf. p. 1]{Hajesmaeel-Gohari2021}. From February 2022 to February 2023, Nurtsch et al. conduct a cross-sectional online-based survey across 350 cancer patients in Germany to determine the acceptance of using video consultations \cite[cf. p. 1]{Nurtsch2024}. About 84\,\% showed high or moderate acceptance, whereas 95.7\,\% had no experience in using video consultations. Female gender, younger age, and low internet anxiety are some of the variables that demonstrate higher acceptance \cite[cf. p. 7 -- 8]{Nurtsch2024}).

In contrast to other research on telehealth use among patients in Germany, such as studies on obesity \cite[cf. p. 1]{Rentrop2022} or general practice during COVID-19 \cite[cf. p. 1]{Esber2023}, mental illness had no effect on the acceptance of video consultations. In the context of psychotherapy, a strong association between mental illness and acceptance may be explained by the fear of stigma, which can be reduced using DMHI \cite[cf. p. 7]{Nurtsch2024}. As shown by Neumann et al. in a systematic review, five of nine studies found that people with higher symptom severity had a lower use rate of teletherapy services during the COVID-19 pandemic \cite[cf. p. 23, 30]{Neumann2023}. In a survey by Meininger et al. during the first quarter of 2020, 73\,\% of patients -- children and adolescents -- switched from \acrfull{FtF} \acrfull{CBT} to teletherapy. Additionally, 66\,\% of the patients' parents and 53\,\% of the therapists intended to use teletherapy in the future \cite[cf. p. 1]{Meininger2022}. Abuyadek et al. quantified the acceptance of teletherapy services among both beneficiaries and providers in a systematic review. Beneficiaries showed a high acceptability rate of 71\,\%, while providers showed a rate of 66\,\%. The review did not distinguish between different technologies, with services delivered through mobile applications, gaming, teleconferencing, video calls, and web-based programs \cite[cf. p. 1, 11]{Abuyadek2024}. This highlights the broad terminology encompassed by telehealth services.

\section{Digital Mental Health Interventions in Germany}
An additional approach to address the increasing prevalence of psychological disorders and the need for prevention is the use of DMHI. Since the advent of the internet and mobile applications, DMHI have offered the potential to improve access to mHealth treatment. These interventions are typically provided via web-based or mobile apps and can be delivered in various formats, including self-guided, with the support of a professional, or blended with FtF treatment \cite[cf. p. 92]{Lattie2022}.

\subsection{Evaluation and Innovation}
In Germany, the 'PraxisBarometer Digitalisierung' is the only representative survey on the digitalization of contract medical and psychotherapeutic professionals \cite[cf. p. 1] {KBV_Praxisbarometer2024}. In their report from 2023, they advocate for digitalization that offers real impact, quality and ease of use for professionals \cite[cf. p. 2] {KBV_Praxisbarometer2024}. Research comparing tele-CBT and internet-based-CBT with FtF-CBT shows promising results, as both digital formats appear to be as effective as traditional treatment \cite[cf. p. 88]{Lattie2022}. In addition to this Lattie et al. give an overview of different digital mHealth services and include recommendations to make them more accessible \cite[cf. p. 87,88]{Lattie2022}. However, the growing number of possibilities and the integration of different technologies must be thoroughly evaluated. Factors such as access, usability, risks, benefits, privacy and data management need to be considered. To support these ethical considerations, Nebeker et al. have developed the Digital Health Checklist \cite[cf. p. 1004]{Nebeker2021}.

Since December 2019, physicians and psychotherapists in Germany have been able to prescribe 'Digital Health Applications' (DiGA), which are covered by the general health insurance \cite[cf. p. 1]{Sauermann2022}. Germany is one of the first innovative players in Europe where the 'Digital Healthcare Act' \acrfull{DVG} enables patients with a diagnosis to access proven and evaluated mobile or web applications. The Federal Institute for Drugs and Medical Devices (\acrshort{BfArM}) evaluates these products and lists them on their website. As of July 2024, there are 64 applications available, with 27 focussing on psychology topics. Of these, 17 are permanently included, nine are provisionally available, and one has been cancelled \cite{BfArM2024}.

\subsection{Awareness and Acceptance of DiGA}
Acceptance and confidence that technological interventions will positively impact patients' well-being are particularly crucial for practitioners. Clinical evidence increases their acceptance, while a lack of evidence can be a major obstacle to prescribing. Additionally, about 40\,\% of physicians cite their limited familiarity with digital innovation as a factor that negatively influences acceptance \cite [cf. p. 5]{Schlieter2024}.

According to a representative study, conducted two years after the integration of DiGA in 2022, more information and the ability to test individual applications are required to increase professional acceptance \cite[cf. p. 1]{Gesundheit2022}. The popularity of DiGA among professionals is increasing. Of the 2,247 physicians surveyed, 36.9\,\% had already used DiGA, 13.9\,\% were willing to use them in the future, 34.7\,\% knew about them but decided not to use them and 14.5\,\% were unaware of DiGA. Additionally, 33.6\,\% of the 2,238 physicians surveyed had already prescribed a DiGA to patients \cite[cf. p. 6 -- 7]{Gesundheit2022}. With 67,8\,\% the topic psyche is the most rated use case, but for treatment of depression or suicidal thoughts, 27.7\,\% of the physicians rate DiGA as ineffective or even counterproductive \cite[cf. p. 9, 13 -- 14]{Gesundheit2022}. Clinical evidence and changing patient needs are the two most important criteria rated for increasing physician acceptance to prescribe DiGA. Conversely, the costs of current regulations (e.\,g. prescription, billing) and data protection issues are the most relevant factors cited for not using or prescribing them \cite[cf. p. 10 -- 11]{Gesundheit2022}.

Even if an application can improve mHealth or general health, acceptance and usability are absolutely required for the success of medical devices and applications. Current surveys show that the willingness to use DiGAs is greatest among doctors in the mHealth sector, and that the available DiGAs can apparently meet existing needs in a meaningful way. The high level of acceptance and frequent use suggests that the existing non-digital therapeutic options may not always meet the expectations of patients and therapists  \cite[cf. p. 4]{Gesundheit2022}.

An additional online survey conducted from March to June 2021 by Frey et al. gathered insights from 150 therapists in the fields of physical therapy, occupational therapy and speech-language pathology regarding the usage and acceptance of DiGAs \cite[cf. p. 1]{Frey2022}. Potential benefits of using DiGAs are seen particularly in the quality improvement of therapy, increased sustainability of therapy and promotion of patients’ health literacy. However, therapists identify barriers such as the lack of technical infrastructure and patients’ insufficient digital health literacy \cite[cf. p. 7]{Frey2022}.

According to the 'DiGA-Report' from the beginning in 2022 until September 2023, about 370,000 DiGA activation codes have been used by thousands of patients \cite[cf. p. 14]{Meskendahl2023}. Additionally, according to one of the three largest statutory health insurance providers in Germany, the \acrfull{TK}, by the end of June 2023, around 86,000 activation codes had been issued to just under 69,000 of the total 11.1 million TK-insured people, which corresponds to just under 0.6\,\% \cite[cf. p. 4]{TK2023}. Most DiGAs are used for mHealth treatment, with about 31\,\% \cite[cf. p. 6]{TK2023}. According to a questionnaire administered to 1,000 German citizen in 2023, 57\,\% of the participants, especially older individuals, were unaware of DiGA \cite[cf. p. 23]{Deloitte2023}.

Nonetheless, the number of digital applications related to mHealth, which are not listed as DiGA, is constantly increasing and being used by the general population. Lack of motivation, individuality, and treatment success can be reasons to discontinue self-guided and unaccompanied treatment. To address this, a combination of DMHI usage and professional companionship should be employed. The integrated and complementary use of traditional and digital tools presents an opportunity to more closely involve patients in therapy and to discover new starting points that previous forms of therapy could only insufficiently address. Especially vulnerable groups would benefit from the combination of innovative technology and human empathy \cite[cf. p. 21]{Meskendahl2023}, which is in addition requested by the professionals \cite[cf. p. 22]{Gesundheit2022}.

\section{Blended Therapy -- The Combination of Digital and Traditional Methods}
Blended psychotherapy, blended care or 'technology-supported care' refers to a combination of FtF and online intervention \cite[cf. p. 447]{Bielinski2021}. A distinction can be made between transforming blends and complementary blends \cite[cf. p. 448 -- 449]{Bielinski2021}. 

\subsection{Key Considerations for Effective Blended Therapy}
Online interventions can be integrated into therapy sessions, allowing patients to first have a FtF intervention and then deepen their understanding and practice using an online tool. The advantage of this approach is that therapists have more time for more patients and can monitor how patients are practicing. However, further research is still necessary to fully understand and optimize this approach. At the overall treatment level, similar to purely online interventions, high drop-out rates and lack of use of the online elements appear to be problematic for blended therapy. To successfully implement blended therapy, the following items are necessary \cite[cf. p. 451]{Bielinski2021}:
\begin{itemize}
    \item Individualization and Adaptability: DMHIs should be tailored based on the patient's ability, preference, severity, and type of problems.
    \item Adjustable FtF to Online Ratio: Professionals should be able to adjust the ratio between FtF and online interventions.
    \item Adequate Information and Communication: There should be thorough information and discussions between the patient and the therapist regarding the treatment operationalization.
    \item Professional Training: More training for professionals is necessary to effectively implement blended therapy.
    \item Advanced Technical Solutions: Highly developed and financially supported technical solutions are required.
\end{itemize}

Like most technologies and innovations, there should be closer checks to ensure the intervention is suitable for both -- the patient and the professional. To facilitate the setup of blended care, a checklist was designed in 2016 by Wentzel et al. in their viewpoint to identify variables that play a role in the desirable reach, use and adherence of online therapy \cite[cf. p. 1]{Wentzel2016}. Additionally, evidence-based and regulated distribution, similar to DiGA, would be necessary to avoid a unmanageable availability of electronic health applications.

\subsection{Advantages and Challenges of Blended Therapy to Enhance Care}
Schuster et al. demonstrated that in countries with low DMHI levels, such as Germany and Austria, the positive attitude towards blended psychotherapy is lower than in countries with high DMHI levels, like Sweden \cite[cf. p. 1]{Schuster2020}. In their online survey conducted in 2019 in Germany and Sweden, they were able to include 165 therapists (51 from Sweden). Although the effects in Austria and Germany were not as strong, they found out that therapists have a more positive attitude towards blended psychotherapy than towards therapies conducted entirely online \cite[cf. p. 1]{Schuster2020}.

Through discussions with a focus group and individual interviews with therapists, Wentzel et al. identified that very few criteria make people fully unfit to benefit from blended mHealth treatment \cite[cf. p. 3]{Wentzel2016}. They found out, that the resources of internet and computer access are absolutely important, as well as the possibility to have a safe and secure space at home for the intervention. Additionally, sufficient cognitive skills, including internet and technical access, as well as self-organization and structure skills, are needed \cite[cf. p. 3]{Wentzel2016}. To facilitate the dialogue between patient and therapist, Wentzel et al. developed a shared decision-making instrument called 'Fit for Blended Care' to jointly decide on the configuration of the blended treatment \cite[cf. p. 3]{Wentzel2016}. This approach helps to consider personal needs and preferences for an optimally personalized treatment, which enhances the self-management of patients and relieves experts \cite[cf. p. 5]{Wentzel2016}.

In an online survey conducted between 2017 and 2018, a total of $1,376$ first-year students from two German universities were asked about their preferences for mHealth care and the acceptance of different therapy delivery modes -- in-person, group, self-help DMHI, or a combination of DMHI and in-person as blended therapy \cite[cf. p. 1]{Kählke2024}. The highest-rated delivery modes were in-person off-campus services (76\,\%), followed by in-person on-campus services (71\,\%). About half of the students (44\,\% -- 48\,\%) were interested in DMHI, but even more accepted a blended delivery mode combining in-person and digital services (57\,\%). Group therapy on- and off-campus was the least accepted delivery mode, with 31\,\% -- 36\,\% acceptance. Students with indicated mental disorders had a significantly higher acceptance of using DMHI \cite[cf. p. 5]{Kählke2024}.

Urech et al. conducted semi-structured interviews with 15 patients suffering from major depression who underwent treatment for 18 weeks with blended CBT \cite[cf. p. 986]{Urech2019}. During their research, they identified 18 advantages and 15 disadvantages of this treatment, which should be considered for further implementations \cite[cf. p. 986]{Urech2019}. On one hand, FtF contact with the therapist was necessary to establish a trustworthy therapeutic relationship and to prevent potential frustration with the online intervention \cite[cf. p. 995]{Urech2019}. On the other hand, the work with the DMHI provided constant availability of care, improved self-efficacy, and empowered patients to work on their disorder independently \cite[cf. p. 995]{Urech2019}. Low motivation or energy to work on the online program, as well as the lack of individualization of the content, were mentioned as disadvantages by the patients \cite[cf. p. 995]{Urech2019}. Additionally, the missing possibilities for the therapist to see the activities of the patient in the DMHI led to its usage as an add-on rather than an integrated component \cite[p. 995]{Urech2019}.

Even before the latest hype of AI, literature shows that blended care treatment can offer synchronous or asynchronous guidance, apply personal or automated feedback and support, with promising effects \cite[cf. p. 2]{Wentzel2016}. Not only to prevent relapses, such as drug consumption, psychological disorders or eating disorders, but also to promote general well-being, DMHIs should be integrated during the local therapy or intervention and not afterwards. For a more successful and sustainable treatment, integration instead of subsequent treatment would be necessary.

\section{AI as Counsellor or Therapist -- Autonomous Interventions}
In response to the need for more individualized and effective mHealth services, the role of digital technology in improving access, engagement, and outcomes of therapeutic treatment is becoming increasingly important. This has led to the development of a wide range of health technologies and applications \cite[cf. p. 2]{Thieme2020}. For several years now, the additional use of AI and technologies such as \acrfull{NLP} or the integration of \acrfull{LLM} in the treatment of mental disorders through telehealth or digital health has been a subject of controversial debate \cite{Benecke2024, Kalam2024, Wong2024}.

\subsection{The Evolution and Integration of Chatbots in Mental Health Care}
The concept of using a chatbot for psychiatric interviews is not all the new. ELIZA, the chatbot developed by Josef Weizenbaum back in the mid-1960s, is not that recent \cite[cf. p. 36]{Weizenbaum1966}. It is the first notable chatbot capable of conducting certain types of natural language conversations between humans and computers. Since then, technology has advanced significantly, and the use of smart speakers, chatbots, and virtual assistants has become commonplace \cite[cf. p. 1]{Carolus2021}. Chatbots based on NLP that serve as therapists, counsellors, or coaches are already available on the market and are being utilized. Woebot is a popular example of a digital chatbot, using NLP and rule-based AI to offer help for depression and anxiety \cite[cf. p. 1]{Fitzpatrick2017}. It is only a matter of time before there are offers with generative AI that conduct versatile and well-defined dialogues.

In November 2022, ChatGPT was introduced to the public by the company OpenAI as one of the first and most powerful generative models \cite[cf. p. 122]{Ray2023}. Generative AI can create new content like text and images based on structures and patterns from existing data used in the training process. Although it is not designed to answer questions related to healthcare or mHealth, users are willing to use it for self-diagnosis \cite[cf. p. 1]{Shahsavar2023}. ChatGPT serves as a first source of information on diagnosis and treatment options that is always available, friendly, and never tired. Some would say it's more of a lazy Google or internet search, but it is accessible to anyone with internet access, and interaction can sometimes feel remarkably similar to interacting with a human \cite[cf. p. 2]{Wong2024}.

In 2024, the creators of Woebot developed principles for integrating LLMs into their platform \cite{WoebotHealth2023}. On one side, this technology can be used to interpret user input and route it to content written by a human. On the other side, LLMs might be used to generate responses shown directly to users. Only in an \acrfull{IRB} regulated study a generative AI is used in Woebot, somewhat as a 'Science in the loop' approach \cite{WoebotHealth2023}. The scientists state that the efficacy of therapy depends on the user's engagement, and LLMs might help to motivate and select the right technique for the right person at the right time \cite{WoebotHealth2023}.

With the growing acceptance of new technologies, such as generative AI in the health sector for self-diagnosis \cite[cf. p. 1]{Shahsavar2023}, it is plausible that CAs have already entered the realm of mHealth. Not only the general population and patients show interest in using this innovative technology. Furthermore, a study from Saudi Arabia on the perceptions and expectations of healthcare sector employees indicates a keen interest in using ChatGPT \cite[cf. p. 1]{Temsah2023}. Similar data is not yet available for Germany, but it can be assumed that this technology is or will be used here too due to the widespread use of ChatGPT.

\subsection{Balancing the Benefits and Risks of AI in Mental Health Care}
Considering the growing number of patients, it is crucial to provide as much relief as possible to healthcare professionals. With the use of AI in healthcare, there is a desire and hope to achieve the quadruple aim: enhance patient experience, improve population health, reduce costs and improve the working lives of healthcare providers. To achieve this goal, it is absolutely essential to use responsible, evidence-based and unbiased AI that is designed and deployed in a way that promotes equity \cite[cf. p. 1]{Crigger2022}. Despite ethical concerns and biases, the practical importance and value of ChatGPT in psychotherapy should not be underestimated. For psychotherapists, ChatGPT can be a useful tool for getting a second opinion on diagnosis, suggestions on appropriate treatment techniques, learning about other psychotherapeutic approaches, and obtaining a second interpretation of patient material \cite[cf. p. 6]{Raile2024}.

In 2019, Pedro et al. conducted an online survey of 1,013 medical doctors to assess their perspectives on AI in medicine in Portugal \cite[cf. p. 1]{Pedro2023}. They focused on the impact of AI on healthcare quality through the extraction and processing of health data, the delegation of clinical procedures to AI tools, the perception of AI's impact on clinical practice and the advantages and disadvantages of using AI in clinical practice \cite[cf. p. 1]{Pedro2023}. Most respondents were optimistic about using AI and were convinced that AI should be part of medical education. More than half of those who took part in the survey agreed or totally agreed that AI could simplify or facilitate patient care \cite[cf. p. 1, 10]{Pedro2023}. Additionally, they view AI as a support instrument rather than a substitute, capable of reducing medical errors, fostering dialogue and proximity between physicians and patients \cite[cf. p. 10]{Pedro2023}. This approach can help to provide more individualized care and planning more appropriate preventive therapeutic measures, especially for isolated populations. Risks and limitations included the potential dehumanization of healthcare, decreased ability to improvise when necessary, and risks to privacy and confidentiality \cite[cf. p. 9, 10]{Pedro2023}. Overall, older and more experienced doctors displayed greater optimism and openness towards incorporating AI in medicine \cite[cf. p. 13 -- 15, 17]{Pedro2023}.

Whether ChatGPT or similar LLMs can serve as a coach, therapist, or counsellor is controversial, but they are already available. When using AI-based CAs for school, university, work, or leisure, the question arises: Are they already or will they be used as counsellors or therapists for dealing with anxiety, sorrow, or psychological disorders? This could eliminate the need to search for a therapist or choose the right DMHI, wait for therapeutic programs, incur additional costs or require installations. In July 2023 Dergaa et al. evaluated the effectiveness, safety, and reliability of ChatGPT in assisting patients in imaginary scenarios \cite[cf. p. 1]{Dergaa2023}. They found out that in more complex cases, ChatGPT generates and provides inappropriate and even dangerous recommendations \cite[cf. p. 1]{Dergaa2023}. Therefore the authors underscore the inability of ChatGPT to ask and interact with the patient to gather more relevant information for creating a diagnosis and managing the patient's clinical condition \cite[cf. p. 1]{Dergaa2023}. One additional point of criticism is that this AI only provides general information and diagnosis, primarily covering CBT, mindfulness-based therapy and skills training \cite[cf. p. 2]{Mayer2022}. But it becomes evident that it does not include topics such as humanistic, systemic, interpersonal and psycho dynamic therapies \cite[cf. p. 2]{Mayer2022}. Furthermore, the lack of non-verbal signals and their interpretation are very important aspects of therapeutic processes, as well as the time between therapy sessions, where the patient has the opportunity to practice and reflect.

In their perspective, Kalam et al. ask whether ChatGPT and mHealth can be friends or foes \cite[cf. p. 1]{Kalam2024}. They mention several criteria that can cause depression or mHealth issues when using ChatGPT \cite[cf. p. 1]{Kalam2024}. Because ChatGPT can create text in a human-like form, using it as a replacement for human interaction, discussions, social communication, and involvement in human-to-human interaction may increase social isolation \cite[cf. p. 2]{Kalam2024}. 
Although there are no official job replacements or reductions due to AI yet, the fear of losing one's job or not finding a new one can lead to a crisis \cite[cf. p. 2]{Kalam2024}. ChatGPT is changing educational progress and productivity, for students, the future may seem bleak if they fall behind in knowledge \cite[cf. p. 2]{Kalam2024}. Additionally, with the current lack of ability to contradict the human counterpart, ChatGPT and other CAs confirm whatever the user wants to hear. This behaviour can lead to risks in mortality and suicidal thoughts \cite{Walker2023}. An increasing and continuing dependency on these models may result in the construction of an information cocoon, restricting users to a limited range of generated material \cite[cf. p. 2]{Kalam2024}. During online searches, users still have to decide which reference, article or website to choose. With an insufficient request to ChatGPT, which can easily happen, users may take the first output and believe what ChatGPT advises or tells them. As an additional risk, it should be mentioned that these technologies may be offered by companies that moreover have to take into account the politics of their country and the interests of other corporations and investors. This may lead to AI alignment, dictating how the CA should react in a political or social direction. Users interacting with these technologies should keep this in mind and better double-check or think critically about the output, which is easy to access but hard to deal with. Additionally, with each interaction, users should think twice about which data and how much privacy to provide to these companies. Educating people to limit the misuse of AI technologies and taking the initiative in advancing knowledge through organizational workshops, counselling services, and training is necessary to reduce the risks and still take advantage of these technologies \cite[cf. p. 3]{Kalam2024}.

An additional limitation is the missing explainability of the decisions the AI creates, making the conversation less trustworthy \cite[p. 1]{Sarkar2023}. Furthermore it must be asked, to which extent these large-scale LLMs like GPT, Gemini, Mistral, or others will be the best fit for health or mHealth-related topics in the future. Perhaps more specialized and trained models on these topics will gain more relevance in the future.

\subsection{Human-AI Interaction -- Satisfaction and Ethical Considerations}
Some studies have reported a high level of user satisfaction with chatbot interactions, with participants expressing a willingness to use such technology for health-related issues in the future \cite{Boucher2021, Lucas2014}. Additionally, some individuals appear to prefer interacting with a chatbot over mHealth professionals, which could potentially encourage those who would not typically seek therapy to do so \cite[cf. p. 40]{Boucher2021}. In clinical interviews with virtual humans, Lucas et al. found out that patients, who believed they were interacting with a computer, reported lower fear of self-disclosure, lower impression management, displayed their sadness more intensely and were rated by observers as more willing to disclose, compared to participants who believed the virtual human was controlled by a human \cite[cf. p. 94]{Lucas2014}. These results suggest that automated virtual humans can help overcome a significant barrier to obtaining truthful patient information \cite[cf. p. 94]{Lucas2014}. On the question why people give more honest response to a computer, the authors were given the answer, that computer-administered assessment formats allow for a "sense of invulnerability to criticism, an illusion of privacy, the impression that responses disappear into the computer" \cite[p. 95]{Lucas2014}. On the other hand, there are ethical concerns that AI may depersonalize medicine, eliminate the need for human interaction, and reduce investment in patient-centred care \cite[cf. p. 17]{Tang2023}.

In a study following a quantitative experimental design, 318 outpatients conducted a psychiatric interview with a virtual medical agent \cite[cf. p. 1]{Philip2020}. In general, the agent was accepted and perceived as trustworthy by the participants and especially older and less-educated patients accepted this virtual agent more than younger and well-educated ones \cite[cf. p. 1]{Philip2020}. Acceptance and trust are in the authors opinion the two major dimensions that consolidate engagement with health technologies. But also age, gender, education, and health conditions influence engagement, acceptance, and trust in technologies, which should be the focus of further studies \cite[cf. p. 1]{Philip2020}.

To understand the behaviour of AI-driven systems, there is a need to control their actions, exploit their benefits, and minimize their damage in the context humans want to use it \cite[cf. p. 477]{Rahwan2019}. Understanding the differences in Human-AI interaction requires considering guidelines to create more human-centred AI-infused applications \cite[cf. p. 12]{Amershi2019}. Müller et al. have developed ten commandments as practical guidelines for AI in medicine to address questions regarding legal responsibility \cite[cf. p. 121]{mueller2021}. These were agreed upon in an online survey of 121 experts (50.4\,\% female), of whom 47\,\% were computer experts and 33\,\% were medical doctors \cite[cf. p. 121]{mueller2021}. The participants confirmed the relevance and importance of these ten commandments \cite[cf. p. 121]{mueller2021}. Among other things, they emphasize the need for recognizability regarding which part of the communication is performed by a CA and that humans should never be deceived by these systems \cite[cf. p. 121]{mueller2021}. Additionally, the responsibility for an AI decision, action, or communicative process must be explainable and transparent, and it must be taken by a competent physical or legal person \cite[cf. p. 121]{mueller2021}. Therefore, if AI "is part of the decision-making process, the patient needs to be appropriately informed. This is achieved not by the phrase 'machine decision,' but, rather, by the specification, an AI-supported decision, diagnostic finding, or treatment proposal" \cite[p. 121]{mueller2021}.

Human-AI interactions in the mHealth environment involve many different stakeholders, including patients, their related persons, clinicians, and developers. Therefore, ethical considerations and guidelines of all parties must be addressed. In their systematic review of published empirical studies on medical AI ethics, Tang et al. aim to map the main approaches, findings and limitations of existing scholarship to inform future practice considerations \cite[cf. p. 1]{Tang2023}. They found out that, generally, clinicians report more ethical concerns related to the use of AI in health than patients \cite[cf. p. 1]{Tang2023}. Nevertheless, the primary concern shared among patients and clinicians regarding the use of AI in health-related environments is the potential reduction of human interaction and trustworthy healthcare communication \cite[cf. p. 19]{Tang2023}.

\section{Human in the Loop -- Securing of Human Oversight and Intervention}
However, AI has its weaknesses, as it is inherently a 'black box' that lacks transparency. The fact that it is unclear how the AI makes decisions and interpretations makes it challenging to identify and quantify errors in the results and to rectify and justify any problems that arise \cite[cf. p. 13765]{Yau2021}. The European ethics guidelines report states that trustworthy AI should be lawful, ethical and robust \cite{EU_EthicGuidlines2019}. It should be based on human-centred design and adhere to ethical principles throughout its life cycle: respect for human autonomy, avoidance of harm, fairness and explainability \cite{EU_EthicGuidlines2019}.

\subsection{Balancing Human and Artificial Intelligence}
Rather than engaging in heated debates about the pros and cons of AI usage, it is essential to explore the shades of gray. One approach to make AI more transparent, explainable and less harmful is to keep a human in the loop as an expert. AuI is one such method, which integrates \acrfull{HI} and AI to capitalize on the strengths of both and mitigate their weaknesses \cite[cf. p. 13764]{Yau2021}. While AI is machine-centred, AuI is a human-centred intelligence system that enables people to expand their existing abilities, creativity and skills or even create new ones. Instead of mimicking HI, AuI should enhance it \cite[cf. p. 773]{Sadiku2021}.

Two methods are commonly used to explain \acrfull{ML} decisions, where humans remain in the loop: 1) Humans having the final say and making the decision; and 2) explaining the data sources responsible for the final decision \cite[cf. p. 246]{Zanzotto2019}. However, involving a person as an expert means losing the advantages of being available at any time for everyone, as well as limiting large-scale scalability and cost reduction \cite[cf. p. 97]{Lattie2022}. Decisions can turn out worse due to unknown problems, biased information, and judgments \cite[cf. p. 2]{Cohen2023}. This underlines, that for each specific use case it is necessary to consider how and whether AI and HI should be combined.

\subsection{Potential Integration of Augmented Intelligence}
Including human support in an intervention is known as a guided intervention \cite[cf. p. 92]{Lattie2022}. The idea is that with the usage of clinician-assisted internet-based CBT, non-specialist professionals can also provide help during these interventions \cite[cf. p. 92]{Lattie2022}. The effects are smaller than those guided by a human FtF, but considering the lack of increasing numbers and parallel reducing offers to get fast FtF therapy, blended therapy might be an option \cite[cf. p. 92]{Lattie2022}. This idea can be enhanced by integrating AI as an expert to enable professionals or non-professionals to support patients while offering proximity to the people. Nonetheless, the effectiveness and acceptability still need to be evaluated for both the population and the professionals.

Haber et al. propose the idea in their viewpoint to integrate a generative AI as an active participant in a psycho therapeutic process \cite[cf. p. 1]{Haber2024}. The AI can "listen to the dialogue, analyse the language and sentiment in real time and provide insights to both the patient and the therapist" \cite[p. 3]{Haber2024}. Besides that, this specific application enables access to a tangible representation of the patients' inner struggle \cite[cf. p. 3]{Haber2024}. With this innovative approach, patients can interact and gain "new perspectives on their internal conflicts, with the therapist guiding the therapeutic process" \cite[p. 3]{Haber2024}. Additionally, the specialists could be relieved, as necessary summaries and reports or proposals can be created during this usage.

In their experimental study, Esmaeilzadeh and colleagues observed the perceptions towards human-AI interaction in healthcare among patients suffering from either a chronic or acute disease \cite[cf. p. 1]{Esmaeilzadeh2021}. In three different modes, patients were asked about their preference to interact with a professional: using AI as a substitute for a physician, AI as a collaborator for the patient and professional interaction, or no AI involved in the total communication \cite[cf. p. 1]{Esmaeilzadeh2021}. In the augmented setting of this study, the patient interacts with the AI, but a physician evaluates the results and recommendations generated by the AI \cite[cf. p. 1]{Esmaeilzadeh2021}. Interestingly, they found in their results that pure AI interaction as well as augmented AI interaction led to a communication barrier \cite[cf. p. 1]{Esmaeilzadeh2021}. Besides concerns about privacy, trust, accountability, liability, and regulatory risks, the best benefits were expected from direct FtF interaction with the physician without any AI included \cite[cf. p. 1]{Esmaeilzadeh2021}. Only respondents with an acute disease showed more benefits to a collaborative AI-professional interaction, which would lead to an acceptable choice of AI in healthcare integration \cite[cf. p. 6]{Esmaeilzadeh2021}.

Aspects that have a significant influence on the use of AuI and AI in general are trust, understanding, and the ability to explain decisions. Currently, there are not many examples and studies available, but with the existing ones, it can be assumed that the acceptance of AI by patients and therapists can be improved if its analyses and suggestions can be explained and thus understood by people, thereby creating trust. A generally understandable and usable interface for AI can serve to strengthen comprehensibility and trust in AI \cite[cf. p. 67]{Kirste2019}. To adopt an approach that is as human-centred as possible, it is necessary to consider the perspectives of both users and experts -- in order to enable a comprehensive assessment of acceptance and distribution.

%% file: content/theoretical-framework.tex
\chapter{Theoretical Framework}
To explore the research objectives, concerning the distribution and acceptance of using CAs within the context of mHealth in Germany, it is necessary to draw upon the following theoretical frameworks. These frameworks will not only guide the formulation of relevant questions for data collection but also assist in addressing the core aspects of interest for patients and professionals through the two conducted online surveys.

\section{Measuring Acceptance and Use of Technology}
Acceptance is a fundamental prerequisite for harnessing the full potential of new technologies, and innovations like DMHI with or without AI. Before implementing or providing new technologies in healthcare or mHealth, a deep understanding of the factors related to technology acceptance among all groups involved is required \cite[cf. p. 17]{Esmaeilzadeh2020}. Acceptance and confidence can be understood as functions of user involvement in system development. Without acceptance and clarification there is no impact on health care, neither for patients nor for specialists. An example are the DiGAs in Germany. If the acceptance is not rated high for therapists, they do not prescribe them \cite[cf. p. 3]{Schlieter2024}. If the acceptance is rated low for patients, they are not asking for it or willing to use it. This might be one out of many reason, why DiGAs, even if they exist since the beginning of 2020, are still not frequently prescribed or used. AI-based CAs have been available for some years and are easy to reach and access. Therefore, not only the frequency, but also the acceptance among the general population and specialists who use them in the context of mHealth, e.\,g. to manage worry and grief, should be measured. This is essential for better understanding, including ethical considerations and providing suitable education.

For measuring the acceptance, the \acrfull{TAM} \cite[cf. p. 319]{Davis1989} and the \acrfull{UTAUT} \cite[cf. p. 425]{Venkatesh2003} are widely used, as well as their further developments TAM2 \cite[cf. p. 186]{Venkatesh2000}, TAM3 \cite[cf. p. 73]{Venkatesh2008} and UTAUT2 \cite[cf. p. 157]{Venkatesh2012}. In the health domain, the \acrfull{HITAM} \cite[ct. p. 1]{Kim2012} provides a more specific focus on the acceptance of health-related technologies, by integrating health status, health beliefs and concerns. Despite its relevance for specification, HITAM has not been able to establish itself. Instead, most publications refer to TAM or UTAUT and adapt these to the relevant environment or use case. This shows the existence of several interpretations associated with technology acceptance and the high number of different and evolving models. In the here cited literature these frameworks are often adjusted according to their considered use case or even completely customized questions are used \cite[cf. p. 4]{Damerau2021}. This makes comparison between different research topics and reproducibility difficult \cite[cf. p. 1]{Nadal2020}. But it furthermore shows the difficulty of providing an abstract model that is suitable for different use cases and incorporates various applications. 

Nevertheless, the UTAUT or one of its extensions can be used to measure the acceptance of telemedicine, DMHI or AI technologies. In 2003, Venkatesh et al. formulated four core determinants of intention and usage. \acrfull{PE}, \acrfull{EE}, \acrfull{SI} and \acrfull{FC} can influence \acrfull{BI} and \acrfull{UB}. Additional drivers for the key relationships are the up to four moderators -- Gender, Age, Experience, Voluntariness to Use \cite [cf. p. 425, 447]{Venkatesh2003}. PE is therefore defined as the degree to which an individual believes that using a system will enhance their job performance, whereas EE is defined as the degree of ease associated with the use of the system \cite[cf. p. 447 -- 450]{Venkatesh2003}. SI refers to an individual’s perception that important others believe that he or she should use the new system and FC refers to an individual’s belief that an organizational and technical infrastructure exists to support the use of the system \cite[cf. p. 451 -- 453]{Venkatesh2003}. The determinants mentioned should more or less influence the BI, the indication that an individual is ready to perform a given behaviour or use the implemented application. At the beginning UTAUT was designed to study information technology adoption behaviours mainly in organizational settings \cite[cf. p. 2]{Schmitz2022}. The in 2012 provided extended version was published to explain the acceptance of technology from the customers' perspective \cite[cf. p. 2]{Schmitz2022}. For this, three additional determinants have been included to achieve better alignment with customer requirements: Habit (H) -- the fun or pleasure derived from using the technology, Hedonic Motivation (HM) and Price Value (PV) \cite[cf. p. 161]{Venkatesh2012}. 

During this time, the TAM model has developed further and additional determinants have been added that affect \acrfull{PU} and \acrfull{PEU}. \acrfull{SN}, Image, Job Relevance, \acrfull{OQ} and \acrfull{RD} have been included next to the two moderators -- Experience and Voluntariness \cite[cf. p.197]{Venkatesh2000}. In TAM3 the anchors -- \acrfull{CS}, \acrfull{PEC}, \acrfull{CoA} and \acrfull{CP} -- have been included, cause the use of a new technology can be "related to individuals’ general beliefs regarding computers and computer use" \cite[p. 6]{Venkatesh2008}. In addition the two adjustments, Perceived Enjoyment and Objective Usability can be included after individuals gain experience with the new technology \cite[cf. p. 6]{Venkatesh2008}.

In this work these models and extensions contribute, among other things, to a broader understanding of user acceptance and adoption of new technologies in different contexts. Adapted questions from the TAM3 and UTAUT2 models will provide information regarding the acceptance of the use of AI in the mHealth context for the population and professionals in Germany. For the population, the acceptance, PE and EE, as well as predictors for the use of AI-based CAs to improve mHealth, is assessed. On the professionals' side, the acceptance as BI, the PE and EE, as well as predictors that could impact the use of this technology during work, are surveyed.

\section{Measuring Mental Health and Well-Being}
In the field of mHealth, well-being and the presence of psychological disorders or anxiety may influence the preferred usage of AI-based CAs \cite{Lucas2014}. Furthermore physicians experiencing poor well-being and excessive workload may be of significance when it comes to the use of this new technology in order to relieve their burden.

\subsection{Psychological Distress and Anxiety}
To measure psychological distress symptoms, anxiety, and the severity of depression, several methods have been provided in the form of short questions and answers using Likert scales \cite[cf. p. 1]{Heyningen2018}.

For assessing psychological distress and understanding individuals' mHealth status, the '\acrlong{K6}' provides a reliable, efficient, short, and commonly used self-report questionnaire \cite[cf. p. 959]{Kessler2002}. Known as K10 or K6 it consists of ten or six questions that assess the frequency and intensity of various symptoms experienced over the past 30 days, including feelings of tiredness, nervousness, sadness, and worthlessness \cite[cf. p. 963]{Kessler2002}. Using a 5-point Likert scale, K6 is widely employed in research and clinical settings, including the WHO World Mental Health Surveys, to screen for mHealth issues and evaluate individuals' psychological well-being \cite[cf. p. 959]{Kessler2002}. As a rule, $\geq$ 14 points indicate a current psychological distress \cite[cf. p. 14]{Staples2019}.

In addition the \acrlong{GAD} 7- and 2-item (\acrshort{GAD}-7 or \acrshort{GAD}-2) scales are reliable and effective methods used to assess the severity of generalized anxiety disorder, which is one of the most prevalent mental disorders \cite[cf. p. 1092]{spitzer2006}. Both brief questionnaires consists of seven or two questions that measure the frequency of anxiety symptoms experienced over the past two weeks. Participants rate their symptoms on a 4-point Likert scale, providing valuable information about the level of anxiety experienced. The GAD-7 scale is widely used in both clinical practice and research settings due to its validity and efficiency in capturing the severity of generalized anxiety disorder \cite[cf. p. 1092]{spitzer2006}. The shorter form with two questions provides insights into whether an individual with $\geq$ 3 points is currently struggling with increased anxiety \cite[cf. p. 14]{Staples2019}.

Both the K6 and the GAD-2 consist of questions that are designed to be easily understood. By answering the questions, individuals can provide valuable insights into their current level of psychological distress and anxiety, allowing for an estimation of their symptom severity. The simplicity and clarity of the questions make these tools practical and effective in assessing psychological well-being and identifying potential areas of concern.

\subsection{Physician Well-Being}
In addition, physicians’ mHealth can adversely affect competency, professionalism and the quality of care they provide for their patients \cite[cf. p. 299]{Lall2019}. Well-being is a complex and multifactorial topic, for this reason Lall et al. created a framework where in total 24 tools for the assessment of physicians' well-being are listed \cite[cf. p. 291]{Lall2019}. All of them were used in physician population and cited in multiple medical literature \cite[cf. p. 297]{Lall2019}.

Unfortunately, existing instruments to evaluate distress are often long, typically measure only one domain of distress, like fatigue or burnout, are cumbersome to analyse, do not identify those with high well-being and are not clearly linked to practice-related risks like the intent to reduce hours or move to a new practice \cite[cf. p. 2]{Dyrbye2010}. To address these limitations, Dyrbye et al. designed a seven-item screening tool to evaluate fatigue, depression, burnout, anxiety/stress and mental/physical \acrfull{QOL} \cite[cf. p. 6]{Dyrbye2010}. At a threshold score of $\geq$ 4, the \acrfull{PWBI} demonstrated a specificity of 85.8\,\% for identifying physicians with low mental QOL, high fatigue or recent suicidal thoughts \cite[cf. p. 421]{Dyrbye2013}. With the intention to use measures of work-related well-being and screening for general distress, the PWBI appears to be a suitable screening tool \cite[cf. p. 8]{Brady2019}.

The benefits of the PWBI include its brevity, taking less than one minute to answer the questions, ease of scoring and the ability to measure multiple dimensions of well-being. Additionally, this instrument is not as widely used as other screening tools such as the PHQ-9, K6 or GAD-2, which are typically used by patients \cite{Staples2019}. In the survey of this work, specialists should not feel as though they are being treated as patients, and their well-being should be more closely related to workload. This is necessary in order to investigate whether there is an influence on the acceptance of the use of this technology. Even though this model is not wide spread, some studies and languages, including Spanish, English, German and French already exist \cite{Robles2022, Dyrbye2013, Zumbrunn2020}.

%% file: content/methods.tex
\chapter{Methods}   
In this section, the characteristics of participants, the materials used and the study procedure for both web-based surveys of this thesis are described. To ensure transparency and reproducibility, the measures and data analysis methods are provided.

\section{Participants}
Sample characteristics, sampling procedures and sample size are described to determine how much the results can be generalized for participants of the German population and mHealth specialists.

\subsection{Population Sample Power and Procedure}
Research with people on a sensitive topic requires careful ethical consideration. The ethical approval for both surveys was granted by the Ethics Committee of the Technical University of Darmstadt (EK 71/2024) and all data were collected in line with the Declaration of Helsinki \cite{DeclarationOfHelsinki2024}.

Depending on the method of recruitment and the interest of the participants, the aim was to collect data from several hundred to thousands of people in total. According to the Demography Portal \cite{DemografiePortal}, $34,230,708$ men and $35,876,414$ women reached the age of majority in Germany in 2022. For a confidence level of 95\,\% and a margin of error of 5\,\%, a total sample size of 385 people would be relevant, as calculated using a sample size calculator \cite{surveymonkeyCalculator}. The total sample size of 444 participants satisfied the requirements. Given that the subject matter of the study also relates to AI, which tends to be a technical, innovative, but also controversial topic, it is assumed that younger groups of people between the ages of 18 and 35 are more likely to feel addressed and participate.

In order to achieve the largest and most random sample size possible, the identical questionnaires were distributed through various channels. Emails were sent to a university-wide mailing list and snowball sampling was employed with family, friends, and colleagues, resulting in 424 visitors, 148 participants and 119 completed surveys that passed the quality criteria. Social media posts and advertisements led to 2,021 views, with 89 participants, of which 68 results could be used. The research platform SurveyCircle attracted 177 visitors, with 162 participants, of which 159 results could be integrated. Finally, through the online panel provider ClickWorker\textregistered GmbH \cite{ClickWorker}, 116 visitors selected the provided URL, with 102 participants and 101 usable results. To provide additional motivation, a note for a donation of 1\,€ to a selected charity organization upon successful completion was given for the approximately ten-minute-long web survey. Additionally, participants received a payment of 2\,€ via the online panel, in accordance with the statutory minimum wage in Germany.

\subsection{Population Participant Characteristics}
The study conducted for the German-speaking population at the legal age included a total of 444 participants. The sample consisted of 196 (44.1\,\%) men, 238 (53.6\,\%) women, 8 (1.8\,\%) non-binary individuals and 2 (0.5\,\%) participants who did not specify their gender. The age of the participants was collected via six cluster groups to ensure anonymity. The first age group, comprising individuals aged 18 to 24 (19\,\%), consisted of young adults who likely grew up with modern technologies and have a high affinity for digital technology solutions. The second group, aged 24 to 34 (32\,\%), included individuals who are often working, tech-savvy and may have a high interest in health and wellness applications. The third group, aged 35 to 44 (14\,\%), consisted of individuals who are often in an established phase of life with a family and career. They may have an interest in health applications that help them manage their own health and that of their loved ones. The fourth group, aged 45 to 54 (11\,\%), may be in a transition phase between working life and retirement. They may have an interest in health applications that help them cope with age-related health problems. The fifth group, aged 55 to 64 (15\,\%), may be retired or approaching retirement. They may have an interest in health applications that help them maintain their health in old age. Finally, the sixth group, aged 65 and above (8\,\%), may be facing health or other life challenges such as loneliness, loss of a partner, relatives or friends.

To identify potential differences based on the level of education and employment status, these data were also collected. Among the respondents, 57\,\% have a degree (Bachelor, Master, or Diploma), 27\,\% have a university of applied sciences or university entrance qualification, 12\,\% have a secondary school qualification, 2\,\% have a general school qualification and 1\,\% each have no qualification or a different qualification level. Regarding employment status, 38\,\% of the respondents are currently employed, 12\,\% self-employed, 32\,\% currently studying, 1\,\% in training, 4\,\% not employed, 2\,\% unable to work and 7\,\% retired.

Since medical and therapeutic support is more difficult to access in rural areas, the place of residence should also be examined more closely. Most of the respondents (47\,\%) live in cities with more than $100,000$ inhabitants, 24\,\% live in cities with more than $20,000$ inhabitants, 11\,\% live in small cities with more than $5,000$ inhabitants and 18\,\% live in rural communities with fewer than $5,000$ inhabitants. Regarding experience with telemedicine, 59\,\% of the participants have never consulted for general medical or therapeutic advice or treatment remotely (e.\,g., via telephone, email, online) with an expert. 34\,\% consulted an expert rarely (once a month or less), 6\,\% occasionally (several times a month), and 1\,\% often (several times a week). 

Finally, feedback on the interest in this topic was collected. Participants were asked if the survey had sparked their interest in AI-based CAs. 19\,\% responded with "yes", 35\,\% with "rather yes", 24\,\% with "either yes nor no", 14\,\% with "rather no" and 8\,\% with "no".

More details regarding gender, age, education level, employment status, living area, frequency of using telemedicine and aroused interest in the studys' topic are provided in \autoref{tab:Socio_And_Mentl_Health_Data}.

\subsection{Specialists Sample Power and Procedure}
Similar to the survey for the general population, ethical approval was granted by the Technical University of Darmstadt. 

According to the German Medical Association physician statistics \cite{BÄK2023}, $12,957$ doctors (54\,\% women) were working in the field of psychiatry and psychotherapy in 2023. Additionally, there are around 11,500 counsellors working in the field of telephone counselling in Germany \cite{NummerGegenKummer_eV, Telefonseelsorge_eV, krisenchat_gGmbH}. With an estimated population size of 25,000 professionals in the field of mHealth, a confidence level of 95\,\%, and a margin of error of 5\,\%, a total number of 379 people for this study is relevant \cite{surveymonkeyCalculator}. The final sample size of 351 narrowly missed the requirements.

To ensure that the topic and questions are relevant and target-orientated, they were reviewed in parallel to the Ethics board by the Bavarian Chamber of Psychotherapists \cite{PTK}, and the counsellor provider Nummer Gegen Kummer e.\,V. \cite{NummerGegenKummer_eV}. To reach specialists in the field of mHealth, a total of 75 organizations related and connected to psychotherapy, psychology, sociotherapy, crisis intervention, DiGA providers (e.\,g. Novego by IVPNetworks GmbH \cite{Novego}), German medical and therapeutic chambers (e.\,g. BDP Verband \cite{BDP}), counselling services and different publishers (e.\,g. Psychosoziale Umschau \cite{PsychiatrieVerlag}) were contacted. In addition, approximately $6,000$ emails with attached leaflet, were sent to various medical and therapeutic professionals with publicly accessible email addresses.

At least a total of 904 individuals visited the online survey, 419 participated, and 351 completed the survey and met the quality criteria. A donation of 1\,€ for each completed survey was given to a charity organization as compensation for the time expenditure of about 10 minutes.

\subsection{Specialists Participant Characteristics}
A total of 351 mHealth professionals participated in the online survey, including 106 (30.2\,\%) men, 240 (68.4\,\%) women, 1 (0.3\,\%) non-binary individual and 4 (1.1\,\%) without specification. The age was collected in similar clusters as described in the previous section for the general population. The age distribution was as follows: 2\,\% were aged 18 to 24, 12\,\% 25 to 34, 23\,\% 35 to 44, 26\,\% between 45 and 54 years, 28\,\% 55 to 64 and 9\,\% were aged 65 or older.

To assess the current employment in the mHealth field, a multi-selection option was enabled. The distribution of professional activities is as follows: 6\,\% of the participants are studying or in training, 7\,\% have a medical activity, 4\,\% a psychiatric activity, 9\,\% have a psychological activity, 57\,\% a psycho therapeutic activity, 4\,\% a counselling activity, 7\,\% a social therapeutic activity and 6\,\% a different activities such as social pedagogy, crisis intervention, social work and neuropsychological activity.

Regarding professional experience, 44\,\% of the respondents have very extensive experience and are likely towards the end of their career. 25\,\% have between 11 and 20 years of considerable experience and are in their mid to late career phase. 19\,\% have 5 to 10 years of experience and are in the early to mid-career phase and 12\,\% have less than 5 years of experience because they are at the beginning of their professional career. The employment status, also with multi-selection, shows that 90\,\% work full-time in the field of mHealth, 3\,\% work part-time, 1\,\% work voluntarily, 4\,\% are studying or in training and 2\,\% have other employment states such as own practice or self-employed.

In terms of location, 54\,\% of the participants work in cities with more than $100,000$ inhabitants, 28\,\% in cities with more than $20,000$ inhabitants, 11\,\% in small cities, and 7\,\% in rural communities with fewer than $5,000$ inhabitants. Regarding remote advising or treatment (e.\,g. by phone, email, online), 11\,\% never offered any, 23\,\% offer it rarely, 35\,\% occasionally, 23\,\% often and 9\,\% very often.

The survey sparked interest in AI-based CAs for 14\,\% who rated "yes", 26\,\% "rather yes", 28\,\% "neither yes nor no", 15\,\% "rather no", and 17\,\%  "no". Additional details about exact numbers are provided in \autoref{tab:Socio_And_Mentl_Health_Data}.

\section{Material}
This part provides an overview of the materials and methods used to conduct the study to answer the research questions and to assure the quality of all measurements.

\subsection{Measurement of the Distribution and Acceptance to Use AI}
\begin{figure}[ht]
    \centering
    \includegraphics[width=1.0\linewidth]{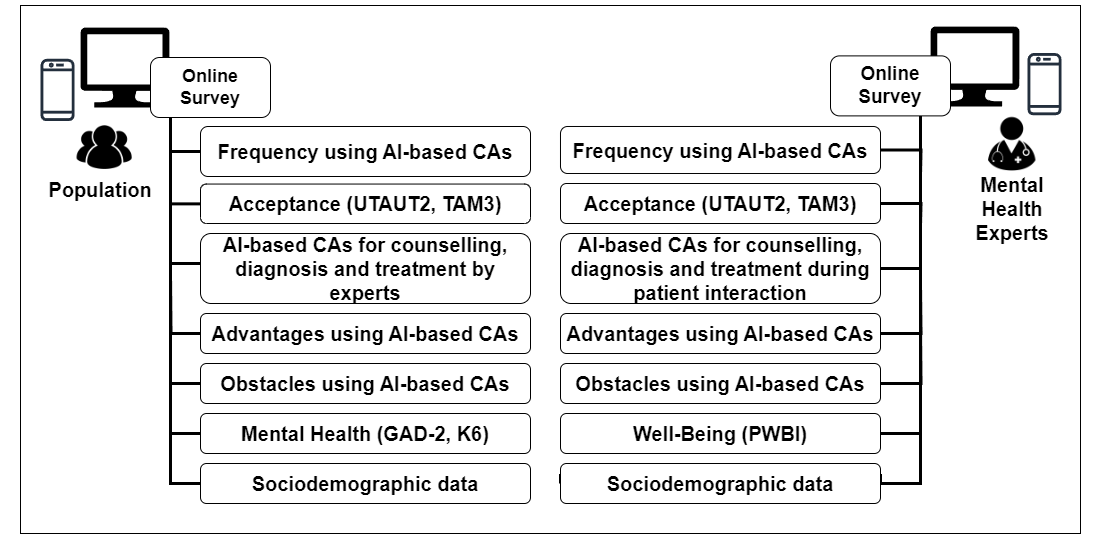}
    \caption{Conceptual diagram of the two surveys}
    \label{fig:conceptual-diagram}
\end{figure}

For the primary outcome, the distribution and acceptance of using AI-based CAs in the field of mHealth are measured, as well as the effect of both on the potential usage of AuI. \autoref{fig:conceptual-diagram} provides an overview of the conceptual framework of the two surveys, which also enables a comparison of the results of the population and the specialists.

To answer the distribution questions, the frequency (4 = very often to 0 = never) of use of AI-based CAs in general and in the area of mHealth was surveyed for both the general population and specialists. People from the general population, who already trusted AI-based CAs with their grief, were asked how satisfied they were with this experience (4 = very satisfied to 0 = very dissatisfied). Professionals, who had already used AI-based CAs in the counselling or treatment of patients, were asked about the level of satisfaction they had experienced. To get an idea of the level of agreement (4 = strongly agree to 0 = strongly disagree) that is comparable between the population and professionals, both groups were asked whether they believe that an AI-based CA (e.\,g. ChatGPT, Siri, Google Bard) can help with mild psychological distress like grief, stress, depression and anxiety. 

\begin{figure}[H]
    \centering
    \includegraphics[width=0.8\linewidth]{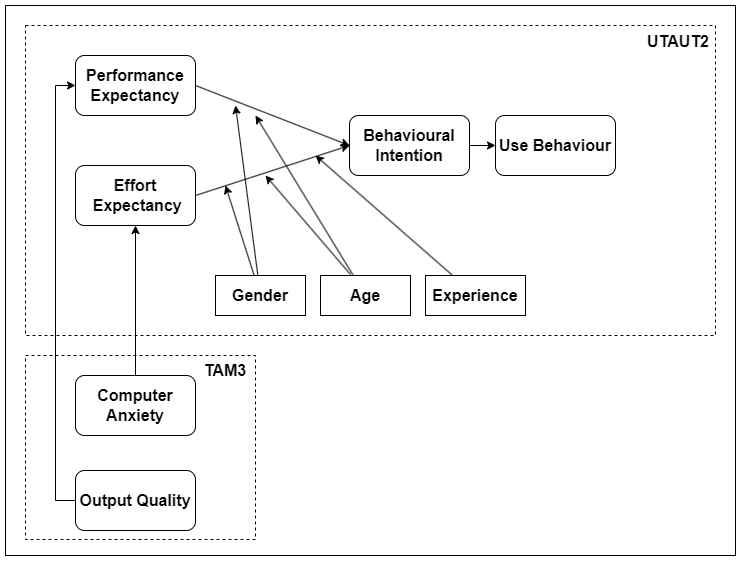}
    \caption{Model with combining elements of UTAUT2 and TAM3 to access the technology acceptance and usage}
    \label{fig:acceptance-model}
\end{figure}

For measuring the acceptance of this technology in the field of mHealth, the models UTAUT2 \cite{Venkatesh2012} and TAM3 \cite{Venkatesh2008} are combined as illustrated in \autoref{fig:acceptance-model}. Acceptance itself is operationalised as BI, which is influenced by the two core predictors PE and EE. In addition QO as a moderator for PE, and CoA as an anchor for EE are introduced. To assess PE, EE, CoA, QO and BI, two questions for each item are asked, and respondents have the possibility to rate on a 5-point Likert scale (5 = totally agree to 1 = totally disagree). The customized questions in \autoref{tab:Acceptance_questions_and_CronbachAlpha} are based on similar acceptance measurements from previous studies and slightly differ between population and specialists \cite{Damerau2021, Hennemann2016, Shahsavar2023}. Cronbach's alpha for both surveys is between acceptable and excellent, indicating good internal consistency.

\begin{table}[H]
    \centering
    \begin{tabular}{ m{5em} p{20em} p{7em}}
    \hline
    \textbf{Variable} & \textbf{Items} & \textbf{Cronbach's $\alpha$}\\
    \hline
     \multirow{2}{5em}{BI Population} & 1. I would like to try a psychological AI-based CA. & .87\\
      & 2. Given that I have access to the system, I predict that I would use it.\\
      \hline
     \multirow{2}{5em}{BI Specialists} & 1. I would like to try an AI-based CA as part of my job as a specialist. & .89\\
      & 2. Assuming I get access to an AI-based CA, I would use it as part of my job as a specialist.\\
      \hline
     \multirow{2}{5em}{PE Population}&  1. A conversation with an AI-based CA could improve my general well-being. & .92\\
     & 2. A conversation with an AI-based CA could help me to improve my personal mental health.\\
     \hline
      \multirow{2}{5em}{PE Specialists}& 1. Using an AI-based CA helps me accomplish things more quickly. & .91\\
     & 2. An AI-based CA could make my work easier.\\
     \hline
     \multirow{2}{5em}{EE} & 1. A conversation with an AI-based CA would be easy to operate and comprehend. & Population: .84\\
     &  2. Learning how to use AI-based CAs would be easy for me. & Specialists: .75\\
     \hline
     \multirow{2}{5em}{QO}& 1. The quality of the output I get from the system is high. & Population: .84\\
     & 2. I have no problem with the quality of the system’s output. & Specialists: .85\\
     \hline
     \multirow{2}{5em}{CoA} & 1. Computers make me feel uncomfortable. & Population: .71  \\
     & 2. Working with a computer makes me nervous. & Specialists: .85\\

    \end{tabular}
    \caption{Adapted items of the UTAUT2 and TAM3 models for the population and the specialists}
    \label{tab:Acceptance_questions_and_CronbachAlpha}
\end{table}

\subsection{Measurement of the Level of Agreement to Use Augmented AI}
\begin{figure}[ht]
    \centering
    \includegraphics[width=0.8\linewidth]{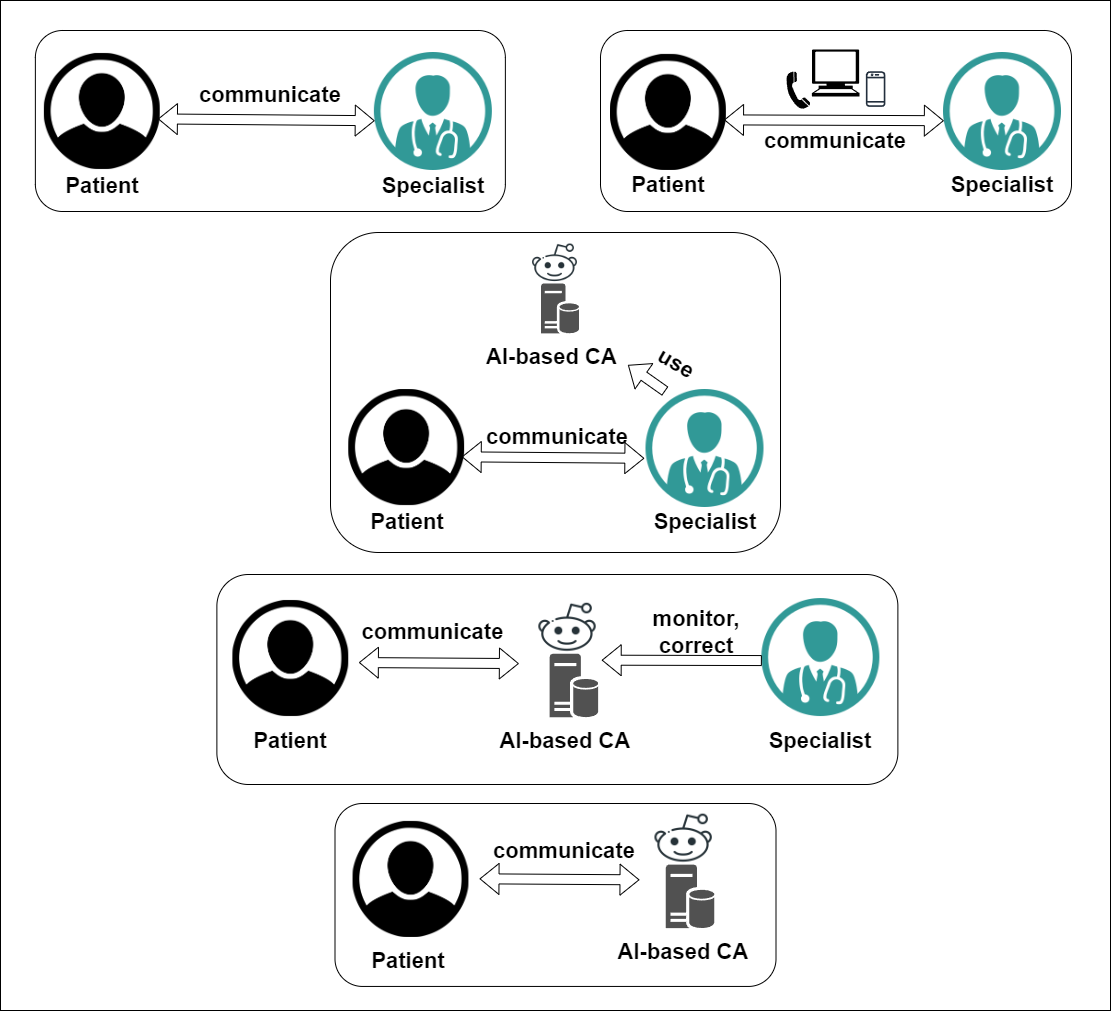}
    \caption{Communication scenarios combining AI and HI for the potential use or offer of counselling, diagnosis and treatment}
    \label{fig:AUI_communication_possibilities}
\end{figure}

In order to gain insights into possible AI and HI combinations as AuI and their acceptance, a total of 15 questions are asked about the favour of the population and the care provided by specialists in order to receive or offer counselling, diagnosis and treatment (5 = completely agree to 1 = completely disagree). The degree of AI integration also varies, based on the survey from 2021 \cite[cf.]{Esmaeilzadeh2021}, and includes:
\begin{itemize}
    \item AI as an additional expert supporting the specialist
    \item AI as companion for the patient, communicating via email or chat, which is controlled by the specialist
    \item AI as a substitute for the specialist
\end{itemize}
These preferences are compared with direct communication and communication from distance with a specialist or patient. \autoref{fig:AUI_communication_possibilities} illustrates all five communication scenarios. 

\subsection{Collection of Drivers and Barriers for the Use of AI}
In order to gain a better understanding of the potential drivers or barriers to the use of this technology for mHealth, eight advantages and six obstacles are provided to the population. Seven potential advantages and seven disadvantages are offered to the specialists (5 = completely agree to 1 = completely disagree). Most of these were based on corresponding surveys \cite[cf.]{Sweeney2021, Palanica2019, Temsah2023} and on feedback received after reviewing both surveys. In order to obtain individual impressions, all participants were each offered an optional free text field to mention further advantages and disadvantages.

\subsection{Measurement of Mental Health and Well-Being}
To access mHealth related data as secondary outcome, the population was asked questions to measure the current GAD-2 and K6. Two questions on a 4-point Likert scale (3 = almost daily to 0 = never) are used to assess generalized anxiety of the past two weeks. The Cronbach's alpha was questionable at .62. Six questions on a 5-point Likert scale (4 = always to 0 = never) are used to assess psychological distress experienced in the past 30 days. The internal consistency was excellent, with a Cronbach's alpha of .90. To access work-related well-being data, the specialists received seven questions from the PWBI with the possibility to answer with "yes" or "no". The Cronbach's alpha of .77 is acceptable and shows good internal consistency.

\subsection{Collection of Sociodemographic Data}
Additional secondary outcome measures include gender, age, education level, employment status, place of residence or work, frequency of telehealth use or provision, and work experience to determine their influence or predictive power on the adoption of AI-based CAs. To ensure anonymity, the possible answers were clustered. As a final question to gain insight into the relevance and interest in this topic, participants were asked whether this survey had sparked their interest in AI-based CAs (4 = yes to 0 = no).

\subsection{Quality of Measurements}
Both surveys were conducted in German. Where available, methods and questions from German literature were used. In most cases, the questions were translated and back-translated from the English literature using the free online translation tool DeepL \cite{Deepl}.

From June 1, 2024 to July 7, 2024 a pre-study was conducted with 34 volunteers of the population and three specialists within the known network to assess usability, technical functionality, comprehension of the questions and timing. The feedback received during the pre-study phase contributed significantly to the refinement of the final survey questions. The main survey was then carried out from July 15, 2024 to September 2, 2024 for a total of 50 days.

\section{Procedure}
This section reports all of the procedures applied for administering the study, processing and analysing the data.

\subsection{Data Collection and Research Design}
To address the research questions, two quantitative, anonymous, web-based, online surveys were conducted among the German population and mHealth specialists with UmfrageOnline \cite{UmfrageOnline}. To ensure the reproducibility of the web-based surveys, as Eysenbach et al. suggests \cite[cf. p. 1]{Eysenbach2004}, their \acrfull{CHERRIES} is used. 

Criteria to participate were knowledge of German, Internet access and legal age. Additionally for the survey of mHealth specialists, participants should be working in a therapeutic, psychological, medical, crisis or counselling capacity. A medical licence or a minimum period of experience was not required in order to gain as many insights and perspectives of specialists as possible.

Both online surveys were carried out in parallel from July 15, 2024 to September 2, 2024. Notes on the survey aim, estimated duration of ten minutes, gender hint, usage of AI in the field of mHealth and guidance for those seeking help and advice were provided in the introduction. Additionally, the contact details of the data protection officer and the responsible scientists were included. Participation was voluntary, anonymous and could be aborted at any time without giving reasons. Every participant had to confirm a declaration of consent. Sociodemographic data, such as age, were clustered to ensure that answers could not be traced back to any individual person. Technical measures taken by the tool provider prevent one and the same person from submitting several contributions within the framework of the online tool used.

\subsection{Data Processing and Diagnostics}
For the subsequent data processing, only completed respondents are used. Also the attention request had to be answered correctly, which was included in both surveys. The collected and coded data is analysed to gain insights into the frequency of using AI-based CAs in general and in connection with mHealth. In addition, the acceptance, drivers and barriers as well as potential interaction possibilities with and without the involvement of specialists are analysed. Sociodemographic and mHealth data are included to understand the demographic characteristics of the participants, to enable comparability with existing studies and to identify possible predictors for the use or rejection of this technology. 

The normal distribution of the coded, ordinal data is analysed using a histogram in Microsoft\textsuperscript{\textregistered} Excel\textsuperscript{\textregistered} Data Analytics. Although the graphical output shows violations of a normal distribution, an approximate normal distribution is assumed due to the large sample size. In addition, the widely used hypothesis tests, t-test and ANOVA appear to be robust against violations of the assumption of a normal distribution \cite[cf. p. xix]{wilcox2011introduction}.

\subsection{Analytic Strategies}
All internal consistencies, descriptive statistics, hypothesis and correlational tests have been calculated using Microsoft\textsuperscript{\textregistered} Excel\textsuperscript{\textregistered} for Microsoft 365 MSO (Version 2409 Build 16.0.18025.20030) 64 Bit. To compare the means of two or more groups, t-test or ANOVA are used with a level of significance set at $\alpha$ = .05 for two-sided tests. Cohen's d is used to get the effect size comparing two groups with large sample size and Levene-Tests is used to test for equal variance. Related to existing research the acceptance as BI is categorized by mean in low (1 -- 2.34), moderate (2.35 -- 3.67), and high (3.68 -- 5) \cite[cf. p. 5]{Damerau2021} \cite[cf. p. 1]{Hennemann2016}. Similar levels of categorization are also used to determine agreement in the use of the provided five communication scenarios.

To answer H1 and RQ1, the mean values of the frequency of use of AI-based CAs in general and in relation to mHealth are compared. In addition, possible differences between the groups in terms of current mHealth status, socio-demographic data and frequency of use or provision of telemedicine are analysed using t-tests and one-way ANOVAs. A t-test with Cohen's d is used to clarify H2 of whether the belief that AI can help with mild mental disorders is higher among the population than among the specialists.

To identify drivers for the acceptance of the use of AI-based CAs for mHealth to address H3, the mean values in low, moderate and high BI are compared between different groups. For this t-tests and one-way ANOVAs are used with regard to the frequency of use of AI-based CAs, mHealth, sociodemographic data, frequency of use or provision of telemedicine. In addition, the Pearson correlation coefficient is used to determine linear correlations between BI and PE, BI and EE, CoA and EE, QO and PE. In order to compare the level of BI, PE and EE between the population and the specialists, t-tests are used to provide the answer for RQ2.

Comparing the means of the agreement on the five communication scenarios will provide answers to the extent of preference for AuI in seeking or providing mHealth counselling, diagnosis or treatment, which addresses H4 to H7. To determine the extent of preference for AuI and address RQ3, the means are compared with the groups of BI, frequency of use of AI-based CAs, mHealth, sociodemographic data, frequency of use or provision of telemedicine using t-tests and one-way ANOVAs.

%% file: content/results.tex
\chapter{Results}
The results of both groups are described below using descriptive statistics, hypothesis tests and correlations in order to provide information on the prevalence and acceptance of the use of AI-based CAs in the field of mHealth. Finally, the extent is determined to which the integration of AI into patient-specialist interaction in the sense of AuI is desired.

\section{Sociodemographics and Mental Health}
To gain a better understanding, all participants are asked finally about certain background characteristics. Sociodemographic data was collected by asking questions about sex, age, education level, employment status, job activity, years of job experience, place of residence or work and experience with telemedicine.

\subsection{Population Characteristics}
A total of 444 results from participants from the German population are analysed. More women (53.6\,\%) than men took part and about 51\,\% of the people are between 18 and 34 years old. 56.8\,\% have a university degree, 32.4\,\% are currently studying, while 50\,\% are employed or self-employed. Most of the respondents (54\,\%) live in a large city with more than $100,000$ inhabitants and almost 60\,\% have never used telemedicine. 45\,\% -- 57\,\% of them are female -- have a threshold score of three or higher for the GAD-2 measure, indicating anxiety. 12\,\% have a threshold value of 14 or higher for K6, which indicates possible psychological distress. Again, more than half of the participants (55\,\%) are female having these symptoms. \autoref{tab:Socio_And_Mentl_Health_Data} provides all sociodemographic details of the participants of the population.

\subsection{Specialists Characteristics}
Around 68\,\% of the 351 participants of the German mHealth specialists are women, and 63\,\% are 45 years old and older. About 57\,\% have a psychotherapeutic occupation and 44\,\% have more than 20 years of experience in their profession, with 90\,\% working full-time. 54\,\% work in large cities and almost 90\,\% provide telemedicine. In terms of their work-related well-being, 11\,\% -- 69\,\% of them are female -- have a threshold score of four or higher, indicating fatigue, depression, burnout, anxiety/stress and mental QOL. All sociodemographic information is shown in \autoref{tab:Socio_And_Mentl_Health_Data}.

\begin{longtable} {m{25em} m{5em} m{5em} }
    \hline
    \centering
    & \textbf{Population} & \textbf{Specialists} \\
    \textbf{Variable} & n [\,\%] & n [\,\%] \\
    \hline
    
   \textbf{Sex} & & \\
    Male & 196 [44.1] &  106 [30.2]\\
    Female & 238 [53.6] & 240 [68.4]\\
    Non-binary & 8 [1.8] & 1 [0.3]\\
    No specification & 2 [0.5]  & 4	[1.1]\\
    
    \textbf{Age in years} & & \\
    18 -- 24 & 85 [19] & 7	[2]\\
    25 -- 34 & 143 [32] & 41 [12]\\
    35 -- 44 & 63 [14] & 80 [23]\\
    45 -- 54 & 48 [11] & 91 [26]\\
    55 -- 64 & 68 [15] & 100 [28]\\
    65+ & 37 [8] & 32 [9]\\
    
    \textbf{Education status } & & \\
    Still in school education & 4 [0.9]& - \\
    Hauptschulabschluss or qualified Hauptschulabschluss & 10 [2.3]& - \\
    Realschulabschluss or qualified Realschulabschluss & 52 [11.7] & - \\
    Advanced technical college or university entrance qualification & 121 [27.3] & - \\
    Bachelor's degree & 142	[32.0] & - \\
    Master's degree & 48 [10.8]& - \\
    State examination, diploma or magister & 62	[14.0] & - \\
    No degree & 1 [0.2]& - \\
    Other & 4 [0.9]& - \\

    \textbf{Employment status } & & \\
    Education   & 4	[0.9] & - \\
    Study  & 144 [32.4]& - \\
    Employment  & 168 [37.8]& - \\
    Self-employed  & 54	[12.2] & - \\
    Not employed, seeking work  & 5	[1.1] & - \\
    Not employed, not seeking work  & 12 [2.7] & - \\
    Unable to work  & 10 [2.3] & - \\
    Retired  & 33 [7.4] & - \\
    Other  & 14	[3.2] & - \\

    \textbf{Activity in the field of mental health (multiple choice)} & & \\
    Training or studies & - & 26 [6.0]\\
    Medical activity & - & 30 [6.9]\\
    Psychiatric activity & - & 18 [4.1]\\
    Psychological activity & - & 41	[9.4]\\
    Psychotherapeutic activity & - & 249 [57.1]\\
    Counselling care & - & 16 [3.7]\\
    Sociotherapeutic activity & - & 29 [6.7]\\
    Other & - & 27 [6.2]\\

    \textbf{Years of job experience} & & \\
    Less than 5 years & - & 41 [12]\\
    Between 5 and 10 years & - & 68	[19]\\
    Between 11 and 20 years & - & 89 [25]\\
    More than 20 years & - & 153 [44]\\

    \textbf{Employment status (multiple choice)} & & \\
    Full-time & - & 326 [90]\\
    Part-time & - & 10 [3]\\
    Voluntary & - & 3 [1]\\
    Training or studies & - & 14 [4]\\
    Other & - & 9 [2]\\

    \textbf{Place of residence / work } & & \\
    Large city & 209 [47] & 190	[54] \\
    Medium-sized town & 107	[24] & 99 [28] \\
    Small town & 49	[11]& 37 [11] \\
    Rural community & 79 [18]& 25 [7]\\

    \textbf{Telemedicine usage / offer frequency } & & \\
    Very often  & 1	[0.2] & 32 [9.1] \\
    Often  & 4 [0.9]& 79 [22.5] \\
    Occasionally & 25 [5.6]& 122 [34.8]\\
    Rarely & 151 [34.0]& 80	[22.8] \\
    Never & 263	[59.2]& 38 [10.8] \\

    \textbf{Generalized Anxiety Disorder} & & \\
    $\geq$ 3 & 200 [45] & -\\
    < 3 & 244 [55] & - \\
    \textbf{Kessler Psychological Distress} & & \\
    $\geq$ 14 & 53 [12]& -\\
    < 14 & 391 [88]& -\\
    \textbf{Physician Well-being Index} & & \\
    $\geq$ 4 &-& 39 [11]\\
    < 4 &-& 312 [89]\\
    \caption{Sociodemographic and mHealth related data of the participating population (n\,=\,444) and the specialists (n\,=\,351) }
    \label{tab:Socio_And_Mentl_Health_Data}
\end{longtable}

\section{Distribution of AI-Based Conversational Agent Usage}
To measure the frequency of use of AI-based CAs, all participants were asked how often they generally use this new technology (4 = very often to 0 = never). In addition, the population participants were asked how often they confide their worries, concerns or fears to CAs (4 = very often to 0 = never). In contrast, the experts were asked how often they use this technology to treat or advise their patients (4 = very often to 0 = never). All participants who did not answer "never" were additionally asked how satisfied they were with this experience (4 = very satisfied to 0 = very dissatisfied). With the last question in this section, all participants were asked whether they belief AI-based CAs could help with mild mental health problems like grief, stress, depression and anxiety (4 = strongly agree to 0 = strongly disagree).

\subsection{General Population}
Already 14\,\% of participants in the German population use AI-based CAs very often, 24\,\% often, 25\,\% occasionally, 22\,\% rarely and 15\,\% never. To cope with grief, worries or fears, this technology is used very often by 1\,\%, often by 5\,\%, occasionally by 7\,\% and rarely by 15\,\% of the general population. However, most of them (73\,\%) have never used this technology for this reason. 11\,\% of the 121 people who have already confided their grief, worries or fears were very satisfied, 52\,\% satisfied, 31\,\% neither satisfied nor dissatisfied, 6\,\% dissatisfied and no one was very dissatisfied. 5\,\% of all participants strongly agree that this technology can help with mild psychological stress, 38\,\% agree, 24\,\% have a neutral opinion, 23\,\% disagree and 10\,\% strongly disagree.
	
The use of two-tailed t-tests and one-way ANOVA shows no significant difference for overall AI-based CA use for sex, place of residence, and frequency or experience of using telemedicine and elevated K6. A slight difference for participants with an increased GAD-2 index seem to use this new technology more often (t\,(428) = 2.196; p = .029). Whereas high significantly younger participants (F\,(5,438) = 25.36; p\,<\,.001), participants with high education level (F\,(8,435) = 10.26; p\,<\,.001) and participants currently studying or being in education (F\,(3,440) = 33.80; p\,<\,.001) seem to use AI-based CAs more often.

Respondents aged 35 -- 44 years (F\,(5,438) = 4.22; p\,<\,.001) and those using telemedicine more frequently (F\,(4,439) = 8.80; p\,<\,.001) show that they use AI-based CAs against grief more frequently. Moreover, participants using AI in general more frequently seem to use it significantly more often (F\,(5,439) = 12.79; p\,<\,.001) for this specific use case. All additional demographics or mHealth related data show no significance in using this technology for help when dealing with grief.

Furthermore, being of young age (F\,(5,438) = 2.87; p\,<\,.01) seems to significantly affect the belief, that AI can support with mild psychological stress. However, not only young participants, but even participants who use AI in general (F\,(4,439) = 7.99; p\,<\,.001) or AI against grief (F\,(4,444) = 7.05; p\,<\,.001) show a significantly stronger belief in it.

AI is indeed commonly used, and the use of AI to manage grief and worry is certainly already being used, but not yet by the majority of respondents, so H1 is not refuted. However, the basic frequency of use of this technology could have an impact on the use for mHealth related issues and on the belief that it can help with mild psychological problems.

\subsection{Mental Health Specialists}
5\,\% of the participating specialists use AI-based CAs very often, 8\,\% often, 12\,\% occasionally, 27\,\% rarely and 48\,\% never. For counselling or treating patients, this technology is never used by 87\,\%. 8\,\% use it rarely, 3\,\% occasionally, and 1\,\% for each, often and very often. This result is consistent with H1, and from the 45 participants who have already used it, 20\,\% were very satisfied, 49\,\% satisfied, 29\,\% neither satisfied nor dissatisfied and 2\,\% very dissatisfied with this experience. 7\,\% of all participants completely agree, that this technology can help with mild psychological stress, 26\,\% agree, 26\,\% have a neutral opinion, 23\,\% do no agree and 19\,\% completely disagree.

Since more than half of the participants are psychotherapists and most of them work full-time, the different answer options for each question are summarized. The use of two-tailed t-tests and one-way ANOVA to compare the mean of AI-based CA usage in general by the indicated groups only shows significance in relation to the age (F\,(5,345) = 3.66; p = .003). In fact, specialists between the ages of 55 and 64 use AI more frequently than their younger colleagues. A small effect on the frequency of use of this technology can additionally be seen among specialists with a higher PWBI index (t\,(45) = 1.80; p = .08). 

Specialists who already have experience in the use of AIs use them significantly more frequently for the counselling or treatment of patients (F\,(4,346) = 36.74; p\,<\,.001), but with a mean value of 1.39 this value is still at the level of rare use. Participants who practice psychotherapy (t\,(130) = 2.04; p = .04), as well as participants who are not frequently offering telemedicine (F\,(4,346) = 3.79; p = .005), seem to reject this technology stronger regarding its use in contact with patients.

Participants who use AI-based CAs more frequently are more likely to believe that AI can help with grief (F\,(4,346) = 14.34; p\,<\,.001), as are experts who use them in their work with patients (t\,(60) = 6.38; p\,<\,.001). Especially young respondents seem to have a significant belief, that AI can help with mild psychological stress (F\,(5,345) = 4.03; p = .001). Similar effect can be observed for those who have less job experience (F\,(3,347) = 3.11; p = .03) and those who are providing telemedicine more frequently (F\,(4,346) = 3.09; p = .02).

\subsection{Comparison of Population and Specialists}
\begin{figure}[H]
    \centering
    \begin{subfigure}{0.8\textwidth}
    \centering
    \includegraphics[width=0.9\linewidth]{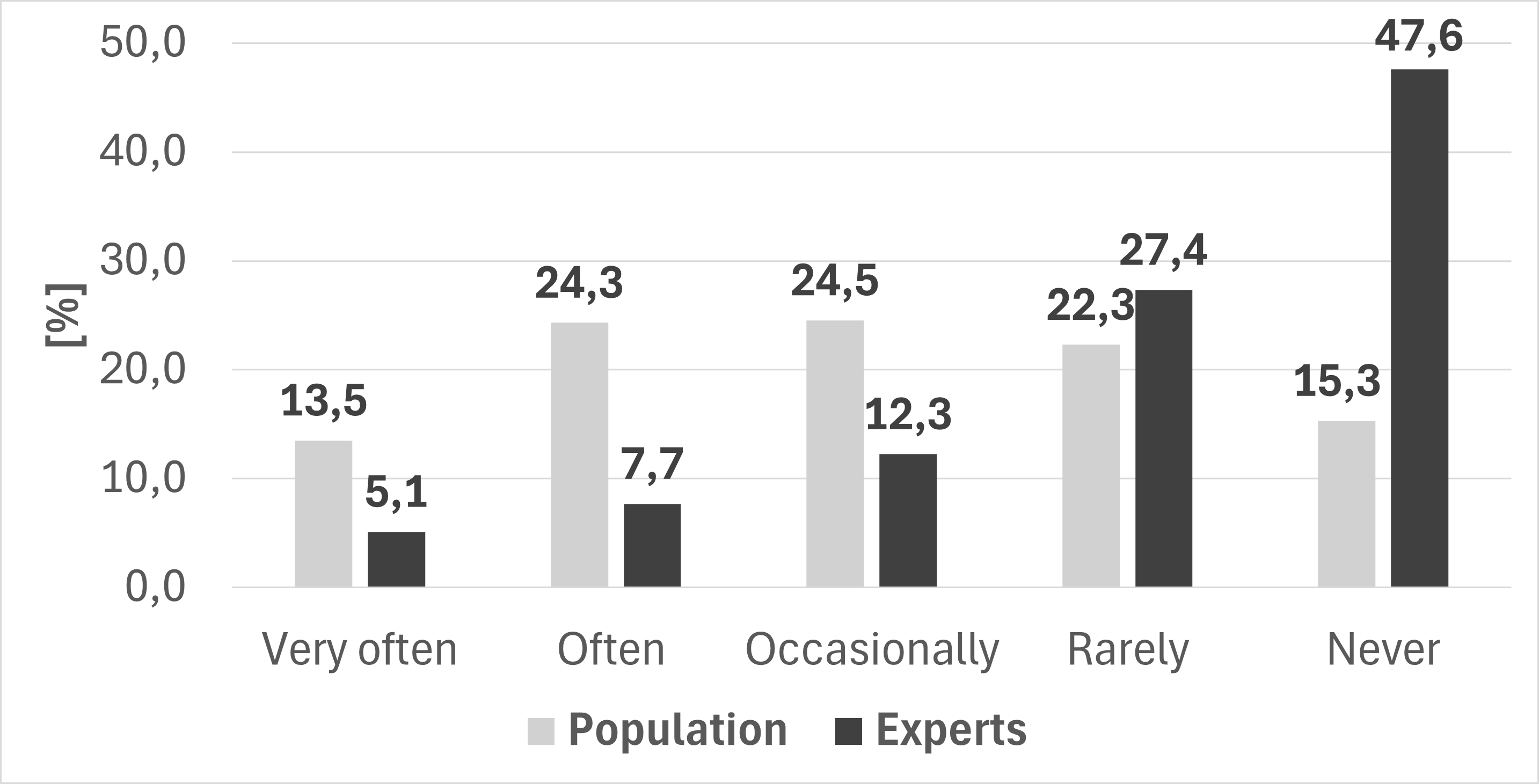} 
    \caption{Frequency of use of general AI by the population and experts}
    \label{fig:general_AI}
    \end{subfigure}
    
    \begin{subfigure}{0.8\textwidth}
    \centering
    \includegraphics[width=0.9\linewidth] {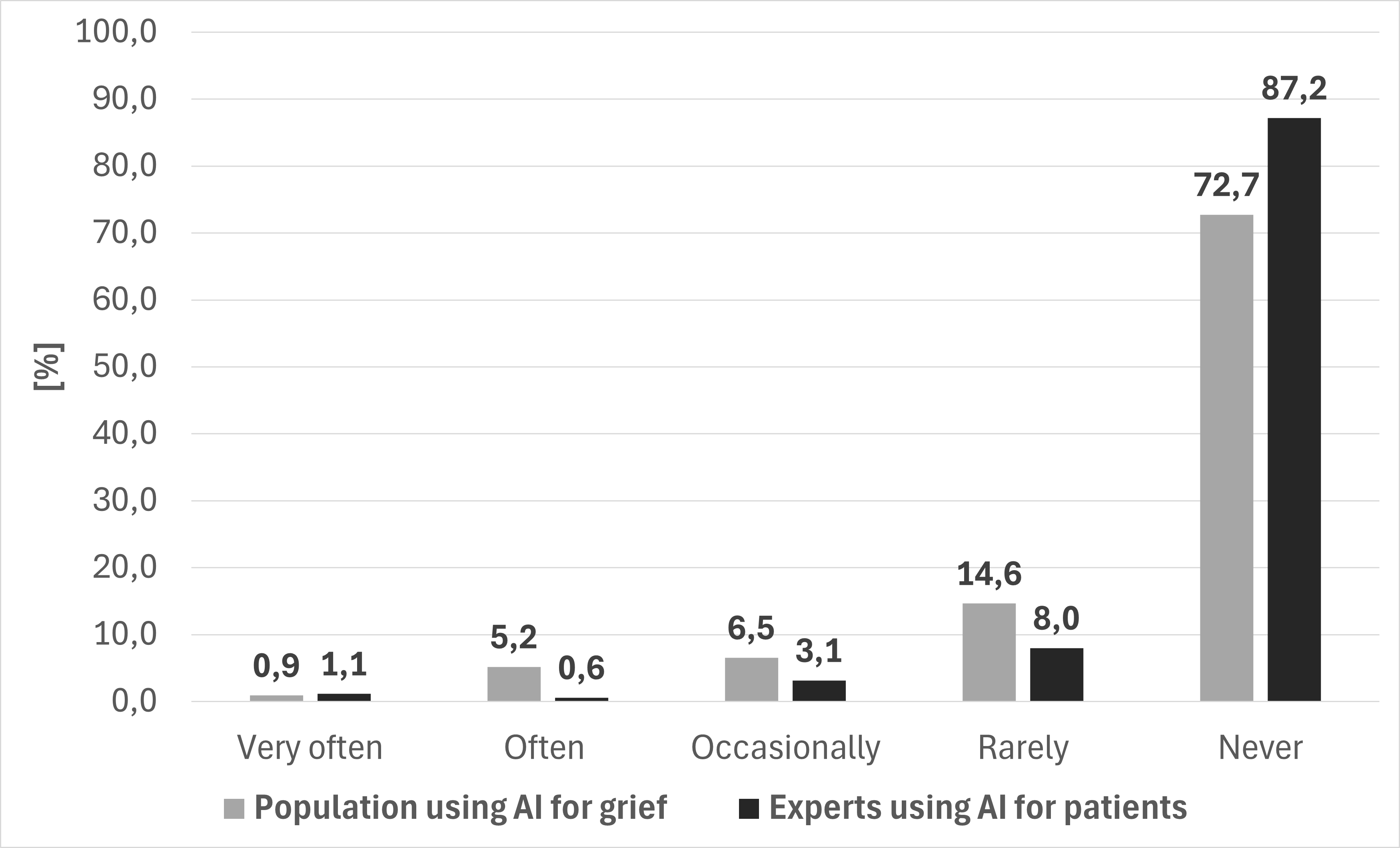}
    \caption{Frequency of use of AI for mHealth by the population and experts}
    \label{fig:mHealth_AI}
    \end{subfigure}
    
    \begin{subfigure}{0.8\textwidth}
    \centering
    \includegraphics[width=0.9\linewidth] {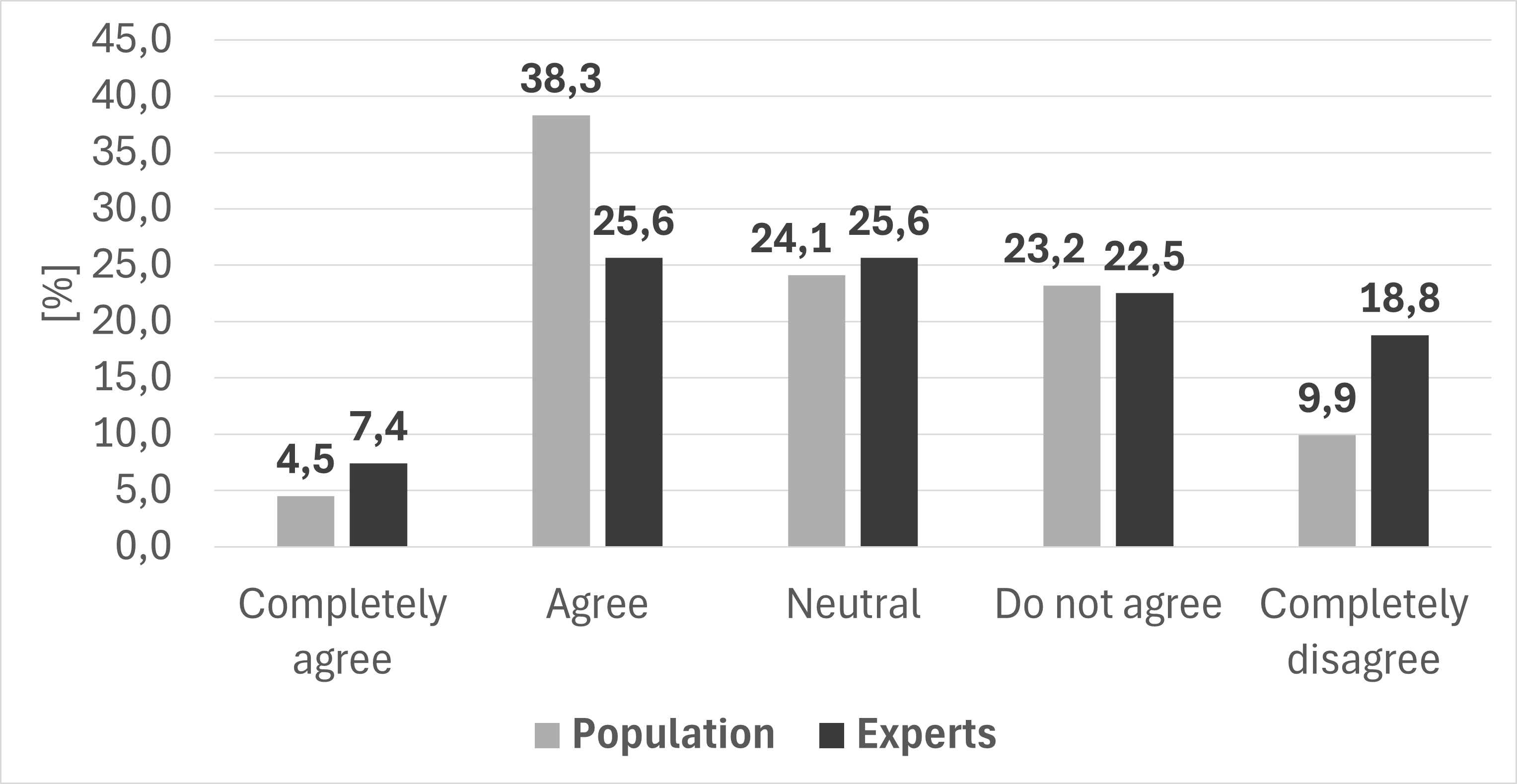}
    \caption{Believe that AI can help with mild psychological stress}
    \label{fig:belief_AI}
    \end{subfigure}

\caption{Comparison of AI use distribution between the population and the specialists}
\label{fig:Comparing_Distribution_AI}
\end{figure}

To address RQ1 and to compare the frequency of using AI for mHealth, the provided answers of the general population and specialists are compared. With regard to the use of this technology in general, a high significance (t\,(774) = 11.804; p\,<\,.001) can be observed between the population and the specialists, as shown in \autoref{fig:general_AI}. In addition, a strong effect size Cohen's d of 1.23 is determined. 

In terms of using AI to support their grief, patients appear to do so more often than professionals who use AI to treat or counsel their patients, as shown in \autoref{fig:mHealth_AI}. And the belief that AI can help with mild psychological distress such as grief, stress, depression and anxiety shows a significant difference (t\,(706) = 2.911; p = .0037; Cohen's d = 1.15) between the two groups, as shown in \autoref{fig:belief_AI}. Finally it is consistent with H2 that the population believes more in AI support than the specialists.

\section{Acceptance of Using AI-Based Conversational Agents}
In order to analyse the acceptance and potential influence of predictors such as sociodemographic data, mHealth, well-being and experience, hypothesis tests are used to compare the different groups. BI, PE, EE, QO and CoA are each measured with two questions on a 5-point Likert scale (5 = strongly agree to 1 = strongly disagree). Because of given internal consistency, the mean values for both items could be used and the significance level is set at $\alpha$=.05 (two-sided tests). Following Hennemann et al. \cite[cf. p. 6]{Hennemann2016} and Damerau et al. \cite[cf. p. 5]{Damerau2021}, the acceptance values were categorized by mean value to describe low (1 -- 2.34), moderate (2.35 -- 3.67) and high (3.68 -- 5) acceptance.

\subsection{Mental Health Related Use in the General Population}
Using Pearson correlation for BI, PE, EE, CoA and QO with two-side t-test, a strong correlation between BI and PE (r\,(442) = .72; p\,<\,.001) is shown, furthermore a weak correlation appears between BI and EE (r\,(442) = .24; p\,<\,.001). CoA and EE have a weak negative correlation (r\,(442) = -.25; p\,<\,.001), whereas QO and PE show a strong correlation (r\,(442) = .52; p\,<\,.001).

CoA is quite low for the general population with a mean of 1.77, and the expectation to use it easily is high with a mean of 3.85, while the expected quality of the response is moderate with a mean of 2.81. This results in a moderate (mean 2.88) expectation that AI-based CAs can improve the overall well-being, which leads to a moderate acceptance (mean 3.13) of using or trying them. For BI 28\,\% of the participants show a low acceptance when it comes to the use of AI-based CAs for mental health, 25\,\% medium and 37\,\% high acceptance. The PE is low for 32\,\%, medium for 43\,\% and high for around 25\,\%. 63\,\% of the participants expect to have a high level of EE, 34\,\% expect a medium level and 4\,\% a low level.

For BI, participants of female sex show a slightly higher acceptance (t\,(407) = 1.86; p = .06) compared to male, whereas age, education status, employment status and place of residence show no impact. In addition a significant higher acceptance is observable for participants, who already use telemedicine (t\,(383) = 2.87; p = .004). This is significant for those, who frequently use AI in general more often (F\,(4,439) = 8.01; p\,<\,.001), or trust AI more often their grief (F\,(4,439) = 12.12; p\,<\,.001). Similar results can be summarised with regard to the 
respondents reaching the threshold of three on GAD-2 (t\,(433) = 3.31; p\,<\,.001) or the threshold of 14 on K6 (t\,(66) = 2.75; p = .008).

As EE and PE correlate with BI, similar statistical analyses are performed for these. Using AI in general more frequently also results in an increased PE (F\,(4,439) = 8.28; p\,<\,.001), as well as having experience in trusting AI in the matter of grief (F\,(4,439) = 10.21; p\,<\,.001). Demographic and mHealth characteristics show no further significant impact. Compared to men, women show a higher EE (t\,(397) = 2.96; p = .003), as does a younger age (F\,(5,438) = 12.78; p\,<\,.001). In addition, there are higher values compared to participants with a lower secondary school leaving certificate and graduation (F\,(8,435) = 4.99; p\,<\,.001). Respondents who are currently studying or in training additionally have a high EE (F\,(3,440) = 14.14; p\,<\,.001), as do respondents who live in large cities (F\,(3,440) = 3.46, p = .017). Frequency of telehealth use does not appear to have a significant impact, whereas more frequent use of AI in general achieves significantly higher EE scores (F\,(4,439) = 15.47; p\,<\,.001), whereas trust in AI in regarding of grief does not have a significant impact. It can be determined that a GAD-2 threshold of three or higher has a small significant effect on EE (t\,(441) = 1.99; p = .05).

To address H3, the main drivers for acceptance are experience in using telemedicine, frequency of using AI, trusting more frequently in AI-based CAs when it comes to coping with grief and increased generalized anxiety or psychological distress. Age and male gender have no significant impact on BI, thus this hypothesis cannot be fully confirmed.

\subsection{Job Related Use by Specialists}
Using Pearson correlation for BI, PE, EE, CoA and QO with two-side t-tests shows a strong correlation between BI and PE (r\,(349) = .76; p\,<\,.001) and a medium correlation between BI and EE (r\,(349) = .37;  p\,<\,.001). CoA and EE have a medium negative correlation (r\,(349) = -.38;  p\,<\,.001) whereas QO and PE show a strong correlation (r\,(349) = .61;  p\,<\,.001). 

With a mean value of 1.84, the CoA is low and the expected quality of the response given by the AI is in the lower middle range with a mean value of 2.39. The willingness to use AI at work is moderate with a mean value of 2.79 and the PE for relief at work is moderate with a mean value of 2.86. A mean value of 3.34 in EE shows a moderate expected ease of use of this technology. A total of 39\,\% of the participants show low acceptance in using this technology in their daily work as specialist, 35\,\% show medium and 26\,\% display high acceptance. Expected relief by using this technology is low for 32\,\%, medium for 41\,\% and high for 21\,\%. Most of the respondents (61\,\%), describe their expected ease of use as moderate, 29\,\% state a high ease and 10\,\% a low one.

The acceptance seems to increase significantly for the following factors: young age (F\,(5,345) = 4.22; p\,<\,.001) -- especially the age 25 -- 34 --, less job experience (F\,(3,347) = 3.62; p = .01), offering more frequently telemedicine (F\,(4,346) = 19.83; p = .02), more frequently using AI in general (F\,(4,346) = 19.83; p\,<\,.001), experience of using AI for interaction with patient (t\,(64) = 6.64; p\,<\,.001) and a PWBI of four or higher (t\,(49) = 2.33; p = .024).

The belief in relief at work due to AI-based CAs is significantly influenced by younger age (F\,(5,345) = 3.96; p = .002), less job experience (F\,(3,347) = 3.55; p = .014), more frequently providing telemedicine (F\,(4,346) = 4.08; p = .003), more frequently using AI in general (F\,(4,346) = 20.24; p\,<\,.001), experience in using AI for patient interaction (t\,(62) = 6.58; p\,<\,.001) and a PWBI of four or higher (t\,(49) = 2.65; p = .01).

Younger age (F\,(5,345) = 8.22; p\,<\,.001), less job experience (F\,(3,347) = 4.78; p = .003), more frequently using AI in general (F\,(4,346) = 12.43; p\,<\,.001), experience in using AI for patient interaction (t\,(58) = 2.94; p = .005) also significantly increase the expectation of the ease of use.

To address H3, the main drivers for the specialist acceptance seem to be young age, less job experience, more frequently providing telemedicine, frequency of using AI and experience in using it for patients as well as an increased PWBI. Only male gender has no significant impact on BI, thus this hypothesis cannot be fully confirmed.

\subsection{Comparison of Population and Specialists}
The acceptance of the use of AI-based CAs in the field of mHealth appears to be significantly higher among the population -- who would use them for their well-being -- than among professionals who would use them for their work. In \autoref{tab:Acceptance_compare_Population_Experts} the population shows a significant higher value on the ease of use and in the trust of the answer quality compared to the specialists. The expectation for providing help to increase the general well-being is medium for the population as well as the expectation to get work faster done by the specialists. There is no significant difference in terms of PE, but according to RQ2, both the population and professionals have a moderate acceptance of the use of AI-based CAs for mHealth, with that of the population being higher.

\begin{table} [h!]
    \centering
    \begin{tabular}{cccccc}
    \hline
     &  \textbf{Population} & \textbf{Specialists} &  & \\
     \hline
    Variable & Mean (SD) & Mean (SD) & Test& P value & Cohen's d\\
    \hline
    BI &  3.13 (1.08) & 2.79 (1.20) & t(712) = 4.21 & <.001 & 1.14\\
    PE &  2.88 (1.02) & 2.86 (1.14) & t(710) = 0.32 & .75 & 1.07\\
    EE &  3.85 (0.81)& 3.34 (0.84) & t(739) = 8.59 & <.001 & 0.82\\
    CoA &  1.77 (0.76)& 1.84 (0.92) & t(671) = 1.29 & .20 & 0.84\\
    QO &  2.81 (0.87)& 2.39 (0.89) & t(742) = 6.70 & .001 & 0.88\\
    \hline
    \end{tabular}
    \caption{UTAUT2 and TAM3 variables comparing the population (n\,=\,444) and the specialists (n\,=\,351)}
    \label{tab:Acceptance_compare_Population_Experts}
\end{table}

\section{Drivers and Barriers for the Use of AI-Based Conversational Agents}
To capture the advantages and disadvantages of using AI-based CAs in mHealth, both groups were given a list of potential advantages on which they could indicate on a 5-point Likert scale whether these apply or not (5 = completely agree to 1 = completely disagree). In addition, optional text fields were provided for extra advantages and obstacles.

\subsection{General Population}
\begin{figure}[H]
    \centering
    \includegraphics[width=1.0\linewidth]{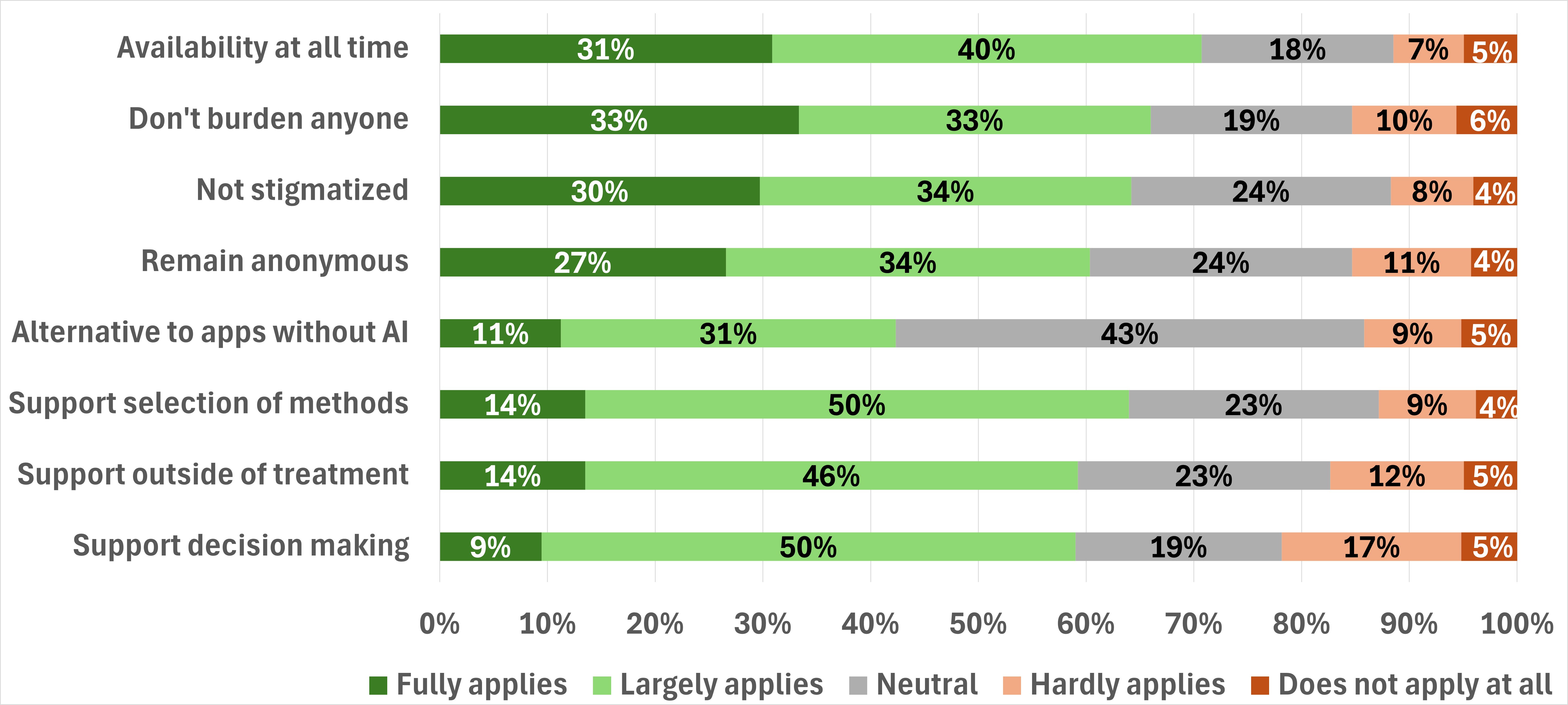}
    \caption{Advantages to the use of AI in the German population}
    \label{fig:population_advantages}
\end{figure}

Eight potential benefits were presented to the population and they were asked to rate their application. As shown in \autoref{fig:population_advantages}, availability at all times is rated highest, as is support in choosing methods and implementations, as well as not burdening anyone or feeling stigmatized. AI as a suitable alternative to apps without this technology is rated lowest. A total of 40 individual statements were made in the text field, including the possibility of using big data, flexibility in terms of location in addition to flexibility in terms of time, objectivity of interaction, low-threshold offer to bridge waiting times, AI does not tire and does not itself have mood swings, patience and the provision of multiple examples and explanations.

\begin{figure}[H]
    \centering
    \includegraphics[width=1.0\linewidth]{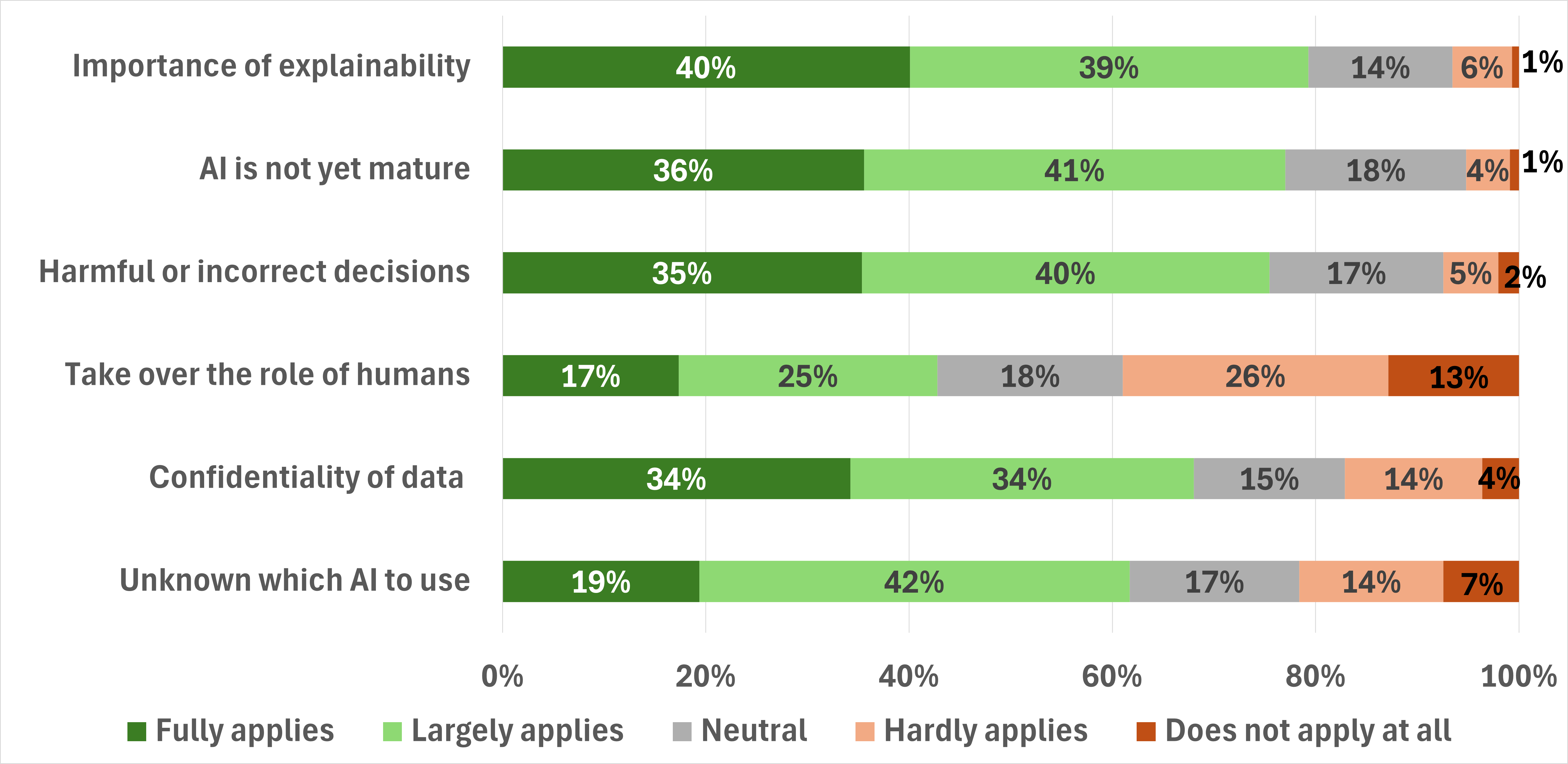}
    \caption{Obstacles to the use of AI in the German population}
    \label{fig:population_disadvantages}
\end{figure}

On a separate page six potential obstacles were provided to rate. Explainability -- understanding how or through what the AI arrives at the answer -- is rated highest as well as that AI is not yet mature and could therefore provide harmful or incorrect decisions. Most of the respondents seem not to be worried that this technology can take over the role of humans in healthcare as visualised in \autoref{fig:population_disadvantages}. In addition a total of 37 individual statements were provided in the text field, including bias, missing empathy, cannot react to gestures and facial expressions, high energy consumption, lack of intuition, does not ask questions or disagree statements, unclear liability in case of harm and hallucination.

\subsection{Mental Health Specialists}
\begin{figure}[H]
    \centering
    \includegraphics[width=1.0\linewidth]{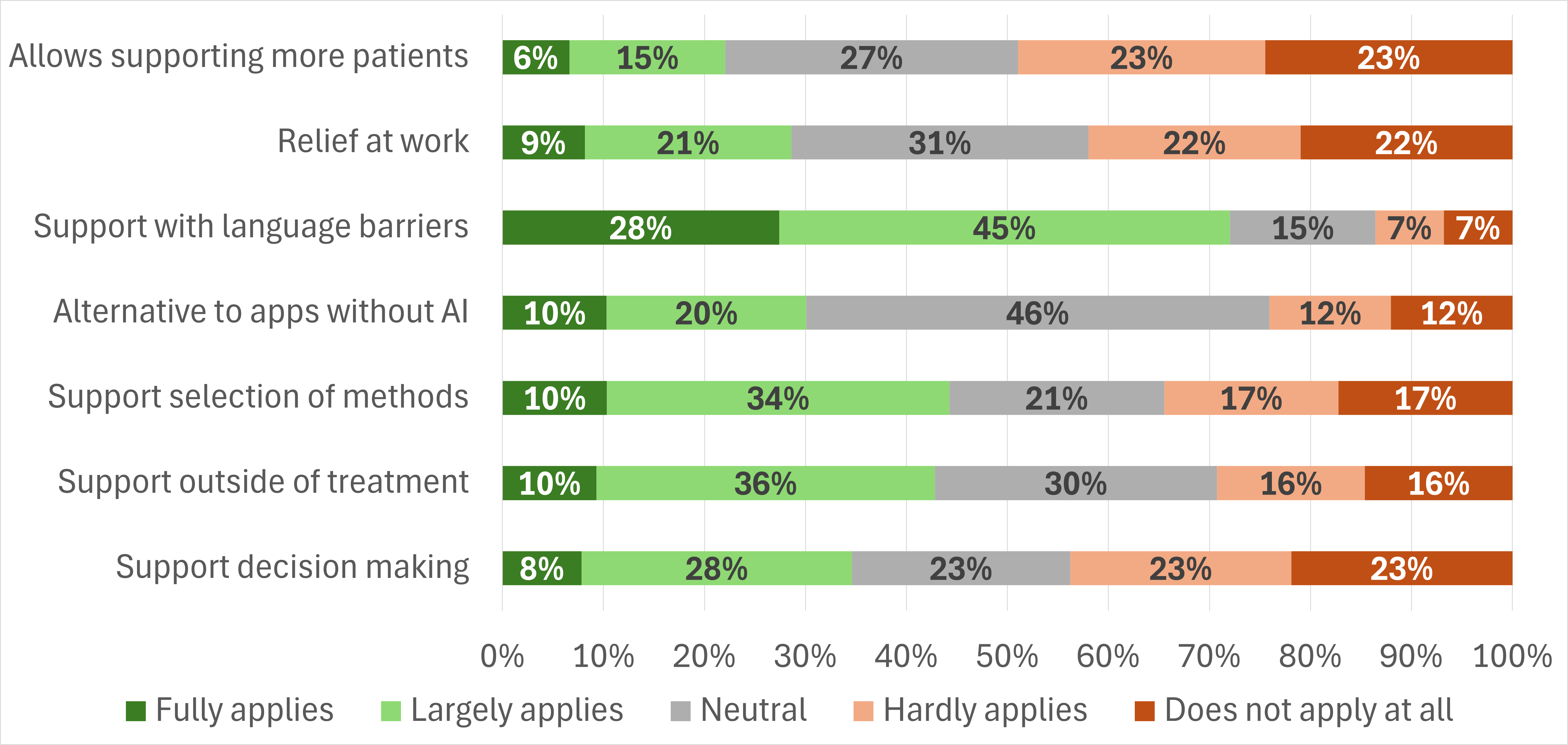}
    \caption{Advantages to the use of AI among the German mHealth specialists}
    \label{fig:experts_advantages}
\end{figure}

Seven possible advantages were offered to the experts for evaluation. The support of AI for language barriers (e.\,g. foreign languages) was rated highest, whereas most of the additional provided advantages did not receive approval, as shown in \autoref{fig:experts_advantages}. Most participants are not in favour of the possibility of using AI to care for more patients or find relief in the workplace. A total of 40 individual statements were made. In addition to references to a lack of personal experience and the naming of disadvantages, some advantages were mentioned. Advantages and other areas of application: objectivity, customized concepts, possible time savings through formulation suggestions for treatment-related specialist texts, expert statements and findings, cost savings for the healthcare system, low-threshold offer, compensation for daily form fluctuations, consideration of intercultural aspects, stimulation of new hypotheses and findings, intervention selection, support in therapy planning, appointment organisation, billing support, feedback addressed to the therapist and for supervision, $24/7$ availability and conducting initial consultations. 

\begin{figure}[H]
    \centering
    \includegraphics[width=1.0\linewidth]{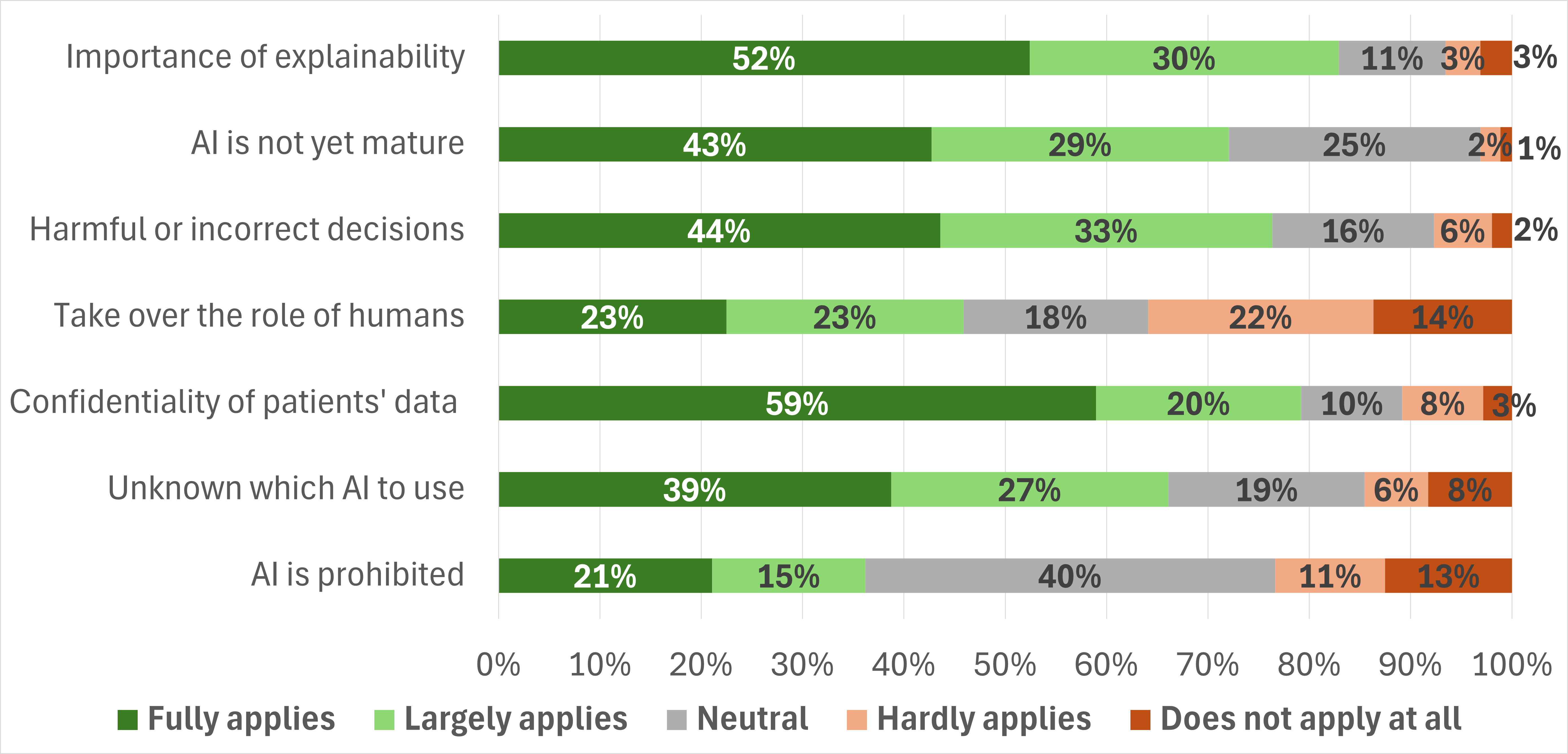}
    \caption{Obstacles to the use of AI among the German mHealth specialists}
    \label{fig:experts_disadvantages}
\end{figure}

In addition seven potential disadvantages were provided to rate. Explainability of the response provided by the AI, as well as confidentiality of patients' data and potential harmful or incorrect decisions are rated highest. AI seems not to be prohibited and most of the experts seem not to worry, that this technology could take over the role of humans as shown in \autoref{fig:experts_disadvantages}. 56 extra comments were provided to mention additional obstacles and risks. Higher frustration regarding therapy success, high costs or subscription models for using generative AI models, disaffection and dehumanization, in-transparent algorithms, unclear what organizations do with the data they collect, missing empathy, missing gesture and mimic interpretation, wrong and harmful suggestions, missing intuition, too much time in front of the computer or on the internet for the patients, removal of jobs, bias, hallucination, ethical considerations, counsellor/therapist becomes ‘lazy to think’, AI as a disruptive factor for the therapeutic relationship, supports isolation, lack of relationship building, missing training and educations as well as AI supports the feeling of worthlessness in patients.

\section{Level of Agreement to Use Augmented AI}
In order to provide an insight into how and whether HI can be combined with digital communication partners for counselling, diagnosis or treatment and the extent to which this is accepted by professionals and the population, five different communication scenarios are used for evaluation. For this purpose, the degree of acceptance or rejection was asked separately for counselling, diagnosis and treatment on a 5-point Likert scale (5 = fully applies to 1 = does not apply at all). Similar to the categorization according to mean values of acceptance, low (1 -- 2.34), medium (2.35 -- 3.67) and high (3.68 -- 5) agreement is used to describe the preference of each single scenario.

\subsection{General Population}
Almost 82\,\% of participants would communicate directly with a specialist to receive advice, and almost 70\,\% would do so remotely, with high levels of approval. If the specialist uses an AI-based CA as an additional expert, only 45.7\,\% would be highly in favour of this, and 50.5\,\% would highly approve this technology as a companion controlled by the specialist. Only interacting with AI for advice is strongly favoured by 33.1\,\%. In \autoref{tab:AuI_number_and_percent_population}, a similar pattern emerges also for the scenarios when requesting a diagnosis or treatment.

\begin{table}[h!]
    \centering
    \begin{tabular}{lccc p{12em} p{7em} p{7em} p{7em}}
        \hline
         & \textbf{Low} & \textbf{Medium}& \textbf{High} \\
        \textbf{Communication with} & n [\,\%] & n [\,\%] & n [\,\%]  \\
        \hline
         \textbf{Counselling} &&& \\
         specialist directly & 44	[9.91]& 38	[8.56]& 362	[81.53] \\
         specialist from distance & 72	[16.22]& 63	[14.19] & 309	[69.59]\\
         specialist using AI & 126	[28.38] & 115	[25.90] & 203	[45.72]\\
         AI controlled by specialist & 126	[28.38]& 94	[21.17] & 224	[50.45]\\
         AI only & 199	[44.82]& 98	[22.07]& 147	[33.11]\\
         
         \textbf{Diagnosis} & && \\
         specialist directly & 36	[8.11] & 27	[6.08]& 381	[85.81]\\
         specialist from distance & 131	[29.50]& 81	[18.24] & 232	[52.25]\\
         specialist using AI & 149	[33.56]& 120	[27.03] & 175	[39.41]\\
         AI controlled by specialist & 157	[35.36] & 103	[23.20]& 184	[41.44]\\
         AI only & 265	[59.68]& 102	[22.97] & 77	[17.34]\\

         \textbf{Treatment} &&& \\
         specialist directly & 27	[6.08]& 33	[7.43]& 384	[86.49]\\
         specialist from distance & 112	[25.23]& 83	[18.69] & 249	[56.08]\\
         specialist using AI & 151	[34.01]& 124	[27.93] & 169	[38.06]\\
         AI controlled by specialist & 154	[34.68] & 106	[23.87]& 184	[41.44]\\
         AI only & 260	[58.56] & 106	[23.87]& 78	[17.57]\\
         \hline      
    \end{tabular}
    \caption{Number and percentage of agreement for the population communication with specialists including AuI (n\,=\,444)}
    \label{tab:AuI_number_and_percent_population}
\end{table}

Comparing the five scenarios in \autoref{tab:AuI_population} with different human-AI compositions, it gets visible that most participants agree to communicate with the expert directly. In this section differences in counselling, diagnosis or treatment can be observed (F\,(2,1329) = 5.88; p = .003). Interacting with an expert from distance is less recognized than direct interaction. In addition, the differences between counselling, diagnosis and treatment (F\,(2,1329) = 13.91; p\,<\,.001) becomes visible.

In the third scenario the expert can use an AI during the interaction (e.\,g. as additional expert), which finds medium approval by the participants and potential patients, but shows no significant difference between counselling, diagnosis and treatment (F\,(2,1329) = 2.70; p = .068). The rather moderate acceptance does not correspond to the hypothesis H6. Directly interacting with AI via email or chat as companion, which is controlled by the specialist as potential ability for interaction out of the therapy, shows a slight increased agreement -- compared to interactions with the experts using AI for counselling. This contradicts the hypothesis H7, because the degree of agreement in using it is higher compared to the degree of the direct interaction with an expert, who uses AI. Interacting with AI only seems to be less accepted by the participants, especially when it comes to the topic of receiving a diagnosis or treatment. Not only does the compare of counselling, diagnosis and treatment show differences in each section. Furthermore a decline regarding the degree of agreement with the distance and inclusion of AI significantly for counselling (F\,(4,2215) = 91.86; p\,<\,.001), diagnosis (F\,(4,2215) = 170.50; p\,<\,.001) and treatment (F\,(4,2215) = 177.02; p\,<\,.001) is noted.

\begin{table}[H]
    \centering
    \begin{tabular}{p{10.5em} p{4.5em} p{4.5em} p{4.5em} p{4em} p{2.5em}}
        \hline
         & \textbf{Counselling} & \textbf{Diagnosis }& \textbf{Treatment} &  &  \\
        \textbf{Communication with} & Mean (SD) & Mean (SD) & Mean (SD) & Test F\,(2,1329) & P value \\
        \hline
         specialist directly &  4.10	(1.02)& 4.29	(0.95) & 4.31	(0.94) & 5.88& .003\\
         specialist from distance & 3.73	(1.11) &  3.33	(1.21)&  3.44	(1.17)& 13.91& <\,.001\\
         specialist using AI & 3.19	(1.16) & 3.05	(1.18) & 3.02	(1.18) & 2.70 & .068\\
         AI controlled by specialist& 3.25	(1.17) & 3.03	(1.21) & 3.04	(1.22) & 4.54 & .011\\
         AI only & 2.75	(1.25) & 2.32	(1.16) & 2.34	(1.18) & 18.79 & <\,.001\\
         \hline      
    \end{tabular}
    \caption{Mean of agreement for the population communication with specialists including AuI (n\,=\,444)}
    \label{tab:AuI_population}
\end{table}

\begin{figure}[H]
    \centering
    \begin{subfigure}{0.9\textwidth}
    \centering
    \includegraphics[width=0.9\linewidth]{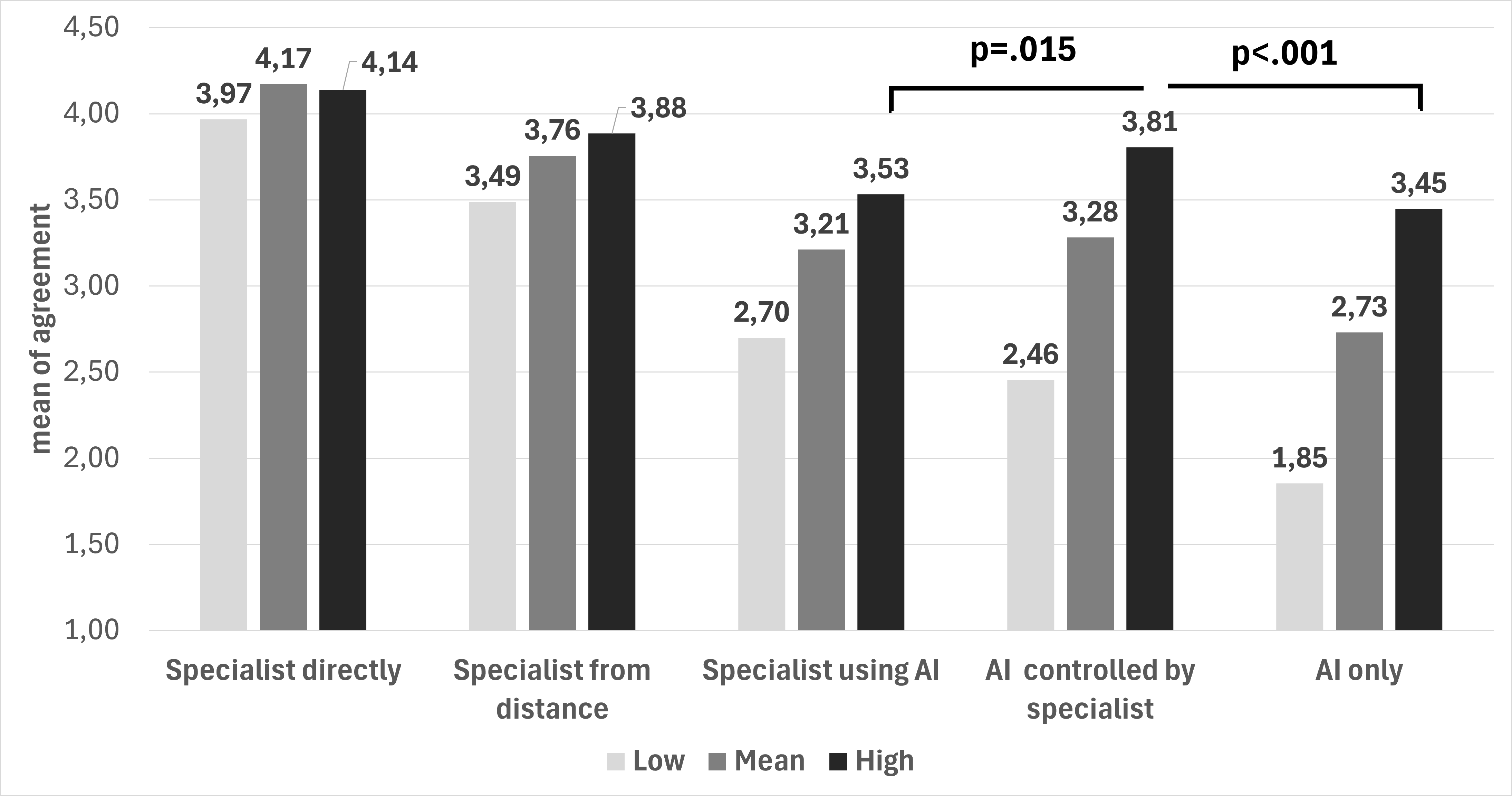} 
    \caption{Mean value of agreement with counselling, grouped by BI for the population}
    \label{fig:AuI_BI_population}
    \end{subfigure}
    
    \begin{subfigure}{1.0\textwidth}
    \centering
    \includegraphics[width=1.0\linewidth] {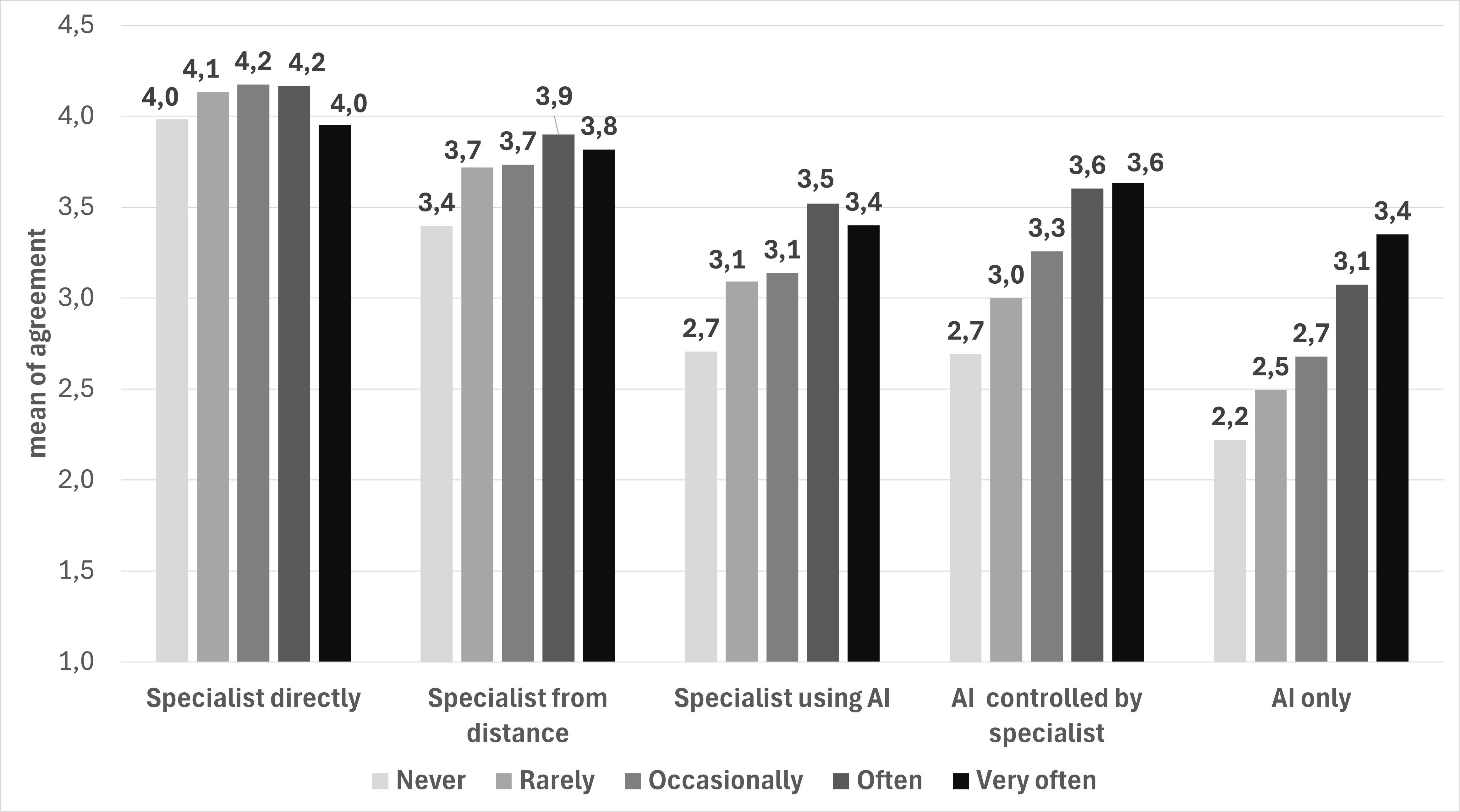}
    \caption{Mean value of agreement with counselling, grouped by frequency of using AI for the population}
    \label{fig:AuI_AI_frequency_population}
    \end{subfigure}
    
    \begin{subfigure}{1.0\textwidth}
    \centering
    \includegraphics[width=0.7\linewidth] {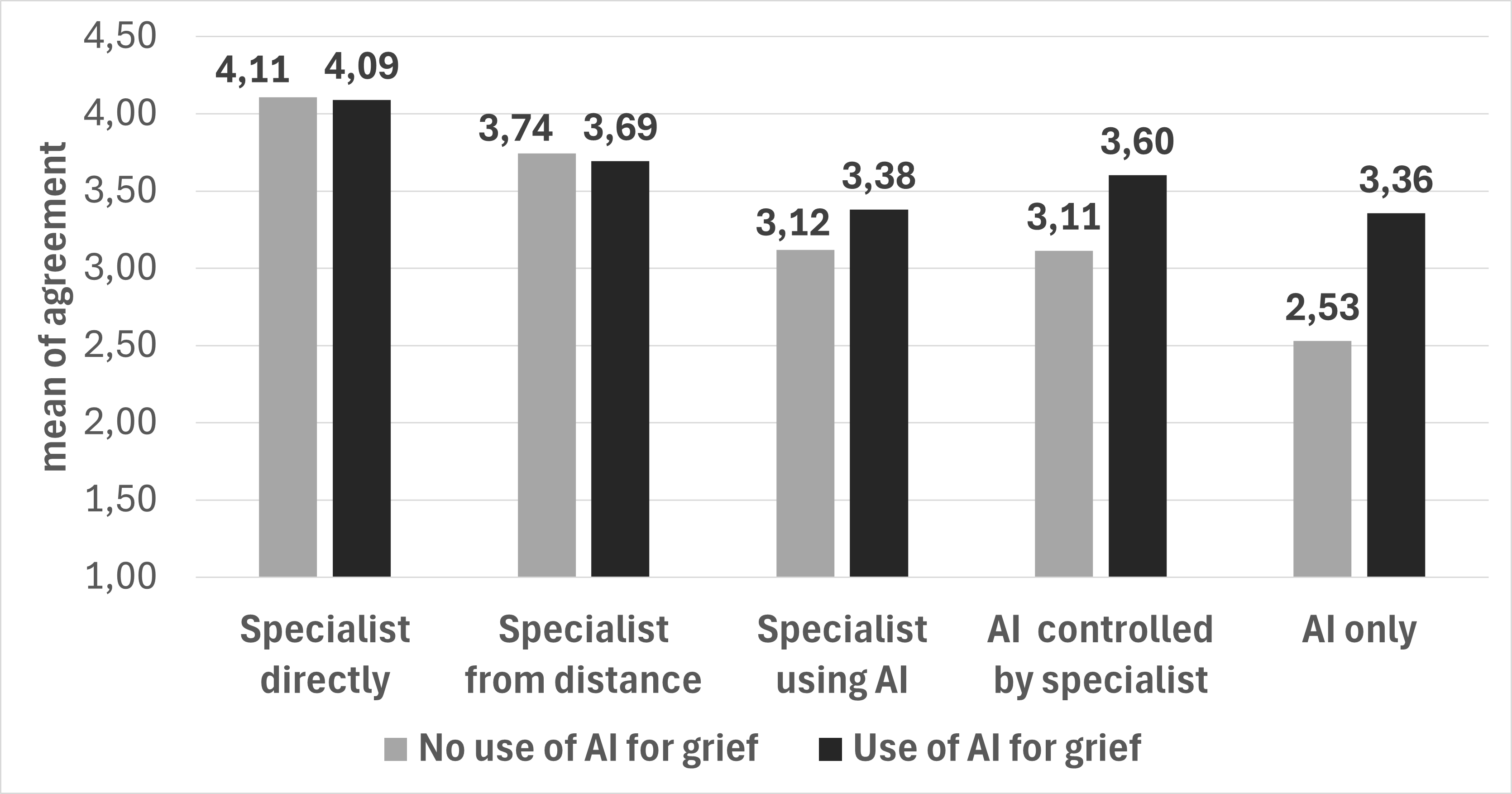}
    \caption{Mean value of agreement with counselling, grouped by experience in using AI against grief for the population}
    \label{fig:AuI_AI_exp_sorrow_population}
    \end{subfigure}
\caption{Comparison of the communication scenarios and mean values of the population's agreement (n\,=\,444)}
\label{fig:Comparing_Distribution_AI}
\end{figure}

For the further analysis, the focus is placed on counselling, as this is where acceptance for the use of AI is strongest. The grouped mean values of the agreement according to participants with low, medium or high BI show a significantly higher agreement in involving AI \autoref{fig:AuI_BI_population}. In particular, the use of AI as a companion is more likely to be accepted than the specialist using AI as an additional expert (t\,(318) = 2.44; p = .015) or the exclusive use of AI (t\,(319) = 3.21; p\,<\,.001). A similar agreement can be observed for the frequency of AI usage in general (\autoref{fig:AuI_AI_frequency_population}) and for grief-related views (\autoref{fig:AuI_AI_exp_sorrow_population}). To address RQ3, the acceptance of integrating AI in different scenarios during counselling, diagnosis or treatment is in most cases medium for the population.

\subsection{Mental Health Specialists}
Most of the specialists (94\,\%) would communicate directly with their patients to provide a consultation. 86\,\% would do this remotely and about 26\,\% would use an AI as an additional expert during this interaction. 20\,\% would accept that a patient communicates directly with an AI-based CA and monitor the communication (e.\,g. email, chat) and correct it if necessary in order to provide a counselling. 11\,\% are in favour of a patient's direct communication with an AI-based CA without the intervention of a specialist during counselling. In order to offer a diagnosis or treatment, the consent to do this remotely decreases, as well as the interest in involving an AI as a companion. \autoref{tab:AuI_number_and_percent_specialists} provides details on the level of consent for each individual scenario.

\begin{table}[H]
    \centering
    \begin{tabular}{lccc p{12em} p{7em} p{7em} p{7em}}
        \hline
         & \textbf{Low} & \textbf{Medium}& \textbf{High} \\
        \textbf{Patient communicates with} & n [\,\%] & n [\,\%] & n [\,\%]  \\
        \hline
         \textbf{Counselling} & && \\
         specialist directly &15	[4.27]& 6	[1.71]& 330	[94.02]\\
         specialist from distance &29	[8.26]&20	[5.70]& 302	[86.04]\\
         specialist using AI &194	[55.27]&67	[19.09]& 90	[25.64]\\
         AI controlled by specialist &224	[63.82]&56	[15.95]& 71	[20.23]\\
         AI only&252	[71.79]&57	[16.24]& 42	[11.97]\\
         
         \textbf{Diagnosis} & && \\
         specialist directly & 14	[3.99]& 13	[3.70]& 324	[92.31]\\
         specialist from distance &194	[55.27]&49	[13.96]& 108	[30.77]\\
         specialist using AI &183	[52.14]&79	[22.51]& 89	[25.36]\\
         AI controlled by specialist&243	[69.23]&61	[17.38]& 47	[13.39]\\
         AI only &308	[87.75]&30	[8.55]& 13	[3.70]\\

         \textbf{Treatment} &&& \\
         specialist directly &12	[3.42]&10	[2.85]& 329	[93.73] \\
         specialist from distance &114	[32.48]&65	[18.52]& 172	[49.00]\\
         specialist using AI &229	[65.24]&65	[18.52]& 57	[16.24]\\
         AI controlled by specialist&262	[74.64]&44	[12.54]& 45	[12.82]\\
         AI only &306	[87.18]&29	[8.26]& 16	[4.56]\\
    \end{tabular}
    \caption{Number and percentage of agreement for specialists' communication with patients including AuI (n\,=\,351)}
    \label{tab:AuI_number_and_percent_specialists}
\end{table}

Communicating with patients directly is highly applied, with no significant difference between counselling, diagnosis or treatment. Interaction from distance, especially for diagnosis and treatment, is high applied for counselling and medium applied for diagnosis and treatment. Using AI as an additional expert is medium applied for counselling and diagnosis but low for treatment, whereas AI controlled by experts for communication with patients is low applied for all offerings. These results are largely consistent with H4 and H5. AI only interaction for patients is not applied by the requested specialists. In total the degree of approval declines with the distance and inclusion of AI significantly for counselling (F\,(4,1750) = 463.45; p\,<\,.001), diagnosis (F\,(4,1750) = 408.83; p\,<\,.001) and treatment (F\,(4,1750) = 517.26; p\,<\,.001) -- as provided in \autoref{tab:Aui_expertsl}.

\begin{figure}[H]
    \centering
    \begin{subfigure}{1.0\textwidth}
    \centering
    \includegraphics[width=0.8\linewidth]{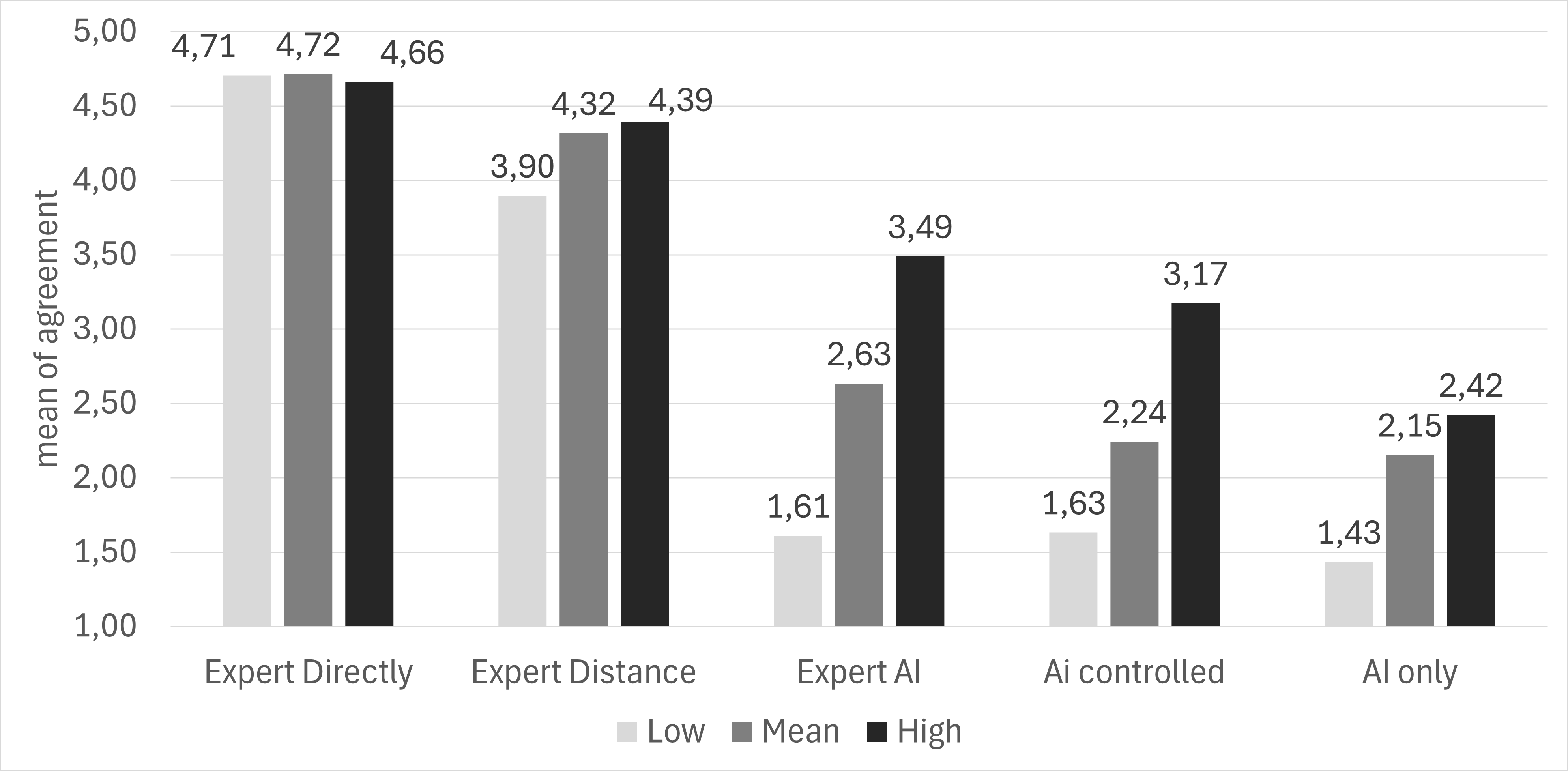}
    \caption{Mean value of agreement with counselling, grouped by BI for the specialists}
    \label{fig:AuI_BI_specialists}
    \end{subfigure}
    
    \begin{subfigure}{1.0\textwidth}
    \centering
    \includegraphics[width=0.9\linewidth]{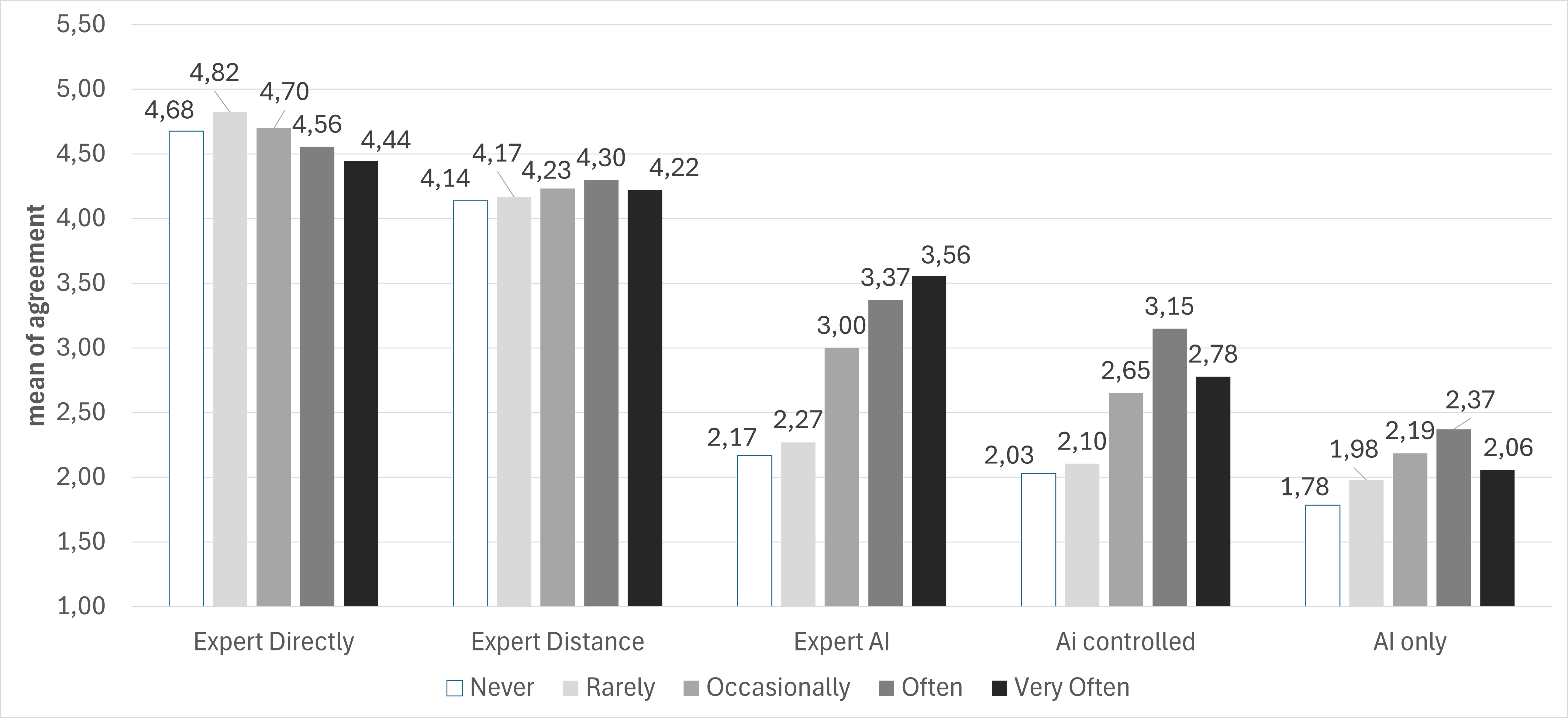}
    \caption{Mean value of agreement with counselling, grouped by frequency of use of AI for the specialists}
    \label{fig:AuI_AI_frequency_specialists}
    \end{subfigure}
    
    \begin{subfigure}{1.0\textwidth}
    \centering
    \includegraphics[width=0.9\linewidth]{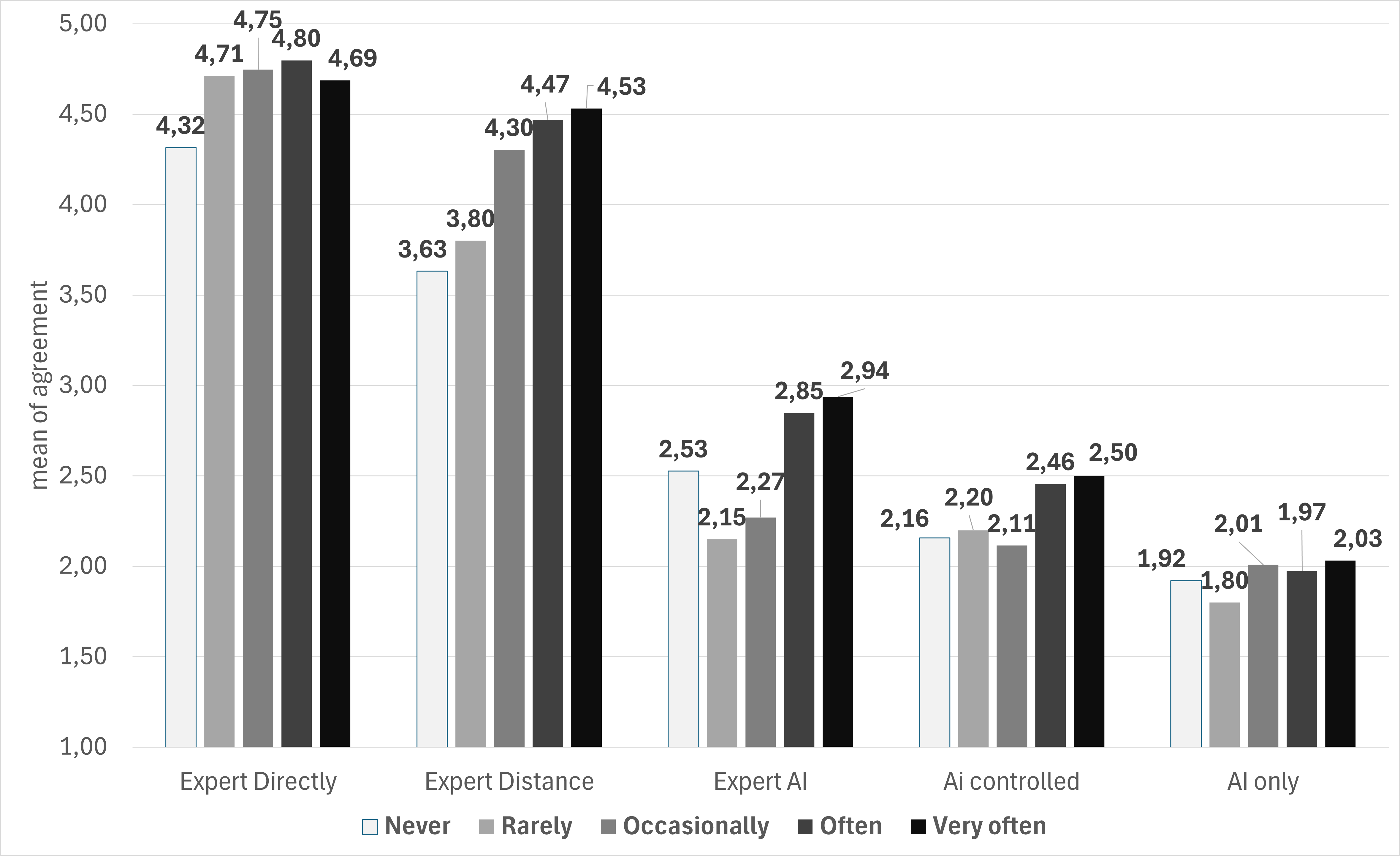}
    \caption{Mean value of agreement with counselling, grouped by frequency of providing telemedicine}
    \label{fig:AuI_Telemedicine_specialists}
    \end{subfigure}
\caption{Comparison of the communication scenarios and mean values of the specialists' agreement (n\,=\,351)}
\label{fig:Comparing_Distribution_AI}
\end{figure}

\begin{table}[H]
    \centering
    \begin{tabular}{p{10.5em} p{4.5em} p{4.5em} p{4.5em} p{4em} p{2.5em}}
        \hline
         & \textbf{Counselling} & \textbf{Diagnosis }& \textbf{Treatment} &  &  \\
        \textbf{Patient communicates with} & Mean (SD) & Mean (SD) & Mean (SD) & Test F\,(2,1050) & P value \\
        \hline
         specialist directly & 4.70	(0.79) & 4.67	(0.82)& 4.75	(0.74)& 0.99 & .37\\
         specialist from distance &4.17	(0.95) & 2.68	(1.30)& 3.22	(1.28) &141.78 & <\,.001 \\
         specialist using AI & 2.46	(1.23) & 2.46	(1.29) & 2.17	(1.18) &6.51& .002 \\
         AI controlled by specialist & 2.25	(1.22)& 2.06	(1.14) & 1.95	(1.12) & 6.17& .002\\
         AI only & 1.95	(1.15) & 1.56	(0.85) & 1.56	(0.88) &18.69& <\,.001\\
         \hline      
    \end{tabular}
    \caption{Mean of agreement for specialists communicating with patients including AuI (n\,=\,351)}
    \label{tab:Aui_expertsl}
\end{table}
Regarding the specialists, they are most likely to agree with the inclusion of AI in interactions with patients during counselling. Comparing the mean values of agreement for counselling the willingness to use AI differs. Grouped by BI (\autoref{fig:AuI_BI_specialists}) and by the frequency of a general use of AI already being in progress (\autoref{fig:AuI_AI_frequency_specialists}) shows increases. Besides that, the frequency in offering telemedicine shows differences in the means of agreement including AI as additional expert (\autoref{fig:AuI_Telemedicine_specialists}). To address RQ3, the acceptance to integrate AI with different scenarios during counselling, diagnosis or treatment is in most cases low for the specialists. Compared to the population, specialists seem also to be less willing to integrate AI as additional expert.

%% file: content/discussion.tex
\chapter{Discussion}
In order to fill the research gap regarding the experience and acceptance of AI-based CAs in the field of mHealth for both involved roles, two quantitative online-surveys were used to access potential patients and specialists in Germany. The collected results of 444 people of the German population and 351 experts were analysed independently, furthermore a comparison between both groups was examined, in order to answer the research questions and hypothesis. In this section, the most relevant principal findings are described and compared with the previous work and literature, beyond that, existing limitations are addressed. Finally a conclusion along with suggestions for future research are given.

\section{Principal Findings}
Mental health is a broad topic that is difficult to define and the topic of AI in this area is controversial, hyped and feared at the same time. In order to reach as many potential affected people and providers in Germany as possible, various media were used to distribute the survey. Advertising via social media, distribution via newsletters and websites of relevant sites on mHealth topics, online panels and student platforms as well as emails were intended to ensure a mix of participants from the population. In addition to the distribution via various organisations, social media and emails, care was taken to disseminate the survey in order to reach a large number of participants from professional circles.

\subsection{Sociodemographics and Mental Health}
The sample of the population indicates a young (51\,\% between 18 and 34 years old), highly educated (56\,\% having a university degree) part of the German population, with 53.6\,\% being female, 44.1\,\% male and 1.8\,\% non-binary. 
The level of anxiety being 45\,\%, measured with GAD-2, is much higher than expected. Besides that, the level of psychological disorder is quite high: 12\,\% of the participants were reaching the threshold. The number of the participants never used telemedicine in general -- not specified for psychology -- is quite high (59\,\%), whereas 89\,\% of the specialist surveyed offer teletherapy, telemedicine or counselling from distance. Participants of the survey requesting mHealth specialists were older (63\,\% are aged 45 or older) compared to the participants of the population. 57\,\% having a psychotherapeutic occupation, with 68\,\% being female and 30\,\% male. 11\,\% of them having a higher PWBI, indicating less job related well-being. All in all, both surveys provide an acceptable sample of the German population and the circle of professionals.

\subsection{Distribution of AI-Based Conversational Agent Usage}
Most (85\,\%) of respondents from the population have been using AI-based CAs in general and already 28\,\% confide their worries, concerns or fears to an AI-based CA -- 15\,\% of them rarely do so. In comparison, 52\,\% of the professionals surveyed use AI-based CAs in general and 13\,\% -- 8\,\% of them rarely -- use this technology for patient counselling or treatment. These observations allow H1 to be confirmed and RQ1 to be answered. This confirms that not only regarding mHealth, but generally speaking, the population in Germany seems to use this technology more often than the German specialists do. Positive favouring factors to use it, e.\,g. to confide grief or worries, are younger age, higher frequency or experience in the usage of telemedicine and higher frequency this technology use in general for the population and potential patients. Using AI for advising or counselling patients seems to be influenced positively by higher frequency of this technology's usage in general and providing more frequently telemedicine.

Not only is the frequency of use of AI-based CAs in general and for mHealth higher among the population than among specialists, but also the belief that AI can help with psychological stress. 43\,\% of the participants of the population belief or strongly belief that AI can help them with mild psychological stress, whereas 33\,\% disagree or strongly disagree. In contrast to this, 33\,\% of the requested specialists strongly agree or agree on this and 42\,\% strongly disagree or disagree. Drivers for agreement are the more frequently use of AI-based CAs, younger age and more often offering telemedicine. This confirms H2 that the belief that AI can help with mild mental disorders is higher among the population than among professionals, and leads to a comparison of acceptance.

\subsection{Acceptance of Using AI-Based Conversational Agents}
The wide spread UTAUT2 and TAM3 are used to measure acceptance. The acceptance to use and the belief to improve mHealth in the population is moderate, but is increased for people who are more frequently using this technology, more frequently use telemedicine or have a increased GAD-2 or K6 index. To use AI-based CAs at work and belief in relief is moderately agreed by the specialists. Younger age, more often provision of telemedicine, more frequent use of AI and a higher PWBI appear to have an significant effect towards a higher acceptance of this technology in the job. This summarised, that not all of the expected influencing factors from H3 were confirmed, like male gender and less job experience, but most of them.

To answer RQ2 regarding the difference in the degree of acceptance between the population and specialists in Germany, the final results are compared. In addition to the frequency and experience of use, the BI and EE of this technology are significantly higher among the population than among experts, as well as the belief and trust in the QO. With mean values of 3.13 in the population and 2.79 among specialists for BI, which correlates positively with PE and EE, acceptance is medium for both groups. These findings are emphasised by the advantages and disadvantages surveyed next.

\subsection{Drivers and Barriers for the Use of AI-Based Conversational Agents}
About 60\,\% of the population believe that this technology can provide support outside of treatment, while 46\,\% of professionals believe the same. The importance of explainability of the provided response is quite similar applied for the population (79\,\%) and the specialists (82\,\%). Furthermore, the obstacle that the provided answer could be incorrect or harmful is agreed by 77\,\% of the specialists and 75\,\% of the population, whereas 68\,\% agree in the issue about confidentiality of their data and 79\,\% of the professionals worry about the patients' data. The fear that AI takes over the humans' role is applied in both groups -- 43\,\% of the specialists and 42\,\% of the population. In the following, the degree of agreement in the combination of HI and AI as AuI is analysed.

\subsection{Level of Agreement to Use Augmented AI}
The previously measured frequency and acceptance of the use of AI also influence the way in which AI could be used in patient-physician interaction. Potential patients and specialists show more agreement in using AI for counselling compared to diagnosis or treatment. Surprisingly it is a bit more favoured by the population to communicate with an AI as companion (mean 3.25), which is controlled by the specialist than if the specialist is using an AI as additional expert (mean 3.19) during the interaction. Interacting with AI only (mean 2.75), is less agreed. This finding rejects H6 and even stands out significantly in the case of high acceptance of this technology. In a somewhat weaker form, this pattern is shown for the more often use of this technology in general, experience in using it to confide grief and experience in using telemedicine.

For professionals in the field of mHealth, the degree of agreement decreases with the degree of involving more AI into the interaction, as expected in H7. Even using this technology as an additional expert is at the lower level of the medium agreement (mean 2.46), which confirms H4. As expected in H5, offering AI as a companion, which is controlled by professionals, finds low acceptance (mean 2.25). Very low agreed (mean 1.95) is the advocacy that patients should interact with AI only.

Regarding the requested extend of using AuI for counselling, diagnosis, and treatment in RQ3, on average, only the sole use of AI in diagnosis and treatment is low accepted by the population, although this is very close to the medium acceptance. All other AI-based CA integrations are moderately agreed, whereas the specialists show only light moderate acceptance towards the usage of AI as an additional expert for counselling and diagnosis. All other communication scenarios involving AI are low agreed.

\section{Comparison With Previous Work}
Still, the prevalence of mHealth problems and mental illness is a significant problem, with worldwide reported rates ranging from 10 -- 20\,\%, and these rates continue to rise \cite{WHO-MentalDisorder2022}. As the WHO states, around 1 in 8 people in the world is currently living with a mental disorder \cite{WHO-MentalDisorder2022}. According to a DAK report, mental illnesses are the third most common group of illnesses in Germany that cause the most days of absence \cite[cf. p. 2]{DAK2022}. Analyses by the DGPPN \cite[cf. p. 1]{DGPPN2023} show that 27.8\,\% of the adult German population are affected by a mental disorder, of which only 18.9\,\% contact service providers each year. ‘The Next Global Pandemic: Mental Health’ is feared and expected, as a blog post from 2021 titles \cite{Clifton2021}. To counteract this fear, various technological offers such as telemedicine, DiGAs and blended therapy options have already been introduced in Germany in recent years. Before these have been properly integrated, accepted, prescribed and offered by the German healthcare system, new solutions such as generative AI or other AI integrations are being propagated and entering the market. Solutions for the integration of AI are already requested in some areas of healthcare and patient interaction, as stated in a position paper published by the German General Practitioners' Association in July 2024 \cite{Hausärzteverband2024}. But instead of AI-based CAs being offered as thus complete solutions, this current study analyses their distribution and acceptance in both groups involved in Germany. 

So far, no comparable study has been found in the existing literature, presumably because this technology is still relatively new. The difficulty of finding enough participants certainly plays a role, as it is a very controversial topic, especially in professional circles and one that is strongly opposed. The previous exchange with organisations, associations and individual professionals involved in mHealth care confirmed this challenge, but furthermore reaffirmed the relevance of this research. The response rates of professionals are low in studies that focus purely on digitalisation and far less controversial topics. For example, $137,388$ doctors were contacted for a survey on DiGAs in 2022, to which only $2,639$ responded validly. This corresponds to a response rate of 1.9\,\% \cite[cf. p. 27 -- 28]{Gesundheit2022}. In order to increase the chance of including more experts and opinions, crisis and telephone counsellors, social therapists and general practitioners were included alongside the usual psychotherapeutic, psychological and psychiatric specialists. With this challenge in mind, the total number of 444 participants of the population and 351 of specialists is very satisfactory.

Overall, the number of participants in the population with elevated GAD-2, who reach the threshold value is quite high at 45\,\%. In the population-based mega-cohort study of the German National Cohort (NAKO), about 5.2\,\% were observed with possible current GAD symptoms \cite[cf.]{erhardt2023}. It is likely that the participants in this survey were biased and people dealing with anxiety, worry and mHealth issues were more interested and engaged with the topic, which was promoted with leaflets and social media adverts. In addition the participants were quite young and it is known that especially adolescents and young adults deal more with generalized anxiety \cite[cf. p. 7, 8]{DAK2022}. 25\,\% of the respondents, who met the threshold of GAD-2 were between 18 and 24 years old and 32\,\% between 25 and 34. The number of participants dealing with psychological disorders was quite high at 12\,\%, of which 36\,\% are between 18 and 24 years old. Young people have grown up with digital and social media, which increases the relevance for the evaluation of new technologies such as AI in sensitive topics like mHealth. In terms of job-related well-being for mHealth specialists, 11\,\% of respondents reached the threshold of the PWBI. 28\,\% of them are between 35 and 44 years old and additional 23\,\% between the ages of 45 and 54. As this method is not used very often, there is not much comparable literature. Only a study in Switzerland from 2019 shows that 20\,\% of the internal medicine residents surveyed, had a reduced sense of well-being, especially older residents \cite[cf. p. 1, 6]{Zumbrunn2020}. Overall, these results indicate that younger people in particular need help or advice in overcoming anxiety, worry and mHealth issues and middle aged and older professionals need relief.

Already a few technologies and innovations have been introduced to reach the quadruple aim, relief specialists and bridge the gap of long waiting lists. Since 2018, specialists who include telemedicine and teletherapy in their portfolio, have been permitted to apply exclusive remote treatment in Germany. Especially the pandemic populated the opportunity to reach a therapist from distance. Although this offer looks promising, it still seems to be underused \cite[cf. p. 1]{Carlo2021}, which is also observed in this survey. Most participants of the population (60\,\%) have no previous experience in telemedicine, although it is offered by most therapists. The acceptance of being virtually advised, treated or diagnosed by a specialist is rated high to medium. Several advantages like the virtual space between patient and specialist can instil a feeling of 'protection', and people like immigrants, asylum seekers or non-native speakers can receive psychiatric assistance in their native language. This language and culture barrier seems to be highly relevant, because 73\,\% of the requested specialists fully applied or applied that AI might help with language support. Beyond that remote therapy brings further advantages for the professionals -- such as increased flexibility in planning appointments as well as the ability to consult additional experts or colleagues remotely \cite[cf. p. 3]{Carlo2021}. In this survey, the mean level of agreement was high for the provision of remote counselling and medium for diagnosis and treatment. The mentioned advantages do also match with the integration of AI-based CAs, like text based companions for the patients or additional experts for the professionals. Nevertheless the disadvantages have to be faced, such as loss of personal contact, decrease of the ability to detect non-verbal cues, loss of therapeutic alliance and potential increase of loneliness. Besides that, data security and protection of sensitive data is an undisputed topic. Teletherapy as well as the integration of AI requires a focused and continuous training and enlightenment of professionals and patients to increase the awareness of the benefits and risks. These and similar advantages and obstacles were mentioned during the conducted study. In both groups surveyed it appeared, that counselling, treatment and diagnosing, when applied from distance, was less agreed on, compared to the scenario of direct contact to the specialist or patient. Due to the fact that the experience is quite low and the terminology quite broad, this might furthermore affect the acceptance.

Not only are terminology and toolkit regarding telemedicine quite broad -- from email, phone, video conference to integration of virtual reality -- the same applies for DMHI. Therefore various formats are available on the market: The range goes from self-guided or supported by specialists to blended. Although in Germany DiGAs are available since 2019, they remain quite unknown and not frequently prescribed or used. This is the reason why no questions regarding frequency of using DiGA or similar DMHI were asked in this study. But during the search for benefits regarding AI-based CAs or similar DMHI, it appeared to be more beneficial to ask, whether these technologies are a suitable alternative to digital health applications without AI (e.\,g. deprexis, Mindable). Most of the population respondents (43\,\%) rated "neutral", as well as 46\,\% of the specialists. This can indicate, that less experience was made with this opportunity. Additional research might be necessary to validate and evaluate the difference of acceptance using DiGAs compared to AI-based CAs for the population and patients. In addition hurdles like the lack of technical infrastructure and the insufficient digital health literacy of the German population can be reflected from DiGAs to AI-based CAs \cite[cf.]{Frey2022}. Possible advantages of using AI-supported CAs over DiGAs without AI could be more individualised and tailored content and language as well as the feeling of accompanied advice and treatment. But these supposed benefits need to be carefully validated in future research. Furthermore, it is important to investigate whether these are really useful for recovery and are what a patient requires. 

The combination of digital and traditional methods in a form of blended therapy is a requested topic. Potentials like patient motivation and involvement in order to increase the success of therapy and thereby reduce the need of long-term therapy and furthermore reducing the relapse quote, are rated high \cite[cf.]{Urech2019}. In case of transforming blends the patient can deepen the understanding and practise after an intervention with the therapist. If there are uncertainties or questions, a chat with an AI -- controlled by the therapist or trained staff -- can provide additional motivation, support or relief between sessions. The integration of AI-based CAs might address the issue of missing individualization and adaptability, mentioned in \cite[cf. p. 7]{Bielinski2021} and \cite[cf. p. 98]{Urech2019}. Still, a training for specialists and patients is necessary as well as easy to learn, understandable and adjustable technology solutions, which are capable of handling low bandwidth and supporting offline functionality. This combination of FtF and blended therapy while using AI-based CAs can bring potentials with it to offer personalized treatment, which enhances the self-management and motivation of the patients and reliefs specialists \cite[cf. p. 104]{Wentzel2016}.

Similar to the study conducted between 2017 and 2018 surveying students from two German universities, the participants of the population rated FtF interaction highest, followed by a combination of FtF and digital/AI-based service, before interacting with DMHI or AI-based CA alone \cite[cf. p. 5]{Kählke2024}. Besides that, in an experimental study by Esmaeilzadeh et al. in 2020 in the United State, the scientists found that a pure AI interaction as well as a AuI led to communication barriers in contrast to a direct FtF interaction without any AI included \cite[cf. p. 1]{Esmaeilzadeh2021}. These findings could be explained by the limited experience with this technology, the lack of possibility of blending DMHI and FtF but also with "privacy concerns, trust issues and communication barriers" \cite[p. 17]{Esmaeilzadeh2021}. In the quantitative experimental study by Philip et al., they found out that elderly or less-educated outpatients' acceptance in favour of a virtual medical agent in a psychiatric interview was higher \cite[cf. p. 1]{Philip2020}. This lies in contrast to this study, where age and education do not influence acceptance in using AI. Experience and frequency of use may lead to this different observation. It is not only the patient's view and acceptance that is important, but the professionals' as well.

In contrast to the conducted study in 2019 in Portugal \cite[cf. p. 17]{Pedro2023} the older and more experienced experts were not greater optimists and not more open towards incorporating AI in their environment. The current study shows, that BI for experts between the age of 25 -- 34 (mean 3.39; SD 1.14) was highest, followed by the age range of 45 -- 54 (mean 2.95; SD 1.18). Whereas PE was highest for the youngest age of 18 -- 24 (mean 3.36; SD 0.85) as well as for age 25 -- 34 (mean 3.18; SD 1.19). In addition participating specialists with low job-experienced rated BI and PE highest. These contrary results to the study from 2019, might be explained through differences in culture, time and especially experience.  

Comparing the expectations regarding the quality of the provided answer given by the AI, the QO is medium rated for both groups, but little higher for the requested population (mean 2.81; SD 0.87) compared to the specialists (mean 2.39; SD 0.89). This observation goes in line with the agreed obstacle that AI can create harmful or incorrect decisions, where 44\,\% of the specialists strongly agreed and 35\,\% of the population as well. This worry and experience in use for healthcare purpose was additionally observed in a study from 2023 in Saudi Arabia towards using ChatGPT for healthcare workers \cite[cf. p. 6]{Temsah2023}. The cultural difference shows that only 25\,\% of the participants in Saudi Arabia worry about the confidentiality of the patients data whereas in this survey 59\,\% of the German respondents strongly agree on this worry. In general professionals report more concerns and less acceptance using AI in health than people from the population do. This is also found in a systematic review of existing studies conducted by Tang et al. in 2023 \cite[cf. p. 1]{Tang2023}. In contrast to the findings of the survey from Sweeney et al. in 2021 \cite[cf.]{Sweeney2021}, the agreement to use AI is not that high, but concerns about data privacy and confidentiality were similarly high. This difference in acceptance might be explained, because during this survey in 2021, chatbots using decision trees and not generative AI were provided as example. This underlines that experience, acceptance and views will change over time. But still, the risks regarding data protection and security remains and makes clear that safety and comfort are difficult to combine.

Introducing AI in the patient-specialist interaction shows the favour of the population using it as a companion. This integration of AI during counselling aligns with the observations made by Mayer et al., that patients prefer a shared decision-making and want to be more involved \cite[cf. p. 14]{Mayer2022}. The results from the different five provided scenarios show, that a human as specialist involved in the interaction with the AI is more preferred than interacting with the AI only. Interestingly the respondents also prefer this against interacting with an specialist using an AI. From these findings it can be observed, that a human as specialist needs to stay in the loop, but patients seem to favour to be involved and not advised or treated by two experts, a human and an AI. This preference towards involving HI during interactions with AI as companion conflicts with the advantage of being independent in time and all-time availability of $24/7$.

Overall, it is clear that before the emergence and hype around AI, other technologies offered potential solutions to bridge the need for therapy places and provide support for mental and psychological health. The advantages -- availability, individualisation, self-guided -- and disadvantages -- internet availability, privacy concerns, costs -- often overlap and there is not one single solution that outshines everything else. Independent of the used technology, involvement of the patients and their individual needs are necessary and can differ from age, time, current health state and experience of technology usage. Moreover, similar to the representative survey of healthcare professionals conducted in Germany in 2022, acceptance increases with clinical evidence and the agreement that patients' wishes and needs are changing \cite[cf. p. 3]{Gesundheit2022}. But also the risks have to be considered. The two most highly rated discrepancies are the costs of prescribing and billing and data protection \cite[cf. p. 3]{Gesundheit2022}. This realisation is not only valid for DiGAs, but certainly in addition for AI-based CAs.

In total, this study can provide comparative insights into the use and acceptance of AI-based CAs in the field of mHealth in Germany of both groups -- the population and the specialists. In addition different interaction possibilities involving AI and HI were analysed for agreement and possible advantages, but also obstacles were collected. This should enable a better understanding for all stakeholders involved as to for whom and how AI could be suitable in the field of mHealth in Germany. However, not only the acceptance, but furthermore the effectiveness and sustainability of counselling or treatment are crucial. Other limitations and future perspectives are covered in the following.

\section{Limitations}
In order to reach as many participants as possible, recruitment was carried out via the internet, which can bias the acceptance rate, as it was predominantly young, tech-savvy and interested people who had access and felt addressed. In addition, AI is an almost omnipresent topic in the media, which leads to a negative attitude and thus to a refusal to participate in this study.

Although both surveys were available for 50 days, the response rate was relatively low. This period included the summer holidays, which meant that many doctors' surgeries were closed for longer periods and therefore fewer specialists were reached. View refusals were received, because the leaflet already contained information on the involvement of AI. Some emails expressed the displeasure and outrage of healthcare professionals about this topic and indicated that the surveys on this topic were getting out of hand. In addition a possible lack of an mHealth definition in the introduction to the survey and in the leaflet cannot be ruled out either, as it was mainly psychotherapists who took part. It can be assumed that many general practitioners did not feel addressed, although around $3,000$ emails were sent to this speciality, but only 7\,\% of respondents practise medicine.

In addition the population's experience with telemedicine is quite low, with almost 60\,\% having never used it before. One possible explanation for this could be that some respondents understood the question to refer to mHealth and not to telemedicine in general. As this possible misunderstanding cannot be ruled out with certainty, the results regarding the experience and frequency of use of telemedicine in the population should be treated with caution.

Furthermore, the measurement of the current mHealth status with GAD-2 shows a weak internal correlation between the two questions. Here it could be advantageous to use other or additional known measurement methods such as QOL, PHQ or Sense of Coherence in order to gain insights into resilience or the mHealth state. Additional requesting experience with counselling, coaching or therapy of the population would have been valuable. Those data could provide insights into the extent to which different degrees of the disease have an impact on acceptance and the extent to which fear of stigmatisation plays a role. Moreover, instead of measuring CoA, it would be of great interest to evaluate how good the access to stable internet is and how much fun it is to interact with a mobile device or computer, in addition to the frequency of use of these.

Most of the provided advantages and obstacles for the use of AI-based CAs in the field of mHealth for the population can be transferred to DMHI without AI. For this purpose, it would be worthwhile to survey the frequency of use of these (DiGA and freely accessible applications), the level of knowledge and training in the field of mHealth in order to find out what the actual advantages of integrated AI could be.

In addition to direct and remote interaction with specialists, three other scenarios were provided to assess the level of agreement, which involve AI to various degrees. Although images (\autoref{fig:AUI_communication_possibilities}) and examples were offered for illustration purposes, it can not completely rule out what the participants imagined. In the 'AI as a companion' scenario in particular, ideas can differ, as there are hardly any applications available in this form in other areas of life either. Moreover, the way of communication -- writing or talking -- can have an impact. Next to research on different communication media, comparative experiments involving this form of blended therapy with and without AI could be used to determine whether and what added value AI can bring at all and if it has an influence on motivation or drop-out rates. 

To overcome the limitations addressed, future research and longitudinal studies are needed to gain insights into how robust and effective the integration of AI-based CAs into patient-specialist interactions could be. To this end, the following section provides additional ideas for future research with a conclusion of this study.
  
\section{Conclusion and Future Research}
Despite the media-popular topics of mHealth and AI, no publications or studies could be found that investigate the use of AI in the field of mHealth in Germany -- both in the population and among specialists. Furthermore, comparisons are often made as to whether this technology is better than an expert. Few applications that combine AI and HI in the form of AuI appear to have been investigated so far. This study closes this gap with two separate, but comparable quantitative online-surveys. This is absolutely necessary as the topic of mHealth is gaining more and more relevance and attention, not only in schools but among employers as well. AI, with the introduction of ChatGPT in 2022, increasingly integrates in the daily life of the population. The use and interaction will certainly become commonplace, not least because more applications such as PDF readers, office products, websites with chatbots and certainly many more digital products are integrating LLaMA, GPT, Mistral or similar LLMs. Perhaps even without the user's knowledge. It is surely only a matter of time before mobile apps including AI are offered as DMHI or coaches provide coaching sessions with AI. This makes a prior understanding of acceptance indispensable in order to reach and educate relevant groups.

Working with AI-based CAs, DMHI or DiGAs requires not only access to a computer or similar device, in addition knowledge of the internet is necessary. In Germany, there are still problems with broadband internet access, especially in rural areas. Because of this, not only questions regarding computer anxiety arise, but also regarding internet familiarity, facilitating possibilities and especially internet anxiety, could have an impact and should be observed in future research. In alignment with this, DMHI without AI should be evaluated regarding acceptance, advantages and risks in more detail for both groups, so as not to overlook potential that is already available, as a lot now revolves around AI.

The frequency of use and acceptance differs between the population and professionals, as analysed in the two research questions RQ1 and RQ2. Only 15\,\% of the requested population have never used AI-based CAs before. But surprisingly, just under 28\,\% have already used them to combat grief, and most (63\,\%) of them were very satisfied or satisfied with this experience. Especially young people, people with experience in telemedicine and those using AI in general more frequently, tend to use it more often for this purpose. This observation leads to the assumption that the numbers could increase in the future. Furthermore, the agreement that it can support with mild psychological stress, which is currently at 43\,\% in the population, could increase. In contrast to this, a percentage of 33\,\% of the requested specialists agree as well. They also use AI less frequently -- 48\,\% have never used AI-based CAs before -- and about 13\,\% have experience in using it for counselling or treating patients. Young age, less job experience and providing more frequently telemedicine are factors increasing this belief as well as experience with this technology.

The frequency and experience influences the acceptance to use AI-based CAs to improve mHealth in the population, and in addition to facilitate work for specialists. BI with a mean of 3.13 is moderate for the population but is increased for those, who are more frequently using this technology, telemedicine or having an increased GAD-2 or K6 threshold. To use AI for work is accepted by the specialists with a mean of 2.79 moderately. This BI is increased significantly for young professionals, those who provide more frequently telemedicine, use AI more often or reach the threshold of the PWBI.

Not only is the belief in the population higher that AI can help with grief or mild depression, but the acceptance of using it is greater as well. Building on this and addressing RQ3, this study examines the extent to which AuI is agreed to use or offer in counselling, diagnosis and treatment. The involvement of AI in the specialist-patient interaction is agreed more on for both groups during the counselling process. Diagnosis and treatment show less agreement. With a mean of 3.25 the population agrees most in using AI as companion, compared to the involvement of AI as additional expert used by the professional with a mean of 3.19. Interacting only with AI is less agreed with a mean of 2.75. The requested specialists show more rejection in involving this technology during the counselling process. Medium agreement with a mean of 2.46 is rated to use AI as additional expert. According to the acceptance, low agreement is rated for involving AI as companion for the patient (mean 2.25) or AI as substitute (mean 1.95). However, direct communication with a professional or patient is still preferred, with most respondents showing a high level of acceptance, followed by remote interaction.

Although the vast majority of the surveyed professionals do not agree with involving AI in the interaction with patients, the collected advantages and obstacles show that AI has the potential to make things easier. Summaries, administrative tasks and reports to the health insurance company could facilitate work easier. Most rated driver for specialists using AI, was the support with language barriers. Future research could try to focus on this and analyse different solutions. In total it becomes clear, that there is a great deal of uncertainty, which AI to use and how to use it without harming personal data. Therefore guidelines and education are necessary. An improved well-being and better relief of specialists also benefits the treatment of patients. Further integration of AI for counselling providers like TelefonSeelsorge\textsuperscript{\textregistered} Deutschland e.\,V. or Nummer gegen Kummer e.\,V. might be a promising opportunity to relief care providers and keep up with the growing demand.

Additional research could specify in which use cases AI could be integrated -- measurement of vital parameters, recordings, behaviour control, addiction prevention -- and whether they offer more benefits compared to digital applications without AI. The validation of systems and architectures involving a HITL during interacting with AI-based CAs could be promising. Ideas from the customer support -- provided by Paikens et al. \cite[cf. p. 279]{Paikens2020} -- can be potentially adapted for mHealth or health support. Future research could try to develop such systems and evaluate, if even specialists with no or low professional training could provide help by taking over light cases of disease treatment. Maybe this could be seen as a stepping stone for healthcare providers to improve their work-life, as potentially untrained experts could support more, or AI could facilitate and simplify illegitimate tasks. In Germany, not only do we have too many patients and too few doctors, but even if a connection between the two sides is established, the relationship or the treatment process may not yet be right. The use of AI as an expert could enable other specialists or appropriate professionals, e.\,g. general practitioners, physician assistants or counsellors, to provide patients with care, even though there is no explicit training in this for mHealth, but they have already established a good or empathic relationship. This combination of HI and AI provides accessible, low-threshold and safe services with the ability to engage patients more fully in the process. In addition personalised advice and treatment can help to improve the patients' experience as part of the quadruple aim.

But still, the risks regarding data protection and security remain, as safety and comfort are difficult to combine. Education is of considerable relevance here in order to protect data and prevent damage to health caused by incorrect advice from the AI. As many models are already freely available on the market, it is important for professionals to familiarise themselves with the topic in order to better inform patients about the risks. The interaction and, in particular, the reflection and evaluation of the response given by an AI should be learned in order to exploit potential and minimise risks. Correct communication is everything, with or without AI, and must be a high priority at all times, especially when it comes to such important topics as mHealth.

%% file: quellen.bib
@online{WHO-MentalHealth2022,
   author = "{World Health Organisation}",
   year = "2022",
   title = {Mental Health -- Concepts in mental health},
   url = {https://www.who.int/news-room/fact-sheets/detail/mental-health-strengthening-our-response},
   note = "[online; accessed 17.11.2024]"
}

@online{WHO-MentalDisorder2022,
   author = "{World Health Organisation}",
   year = "2022",
   title = {Mental disorders},
   url = {https://www.who.int/news-room/fact-sheets/detail/mental-disorders},
   note = "[online; accessed 22.11.2024]"
}

@online{Novego,
   author = "{IVPNetworks GmbH}",
   title = {Fühl dich besser -- Novego},
   url = {https://www.novego.de/},
   note = "[online; accessed 22.11.2024]"
}

@online{BDP,
   author = "{Berufsverband Deutscher Psychologinnen und Psychologen e.V.}",
   title = {Berufsverband Deutscher Psychologinnen und Psychologen},
   url = {https://www.bdp-verband.de/},
   note = "[online; accessed 22.11.2024]"
}

@online{PsychiatrieVerlag,
   author = "{Psychiatrie Verlag}",
   title = {Psychosoziale Umschau},
   url = {https://psychiatrie-verlag.de/series/psychosoziale-umschau/},
   note = "[online; accessed 22.11.2024]"
}

@article{Weizenbaum1966,
author = {Weizenbaum, Joseph},
title = {ELIZA  -- a Computer Program for the Study of Natural Language Communication between Man and Machine},
year = {1966},
issue_date = {Jan. 1966},
publisher = {Association for Computing Machinery},
address = {New York, NY, USA},
volume = {9},
number = {1},
issn = {0001-0782},
url = {https://doi.org/10.1145/365153.365168},
doi = {10.1145/365153.365168},
journal = {Commun. ACM},
month = {1},
pages = {36 -– 45},
}

@article{DGPPN2023,
author = "DGPPN",
   title = {Basisdaten - Psychische Erkrankungen},
year = {2023},
url = "https://www.dgppn.de/_Resources/Persistent/6c85d23473cbf71340bd7bff788ad55851cf3982/20231108_Factsheet_Kennzahlen.pdf",
note = "[online; in German; accessed 03.11.2024]",
}

@article{Jacobi2016,
   author = {F. Jacobi and M. Höfler and J. Strehle and S. Mack and A. Gerschler and L. Scholl and M. A. Busch and U. Maske and U. Hapke and W. Gaebel and W. Maier and M. Wagner and J. Zielasek and H. U. Wittchen},
   doi = {10.1007/s00115-015-4458-7},
   issn = {14330407},
   issue = {1},
   journal = {Nervenarzt},
   month = {1},
   pages = {88 -- 90},
   publisher = {Springer Verlag},
   title = {Erratum to Mental disorders in the general population. Study on the health of adults in Germany and the additional module mental health},
   volume = {87},
   year = {2016},
}

@article{Shahsavar2023,
  title={User Intentions to Use ChatGPT for Self-Diagnosis and Health-Related Purposes: Cross-sectional Survey Study},
  author={Shahsavar, Yeganeh and Choudhury, Avishek and others},
  journal={JMIR Human Factors},
  volume={10},
  number={1},
  year={2023},
  publisher={JMIR Publications Inc., Toronto, Canada}
}

@online{DAK2022,
author = "DAK-Gesundheit",
title = {DAK-Psychreport},
url ={https://www.dak.de/dak/bundesthemen/erneuter-hoechststand-bei-psychisch-bedingten-fehltagen-2609614.html#/},
year = "2022",
note = "[online; in German; accessed 03.11.2024]",
}

@online{DStatist2022,
author = "Statistisches Bundesamt",
title = {Anzahl der Gestorbenen nach Kapiteln der ICD-10 und nach Geschlecht für 2022},
url ={https://www.destatis.de/DE/Themen/Gesellschaft-Umwelt/Gesundheit/Todesursachen/Tabellen/gestorbene_anzahl.html},
year = "2022",
note = "[online; in German; accessed 03.11.2024]",
}

@online{DStatist2022_sucide,
author = "Statistisches Bundesamt",
title = {Im Jahr 2022 starben in Deutschland insgesamt 10 119 Menschen durch Suizid},
url ={https://www.destatis.de/DE/Themen/Gesellschaft-Umwelt/Gesundheit/Todesursachen/suizid.html},
year = "2022",
note = "[online; in German; accessed 03.11.2024]",
}

@article{Thieme2020,
   author = {Anja Thieme and Danielle Belgrave and Gavin Doherty},
   doi = {10.1145/3398069},
   issn = {15577325},
   issue = {5},
   journal = {ACM Transactions on Computer-Human Interaction},
   month = {10},
   publisher = {Association for Computing Machinery},
   title = {Machine Learning in Mental Health: A systematic review of the HCI literature to support the development of effective and implementable ML Systems},
   volume = {27},
   year = {2020},
}

@article{Temsah2023,
   doi = {10.3390/healthcare11131812},
   issn = {22279032},
   issue = {13},
   journal = {Healthcare (Switzerland)},
   month = {7},
   publisher = {Multidisciplinary Digital Publishing Institute (MDPI)},
   title = {ChatGPT and the Future of Digital Health: A Study on Healthcare Workers’ Perceptions and Expectations},
   volume = {11},
   year = {2023},
}

@article{Carolus2021,
   author = {Astrid Carolus and Carolin Wienrich and Anna Törke and Tobias Friedel and Christian Schwietering and Mareike Sperzel},
   doi = {10.3389/fcomp.2021.682982},
   issn = {26249898},
   journal = {Frontiers in Computer Science},
   month = {5},
   publisher = {Frontiers Media S.A.},
   title = {Alexa, I feel for you! Observers’ Empathetic Reactions towards a Conversational Agent},
   volume = {3},
   year = {2021},
}

@article{Nadal2020,
   author = {Camille Nadal and Corina Sas and Gavin Doherty},
   doi = {10.2196/17256},
   issn = {14388871},
   issue = {7},
   journal = {Journal of Medical Internet Research},
   month = {7},
   pmid = {32628122},
   publisher = {JMIR Publications Inc.},
   title = {Technology acceptance in mobile health: Scoping review of definitions, models, and measurement},
   volume = {22},
   year = {2020},
}

@article{Sadiku2021, 
author = {Matthew N. O. Sadiku and Tolulope J. Ashaolu and Abayomi Ajayi-Majebi and Sarhan M. Musa},
doi = {10.51542/ijscia.v2i5.17}, 
issue = {5},
journal = {International Journal Of Scientific Advances},
publisher = {International Journal of Scientific Advances},
title = {Augmented Intelligence},
volume = {2},
year = {2021},
}

@online{krisenchat_gGmbH,
author = "krisenchat gGmbH",
title = {krisenchat -- 24/7 Krisenberatung per Chat},
url = {https://krisenchat.de/},
year = {2024},
note = "[online; accessed 10.11.2024]"
}

@online{NummerGegenKummer_eV,
author = "Nummer gegen Kummer e. V.",
title = {Kostenfreie Beratung für Eltern, Kinder und Jugendliche},
url = {https://www.nummergegenkummer.de/},
year = {2024},
note = "[online; accessed 10.11.2024]"
}

@online{Telefonseelsorge_eV,
author = "TelefonSeelsorge® Deutschland e. V. – Ökumenischer Verein für TelefonSeelsorge und Offene Tür in Deutschland",
title = {TelefonSeelsorge® Deutschland | Sorgen kann man teilen. 0800/1110111 · 0800/1110222 · 116123. Ihr Anruf ist kostenfrei.},
url = {https://www.telefonseelsorge.de/},
year = {2024},
note = "[online; accessed 10.11.2024]"
}

@article{Mayer2022,
   author = {Gwendolyn Mayer and Svenja Hummel and Neele Oetjen and Nadine Gronewold and Stefan Bubolz and Kim Blankenhagel and Mathias Slawik and Rüdiger Zarnekow and Thomas Hilbel and Jobst Hendrik Schultz},
   doi = {10.1177/20552076221091353},
   issn = {20552076},
   journal = {Digital Health},
   month = {4},
   publisher = {SAGE Publications Inc.},
   title = {User experience and acceptance of patients and healthy adults testing a personalized self-management app for depression: A non-randomized mixed-methods feasibility study},
   volume = {8},
   year = {2022},
}

@article{Nov2021,
   author = {Oded Nov and Yindalon Aphinyanaphongs and Yvonne W. Lui and Devin Mann and Maurizio Porfiri and Mark Riedl and John Ross Rizzo and Batia Wiesenfeld},
   doi = {10.1145/3417518},
   issn = {15577317},
   journal = {Communications of the ACM},
   month = {3},
   pages = {46 -- 48},
   publisher = {Association for Computing Machinery},
   title = {The transformation of patient-clinician relationships with AI-based medical advice},
   volume = {64},
   year = {2021},
}

@article{Sweeney2021,
  title={Can chatbots help support a person’s mental health? Perceptions and views from mental healthcare professionals and experts},
  author={Sweeney, Colm and Potts, Courtney and Ennis, Edel and Bond, Raymond and Mulvenna, Maurice D and O’neill, Siobhan and Malcolm, Martin and Kuosmanen, Lauri and Kostenius, Catrine and Vakaloudis, Alex and others},
  journal={ACM Transactions on Computing for Healthcare},
  volume={2},
  number={3},
  pages={1 -- 15},
  year={2021},
  publisher={ACM New York, NY, USA}
}

@article{mueller2021,
  title={The Ten Commandments of Ethical Medical AI.},
  author={M{\"u}ller, Heimo and Mayrhofer, Michaela Theresia and Van Veen, Evert-Ben and Holzinger, Andreas},
  journal={Computer},
  volume={54},
  number={7},
  pages={119 -- 123},
  year={2021}
}

@article{Palanica2019,
  title={Physicians’ perceptions of chatbots in health care: cross-sectional web-based survey},
  author={Palanica, Adam and Flaschner, Peter and Thommandram, Anirudh and Li, Michael and Fossat, Yan},
  journal={Journal of medical Internet research},
  volume={21},
  number={4},
  year={2019},
  publisher={JMIR Publications Toronto, Canada}
}

@article{Bodenheimer2014,
   author = {Thomas Bodenheimer and Christine Sinsky},
   doi = {10.1370/afm.1713},
   issn = {15441717},
   issue = {6},
   journal = {Annals of Family Medicine},
   month = {11},
   pages = {573 -- 576},
   pmid = {25384822},
   publisher = {Annals of Family Medicine, Inc},
   title = {From triple to Quadruple Aim: Care of the patient requires care of the provider},
   volume = {12},
   year = {2014},
}

@article{Crigger2022,
author = {Elliott Crigger and Karen Reinbold and Chelsea Hanson and Audiey Kao and Kathleen Blake and Mira Irons},
doi = {10.1007/s10916-021-01790-z},
issn = {1573689X},
issue = {2},
journal = {Journal of Medical Systems},
month = {2},
pmid = {35020064},
publisher = {Springer},
title = {Trustworthy Augmented Intelligence in Health Care},
volume = {46},
year = {2022},
}

@article{Dingler2021,
   author = {Tilman Dingler and Dominika Kwasnicka and Jing Wei and Enying Gong and Brian Oldenburg},
   doi = {10.1055/s-0041-1726510},
   issn = {23640502},
   journal = {Yearbook of medical informatics},
   month = {8},
   pages = {191 -- 199},
   pmid = {34479391},
   publisher = {NLM (Medline)},
   title = {The Use and Promise of Conversational Agents in Digital Health},
   volume = {30},
   year = {2021},
}

@article{Klinge2021,
  title={Videosprechstunde -- Rechtlicher Rahmen, Einsatzgebiete und Limitationen},
  author={Klinge, Kay and Bleckwenn, Markus},
  journal={Der Deutsche Dermatologe},
  volume={69},
  number={12},
  pages={1002},
  year={2021},
  publisher={Nature Publishing Group}
}

@article{Carlo2021,
   author = {Francesco Di Carlo and Antonella Sociali and Elena Picutti and Mauro Pettorruso and Federica Vellante and Valeria Verrastro and Giovanni Martinotti and Massimo di Giannantonio},
   doi = {10.1111/ijcp.13716},
   issn = {17421241},
   issue = {1},
   journal = {International Journal of Clinical Practice},
   month = {1},
   pmid = {32946641},
   publisher = {Blackwell Publishing Ltd},
   title = {Telepsychiatry and other cutting-edge technologies in COVID-19 pandemic: Bridging the distance in mental health assistance},
   volume = {75},
   year = {2021},
}

@online{KBV_Praxisbarometer2024,
   author = "{Kassenärztliche Bundesvereinigung}",
   year = "2024",
   title = {Schlussfolgerungen, Praxisbarometer Digitalisierung 2023},
   url = {https://www.kbv.de/html/praxisbarometer.php},
   note = "[online; in German; accessed 20.11.2024]"
}

@article{Ray2023,
   author = {Partha Pratim Ray},
   doi = {10.1016/j.iotcps.2023.04.003},
   issn = {26673452},
   journal = {Internet of Things and Cyber-Physical Systems},
   month = {1},
   pages = {121 -- 154},
   publisher = {KeAi Communications Co.},
   title = {ChatGPT: A comprehensive review on background, applications, key challenges, bias, ethics, limitations and future scope},
   volume = {3},
   year = {2023},
}

@article{Boucher2021,
   author = {Eliane M. Boucher and Nicole R. Harake and Haley E. Ward and Sarah Elizabeth Stoeckl and Junielly Vargas and Jared Minkel and Acacia C. Parks and Ran Zilca},
   doi = {10.1080/17434440.2021.2013200},
   issn = {17452422},
   journal = {Expert Review of Medical Devices},
   pages = {37 -- 49},
   pmid = {34872429},
   publisher = {Taylor and Francis Ltd.},
   title = {Artificially intelligent chatbots in digital mental health interventions: a review},
   volume = {18},
   year = {2021},
}

@article{Raile2024,
   author = {Paolo Raile},
   doi = {10.1057/s41599-023-02567-0},
   issn = {26629992},
   issue = {1},
   journal = {Humanities and Social Sciences Communications},
   month = {12},
   publisher = {Springer Nature},
   title = {The usefulness of ChatGPT for psychotherapists and patients},
   volume = {11},
   year = {2024},
}

@article{Tang2023,
   author = {Lu Tang and Jinxu Li and Sophia Fantus},
   doi = {10.1177/20552076231186064},
   issn = {20552076},
   journal = {Digital Health},
   month = {1},
   publisher = {SAGE Publications Inc.},
   title = {Medical artificial intelligence ethics: A systematic review of empirical studies},
   volume = {9},
   year = {2023},
}

@article{Kirste2019,
  title={Augmented Intelligence -- wie Menschen mit KI zusammen arbeiten},
  author={Kirste, Moritz},
  journal={K{\"u}nstliche Intelligenz: Technologie| Anwendung| Gesellschaft},
  pages={58 -- 71},
  year={2019},
  isbn = {978-3-662-58041-7},
  doi = {10.1007/978-3-662-58042-4},
  publisher={Springer Berlin Heidelberg}
}

@article{Li2023,
   author = {Han Li and Renwen Zhang and Yi Chieh Lee and Robert E. Kraut and David C. Mohr},
   doi = {10.1038/s41746-023-00979-5},
   issn = {23986352},
   issue = {1},
   journal = {npj Digital Medicine},
   month = {12},
   publisher = {Nature Research},
   title = {Systematic review and meta-analysis of AI-based conversational agents for promoting mental health and well-being},
   volume = {6},
   year = {2023},
}

@article{Dergaa2023,
   author = {Ismail Dergaa and Feten Fekih-Romdhane and Souheil Hallit and Alexandre Andrade Loch and Jordan M. Glenn and Mohamed Saifeddin Fessi and Mohamed Ben Aissa and Nizar Souissi and Noomen Guelmami and Sarya Swed and Abdelfatteh El Omri and Nicola Luigi Bragazzi and Helmi Ben Saad},
   doi = {10.3389/fpsyt.2023.1277756},
   issn = {16640640},
   journal = {Frontiers in Psychiatry},
   publisher = {Frontiers Media SA},
   title = {ChatGPT is not ready yet for use in providing mental health assessment and interventions},
   volume = {14},
   year = {2023},
}

@article{Rahwan2019,
   author = {Iyad Rahwan and Manuel Cebrian and Nick Obradovich and Josh Bongard and Jean François Bonnefon and Cynthia Breazeal and Jacob W. Crandall and Nicholas A. Christakis and Iain D. Couzin and Matthew O. Jackson and Nicholas R. Jennings and Ece Kamar and Isabel M. Kloumann and Hugo Larochelle and David Lazer and Richard McElreath and Alan Mislove and David C. Parkes and Alex ‘Sandy’ Pentland and Margaret E. Roberts and Azim Shariff and Joshua B. Tenenbaum and Michael Wellman},
   doi = {10.1038/s41586-019-1138-y},
   issn = {14764687},
   issue = {7753},
   journal = {Nature},
   month = {4},
   pages = {477 -- 486},
   pmid = {31019318},
   publisher = {Nature Publishing Group},
   title = {Machine behaviour},
   volume = {568},
   year = {2019},
}

@inproceedings{Amershi2019,
   author = {Saleema Amershi and Dan Weld and Mihaela Vorvoreanu and Adam Fourney and Besmira Nushi and Penny Collisson and Jina Suh and Shamsi Iqbal and Paul N. Bennett and Kori Inkpen and Jaime Teevan and Ruth Kikin-Gil and Eric Horvitz},
   doi = {10.1145/3290605.3300233},
   isbn = {9781450359702},
   journal = {Conference on Human Factors in Computing Systems - Proceedings},
   month = {5},
   publisher = {Association for Computing Machinery},
   title = {Guidelines for human-AI interaction},
   year = {2019},
}

@article{Yau2021,
   author = {Kok Lim Alvin Yau and Heejeong Jasmine Lee and Yung Wey Chong and Mee Hong Ling and Aqeel Raza Syed and Celimuge Wu and Hock Guan Goh},
   doi = {10.1109/ACCESS.2021.3115494},
   issn = {21693536},
   journal = {IEEE Access},
   pages = {136744 -- 136761},
   publisher = {Institute of Electrical and Electronics Engineers Inc.},
   title = {Augmented Intelligence: Surveys of Literature and Expert Opinion to Understand Relations between Human Intelligence and Artificial Intelligence},
   volume = {9},
   year = {2021},
}

@online{EU_EthicGuidlines2019,
author = "European Commission",
title = {Ethics guidelines for trustworthy AI},
url = {https://digital-strategy.ec.europa.eu/en/library/ethics-guidelines-trustworthy-ai},
year = {2019},
note = "[online; accessed 03.09.2024]"
}

@article{Zanzotto2019,
   author = {Fabio Massimo Zanzotto},
   journal = {Journal of Artificial Intelligence Research},
   pages = {243 -- 252},
   title = {Viewpoint: Human-in-the-loop Artificial Intelligence},
   volume = {64},
   year = {2019},
}

@article{Cohen2023,
   author = {I. Glenn Cohen and Boris Babic and Sara Gerke and Qiong Xia and Theodoros Evgeniou and Klaus Wertenbroch},
   doi = {10.1038/s41746-023-00906-8},
   issn = {23986352},
   issue = {1},
   journal = {npj Digital Medicine},
   month = {12},
   publisher = {Nature Research},
   title = {How AI can learn from the law: putting humans in the loop only on appeal},
   volume = {6},
   year = {2023},
}

@article{Sarkar2023,
   author = {Surjodeep Sarkar and Manas Gaur and Lujie Karen Chen and Muskan Garg and Biplav Srivastava},
   doi = {10.3389/frai.2023.1229805},
   issn = {26248212},
   journal = {Frontiers in Artificial Intelligence},
   publisher = {Frontiers Media SA},
   title = {A review of the explainability and safety of conversational agents for mental health to identify avenues for improvement},
   volume = {6},
   year = {2023},
}

@article{Kim2012,
   author = {Jeongeun Kim and Hyeoun Ae Park},
   doi = {10.2196/jmir.2143},
   issn = {14388871},
   issue = {5},
   journal = {Journal of Medical Internet Research},
   pmid = {23026508},
   publisher = {JMIR Publications Inc.},
   title = {Development of a health information technology acceptance model using consumers' health behavior intention},
   volume = {14},
   year = {2012},
}

@article{Heyningen2018,
   author = {Thandi Van Heyningen and Simone Honikman and Mark Tomlinson and Sally Field and Landon Myer},
   doi = {10.1371/journal.pone.0193697},
   issn = {19326203},
   issue = {4},
   journal = {PLoS ONE},
   month = {4},
   pmid = {29668725},
   publisher = {Public Library of Science},
   title = {Comparison of mental health screening tools for detecting antenatal depression and anxiety disorders in South African women},
   volume = {13},
   year = {2018},
}

@article{Kessler2002,
   author = {R. C. Kessler and G. Andrews and L. J. Colpe and E. Hiripi and D. K. Mroczek and S. L.T. Normand and E. E. Walters and A. M. Zaslavsky},
   doi = {10.1017/S0033291702006074},
   issn = {00332917},
   issue = {6},
   journal = {Psychological Medicine},
   month = {8},
   pages = {959 -- 976},
   pmid = {12214795},
   title = {Short screening scales to monitor population prevalences and trends in non-specific psychological distress},
   volume = {32},
   year = {2002},
}

@article{spitzer2006,
  title={A brief measure for assessing generalized anxiety disorder: the GAD-7},
  author={Spitzer, Robert L and Kroenke, Kurt and Williams, Janet BW and L{\"o}we, Bernd},
  journal={Archives of internal medicine},
  volume={166},
  number={10},
  pages={1092 -- 1097},
  year={2006},
  publisher={American Medical Association}
}

@article{Damerau2021,
   author = {Mirjam Damerau and Martin Teufel and Venja Musche and Hannah Dinse and Adam Schweda and Jil Beckord and Jasmin Steinbach and Kira Schmidt and Eva Maria Skoda and Alexander Bäuerle},
   doi = {10.2196/27436},
   issn = {2561326X},
   issue = {7},
   journal = {JMIR Formative Research},
   month = {7},
   publisher = {JMIR Publications Inc.},
   title = {Determining acceptance of e-mental health interventions in digital psychodiabetology using a quantitative web-based survey: Cross-sectional study},
   volume = {5},
   year = {2021},
}

@article{Eysenbach2004,
   author = {Gunther Eysenbach},
   doi = {10.2196/jmir.6.3.e34},
   issn = {14388871},
   issue = {3},
   journal = {Journal of Medical Internet Research},
   pmid = {15471760},
   publisher = {JMIR Publications Inc.},
   title = {Improving the quality of web surveys: The Checklist for Reporting Results of Internet E-Surveys (CHERRIES)},
   volume = {6},
   year = {2004},
}

@article{Lall2019,
   author = {Michelle D. Lall and Theodore J. Gaeta and Arlene S. Chung and Sneha A. Chinai and Manish Garg and Abbas Husain and Cara Kanter and Sorabh Khandelwal and Caitlin S. Rublee and Ramin R. Tabatabai and James Kimo Takayesu and Mohammad Zaher and Nadine T. Himelfarb},
   doi = {10.5811/westjem.2019.1.39666},
   issn = {19369018},
   issue = {2},
   journal = {Western Journal of Emergency Medicine},
   month = {3},
   pages = {291 -- 304},
   pmid = {30881549},
   publisher = {eScholarship},
   title = {Assessment of physician well-being, part two: Beyond burnout},
   volume = {20},
   year = {2019},
}

@article{Dyrbye2013,
   author = {Liselotte N. Dyrbye and Daniel Satele and Jeff Sloan and Tait D. Shanafelt},
   doi = {10.1007/s11606-012-2252-9},
   issn = {08848734},
   issue = {3},
   journal = {Journal of General Internal Medicine},
   month = {3},
   pages = {421 -- 427},
   pmid = {23129161},
   title = {Utility of a Brief Screening Tool to Identify Physicians in Distress},
   volume = {28},
   year = {2013},
}

@article{Dyrbye2010,
  title={Development and preliminary psychometric properties of a well-being index for medical students},
  author={Dyrbye, Liselotte N. and Szydlo, Daniel W. and Downing, Steven M and Sloan, Jeff A and Shanafelt, Tait D},
  journal={BMC medical education},
  volume={10},
  pages={1 -- 9},
  year={2010},
  publisher={Springer}
}

@article{Haber2024,
   author = {Yuval Haber and Inbar Levkovich and Dorit Hadar-Shoval and Zohar Elyoseph},
   doi = {10.2196/54781},
   issn = {23687959},
   journal = {JMIR Mental Health},
   publisher = {JMIR Publications Inc.},
   title = {The Artificial Third: A Broad View of the Effects of Introducing Generative Artificial Intelligence on Psychotherapy},
   volume = {11},
   year = {2024},
}

@article{Mason2024,
   author = {Ashley E. Mason and Patrick Kasl and Severine Soltani and Abigail Green and Wendy Hartogensis and Stephan Dilchert and Anoushka Chowdhary and Leena S. Pandya and Chelsea J. Siwik and Simmie L. Foster and Maren Nyer and Christopher A. Lowry and Charles L. Raison and Frederick M. Hecht and Benjamin L. Smarr},
   doi = {10.1038/s41598-024-51567-w},
   issn = {20452322},
   issue = {1},
   journal = {Scientific Reports},
   month = {12},
   pmid = {38316806},
   publisher = {Nature Research},
   title = {Elevated body temperature is associated with depressive symptoms: results from the TemPredict Study},
   volume = {14},
   year = {2024},
}

@article{Bundo2023,
   author = {Marvin Bundo and Martin Preisig and Kathleen Merikangas and Jennifer Glaus and Julien Vaucher and Gérard Waeber and Pedro Marques-Vidal and Marie Pierre F. Strippoli and Thomas Müller and Oscar Franco and Ana Maria Vicedo-Cabrera},
   doi = {10.1186/s12940-023-01003-9},
   issn = {1476069X},
   issue = {1},
   journal = {Environmental Health: A Global Access Science Source},
   month = {12},
   pmid = {37430261},
   publisher = {BioMed Central Ltd},
   title = {How ambient temperature affects mood: an ecological momentary assessment study in Switzerland},
   volume = {22},
   year = {2023},
}

@online{BfArM2024,
   author = {Bundesinstitut für Arzneimittel und Medizinprodukte (BfArM)},
   title = {Das DiGA-Verzeichnis Antworten zur Nutzung von DiGA},
   url = {https://diga.bfarm.de/de},
   year = {2023},
   note = "[online; accessed 18.07.2024]"
}

@article{Schlieter2024,
   author = {Hannes Schlieter and Maren Kählig and Emily Hickmann and Daniel Fürstenau and Ali Sunyaev and Peggy Richter and Rüdiger Breitschwerdt and Christian Thielscher and Martin Gersch and Wolfgang Maaß and Melanie Reuter-Oppermann and Lena Wiese},
   doi = {10.1007/s00103-023-03804-2},
   issn = {14371588},
   issue = {1},
   journal = {Bundesgesundheitsblatt -- Gesundheitsforschung -- Gesundheitsschutz},
   month = {1},
   pages = {107 -- 114},
   pmid = {38086924},
   publisher = {Springer Science and Business Media Deutschland GmbH},
   title = {Digital health applications (DiGA) in the area of tension between progress and criticism: Discussion paper from the “Digital health” specialist group of the German Informatics Society},
   volume = {67},
   year = {2024},
}

@misc{Gesundheit2022,
   author = {Stiftung Gesundheit},
   title = {Eine repräsentative deutschlandweite Befragung von Leistungserbringer:innen durch die Stiftung Gesundheit in Zusammenarbeit mit der Informationsgesellschaft DiGA info STIFTUNG GESUNDHEIT Wissen ist die beste Medizin},
    url = {https://www.stiftung-gesundheit.de/pdf/studien/aerzte-im-zukunftsmarkt-gesundheit_2022_barrierefrei.pdf},
   note = "[online; in German, accessed 19.11.2024]",
   year = {2022},
}

@article{Sauermann2022,
   author = {Sven Sauermann and Julia Herzberg and Solveig Burkert and Susanne Habetha},
   doi = {10.1080/21614083.2021.2014047},
   issue = {1},
   journal = {Journal of European CME},
   month = {12},
   publisher = {Informa UK Limited},
   title = {DiGA – A Chance for the German Healthcare System},
   volume = {11},
   year = {2022},
}

@online{WoebotHealth2023,
   author = {Joe Gallagher and Sharanya Srinivasan},
   title = {AI at Woebot Health – Our Core Principles},
   url = {https://woebothealth.com/ai-core-principles/},
   year = {2023},
   note = "[online; accessed 19.07.2024]"
}

@article{Frey2022,
   author = {Silke Frey and Linda Kerkemeyer},
   doi = {10.1177/20552076221131142},
   issn = {20552076},
   journal = {Digital Health},
   month = {1},
   publisher = {SAGE Publications Inc.},
   title = {Acceptance of digital health applications in non-pharmacological therapies in German statutory healthcare system: Results of an online survey},
   volume = {8},
   year = {2022},
}

@article{Kählke2024,
   author = {Fanny Kählke and Penelope Hasking and Ann Marie Küchler and Harald Baumeister},
   doi = {10.3389/fdgth.2024.1284661},
   issn = {2673253X},
   journal = {Frontiers in Digital Health},
   publisher = {Frontiers Media SA},
   title = {Mental health services for German university students: acceptance of intervention targets and preference for delivery modes},
   volume = {6},
   year = {2024},
}

@online{Meskendahl2023,
   author = {Diana Meskendahl and Tobias Bachmann},
   pages = {1 -- 24},
   title = {Marktentwicklung digitaler Gesundheitsanwendungen
(DiGA-Report)},
   url = {https://digitalversorgt.de/wp-content/ uploads/2024/01/DiGA-Report-2023.pdf},
   year = {2023},
   note = "[online; in German; accessed 08.11.2024]",
}

@online{TK2023,
   author = {Techniker Krankenkasse},
   title = {DiGA-Report II},
   url = {https://www.tk.de/resource/blob/2170850/e7eaa59ecbc0488b415409d5d3a354cf/tk-diga-report-2-2024-data.pdf},
   year = {2023},
   note = "[online; accessed 08.11.2024]",
}

@misc{Deloitte2023,
   author = {Deloitte},
   title = {Gegenwind für die Digitalisierung im Gesundheitswesen -- Bürger fühlen sich nicht ausreichend informiert},
   year = {2023},
url = {https://cloud.marketing.deloitte.de/RegistrationPage?eventname=Digitalisierung%20im%20Gesundheitswesen%202023&locale=de&mid2=MID-19805&category=Publications&eventCampaignId=7015p0000016dKxAAI}, 
note = "[online; accessed 23.11.2024]"
}

@article{Bielinski2021,
   author = {Laura Luisa Bielinski and Leonie Trimpop and Thomas Berger},
   doi = {10.1007/s00278-021-00524-3},
   issn = {14322080},
   issue = {5},
   journal = {Psychotherapeut},
   month = {9},
   pages = {447 -- 454},
   publisher = {Springer Medizin},
   title = {All in the mix? Blended psychotherapy as an example of digitalization in psychotherapy},
   volume = {66},
   year = {2021},
}

@article{Wentzel2016,
   author = {Jobke Wentzel and Rosalie Van der Vaart and Ernst T. Bohlmeijer and Julia E.W.C. Van Gemert-Pijnen},
   doi = {10.2196/mental.4534},
   issn = {23687959},
   issue = {1},
   journal = {JMIR Mental Health},
   month = {1},
   pmid = {26860537},
   publisher = {JMIR Publications Inc.},
   title = {Mixing online and face-to-face therapy: How to benefit from blended care in mental health care},
   volume = {3},
   year = {2016},
}

@article{Schuster2020,
   author = {Raphael Schuster and Naira Topooco and Antonia Keller and Ella Radvogin and Anton Rupert Laireiter},
   doi = {10.1016/j.invent.2020.100326},
   issn = {22147829},
   journal = {Internet Interventions},
   month = {9},
   publisher = {Elsevier B.V.},
   title = {Advantages and disadvantages of online and blended therapy: Replication and extension of findings on psychotherapists' appraisals},
   volume = {21},
   year = {2020},
}

@article{Urech2019,
   author = {Antoine Urech and Tobias Krieger and Laura Möseneder and Adriana Biaggi and Alessia Vincent and Christine Poppe and Björn Meyer and Heleen Riper and Thomas Berger},
   doi = {10.1080/10503307.2018.1430910},
   issn = {14684381},
   issue = {8},
   journal = {Psychotherapy Research},
   month = {11},
   pages = {986 -- 998},
   pmid = {29385964},
   publisher = {Routledge},
   title = {A patient post hoc perspective on advantages and disadvantages of blended cognitive behaviour therapy for depression: A qualitative content analysis},
   volume = {29},
   year = {2019},
}

@article{Venkatesh2003,
 author = {Viswanath Venkatesh and Michael G. Morris and Gordon B. Davis and Fred D. Davis},
 journal = {MIS Quarterly},
 issn = {02767783},
 number = {3},
 pages = {425 -- 478},
 publisher = {Management Information Systems Research Center, University of Minnesota},
 title = {User Acceptance of Information Technology: Toward a Unified View},
 volume = {27},
 year = {2003}
}

@article{Davis1989,
   author = {Fred D. Davis},
   doi = {10.2307/249008},
   issn = {02767783},
   issue = {3},
   journal = {MIS Quarterly: Management Information Systems},
   pages = {319 -- 339},
   publisher = {Management Information Systems Research Center},
   title = {Perceived usefulness, perceived ease of use, and user acceptance of information technology},
   volume = {13},
   year = {1989},
}

@article{Venkatesh2012,
  title={Consumer acceptance and use of information technology: extending the unified theory of acceptance and use of technology},
  author={Venkatesh, Viswanath and Thong, James YL and Xu, Xin},
  journal={MIS quarterly},
  pages={157 -- 178},
  year={2012},
  publisher={JSTOR}
}

@article{Venkatesh2000,
   author = {Viswanath Venkatesh and Fred D. Davis},
   doi = {10.1287/mnsc.46.2.186.11926},
   issn = {00251909},
   issue = {2},
   journal = {Management Science},
   pages = {186 -- 204},
   publisher = {INFORMS},
   title = {Theoretical extension of the Technology Acceptance Model: Four longitudinal field studies},
   volume = {46},
   year = {2000},
}

@article{Venkatesh2008,
   author = {Viswanath Venkatesh and Hillol Bala},
   journal = {The Author Journal compilation C},
   publisher = {Decision Sciences Institute},
   title = {Technology Acceptance Model 3 and a Research Agenda on Interventions},
   volume = {39},
   year = {2008},
}

@article{Werdecker2021,
   author = {Lena Werdecker and Tobias Esch},
   doi = {10.1371/journal.pone.0253447},
   issn = {19326203},
   issue = {6 June},
   journal = {PLoS ONE},
   month = {6},
   pmid = {34143849},
   publisher = {Public Library of Science},
   title = {Burnout, satisfaction and happiness among German general practitioners (GPs): A cross-sectional survey on health resources and stressors},
   volume = {16},
   year = {2021},
}

@article{Tawfik2019,
   author = {Daniel S. Tawfik and Jochen Profit and Sarah Webber and Tait D. Shanafelt},
   doi = {10.1007/s40746-019-00147-6},
   issn = {21986088},
   issue = {1},
   journal = {Current Treatment Options in Pediatrics},
   month = {3},
   pages = {11 -- 25},
   publisher = {Springer International Publishing},
   title = {Organizational Factors Affecting Physician Well-Being},
   volume = {5},
   year = {2019},
}

@article{Brady2019,
   author = {Keri J.S Brady and Lewis E. Kazis and R. Christopher Sheldrick and Pengsheng Ni and Mickey T. Trockel},
   doi = {10.1016/j.cppeds.2019.100662},
   issn = {15383199},
   issue = {12},
   journal = {Current Problems in Pediatric and Adolescent Health Care},
   month = {12},
   pmid = {31562054},
   publisher = {Mosby Inc.},
   title = {Selecting physician well-being measures to assess health system performance and screen for distress: Conceptual and methodological considerations},
   volume = {49},
   year = {2019},
}

@article{Zumbrunn2020,
   author = {Brigitta Zumbrunn and Odile Stalder and Andreas Limacher and Peter E. Ballmer and Stefano Bassetti and Edouard Battegay and Jürg Hans Beer and Michael Brändle and Daniel Genné and Daniel Hayoz and Christoph Henzen and Lars Christian Huber and Pierre Auguste Petignat and Jean Luc Reny and Peter Vollenweider and Drahomir Aujesky},
   doi = {10.4414/smw.2020.20255},
   issn = {14243997},
   issue = {23-24},
   journal = {Swiss Medical Weekly},
   month = {6},
   pmid = {32557425},
   publisher = {EMH Schweizerischer Arzteverlag AG},
   title = {The well-being of Swiss general internal medicine residents},
   volume = {150},
   year = {2020},
}

@article{Robles2022,
   author = {Rebeca Robles and Ana Fresán and Natasha Alcocer-Castillejos and Janet Real-Ramírez and Silvia Morales-Chainé},
   doi = {10.3390/ijerph19159451},
   issn = {16604601},
   issue = {15},
   journal = {International Journal of Environmental Research and Public Health},
   month = {8},
   pmid = {35954808},
   publisher = {MDPI},
   title = {Brief Screening for Distress among Healthcare Professionals: Psychometric Properties of the Physician Well-Being Index -- Spanish Version},
   volume = {19},
   year = {2022},
}

@article{Esmaeilzadeh2021,
author = {Esmaeilzadeh, Pouyan and Mirzaei, Tala and Dharanikota, Spurthy},
doi = {10.2196/25856},
issn = {14388871},
journal = {Journal of Medical Internet Research},
month = {11},
number = {11},
pmid = {34842535},
publisher = {JMIR Publications Inc.},
title = {{Patients' perceptions toward human-artificial intelligence interaction in health care: Experimental study}},
volume = {23},
year = {2021}
}

@article{Esmaeilzadeh2020,
   author = {Pouyan Esmaeilzadeh},
   doi = {10.1186/s12911-020-01191-1},
   issn = {14726947},
   issue = {1},
   journal = {BMC Medical Informatics and Decision Making},
   month = {7},
   pmid = {32698869},
   publisher = {BioMed Central Ltd},
   title = {Use of AI-based tools for healthcare purposes: A survey study from consumers' perspectives},
   volume = {20},
   year = {2020},
}

@article{Hajesmaeel-Gohari2021,
author = {Hajesmaeel-Gohari, Sadrieh and Bahaadinbeigy, Kambiz},
doi = {10.1186/s12911-021-01407-y},
issn = {14726947},
journal = {BMC Medical Informatics and Decision Making},
month = {12},
number = {1},
pmid = {33531013},
publisher = {BioMed Central Ltd},
title = {{The most used questionnaires for evaluating telemedicine services}},
volume = {21},
year = {2021}
}

@article{Pedro2023,
author = {Pedro, Ana Rita and Dias, Michelle B. and Laranjo, Liliana and Cunha, Ana Soraia and Cordeiro, Jo{\~{a}}o V.},
doi = {10.1371/journal.pone.0290613},
issn = {19326203},
journal = {PLoS ONE},
month = {9},
pmid = {37676884},
publisher = {Public Library of Science},
title = {{Artificial intelligence in medicine: A comprehensive survey of medical doctor's perspectives in Portugal}},
volume = {18},
year = {2023}
}

@article{Nurtsch2024,
author = {Nurtsch, Angelina and Teufel, Martin and Jahre, Lisa Maria and Esber, Andr{\'{e}} and Rausch, Raya and Tewes, Mitra and Sch{\"{o}}bel, Christoph and Palm, Stefan and Schuler, Martin and Schadendorf, Dirk and Skoda, Eva Maria and B{\"{a}}uerle, Alexander},
doi = {10.1177/20552076231222108},
issn = {20552076},
journal = {Digital Health},
month = {1},
publisher = {SAGE Publications Inc.},
title = {{Drivers and barriers of patients' acceptance of video consultation in cancer care}},
volume = {10},
year = {2024}
}

@article{Rentrop2022,
author = {Rentrop, Vanessa and Damerau, Mirjam and Schweda, Adam and Steinbach, Jasmin and Schuren, Lynik Chantal and Niedergethmann, Marco and Skoda, Eva Maria and Teufel, Martin and B{\"{a}}uerle, Alexander},
doi = {10.2196/31229},
issn = {2561326X},
journal = {JMIR Formative Research},
number = {3},
pages = {1 -- 16},
title = {{Predicting Acceptance of e-Mental Health Interventions in Patients With Obesity by Using an Extended Unified Theory of Acceptance Model: Cross-sectional Study}},
volume = {6},
year = {2022}
}

@article{Esber2023,
author = {Esber, Andr{\'{e}} and Teufel, Martin and Jahre, Lisa and in der Schmitten, J{\"{u}}rgen and Skoda, Eva Maria and B{\"{a}}uerle, Alexander},
doi = {10.1177/20552076221149317},
issn = {20552076},
journal = {Digital Health},
title = {{Predictors of patients' acceptance of video consultation in general practice during the coronavirus disease 2019 pandemic applying the unified theory of acceptance and use of technology model}},
volume = {9},
year = {2023}
}

@article{Neumann2023,
author = {Neumann, Ariana and K{\"{o}}nig, Hans Helmut and Bokermann, Josephine and Hajek, Andr{\'{e}}},
doi = {10.2196/46148},
issn = {23687959},
journal = {JMIR Mental Health},
pages = {1 -- 36},
title = {{Determinants of Patient Use and Satisfaction With Synchronous Telemental Health Services During the COVID-19 Pandemic: Systematic Review}},
volume = {10},
year = {2023}
}

@article{Abuyadek2024,
author = {Abuyadek, Rowan M. and Hammouda, Esraa Abdellatif and Elrewany, Ehab and Elmalawany, Dina Hussein and Ashmawy, Rasha and Zeina, Sally and Gebreal, Assem and Ghazy, Ramy Mohamed},
doi = {10.1186/s12889-024-18436-7},
issn = {14712458},
journal = {BMC Public Health},
number = {1},
pages = {1 -- 18},
pmid = {38658881},
title = {{Acceptability of Tele-mental Health Services Among Users: A Systematic Review and Meta-analysis}},
volume = {24},
year = {2024}
}

@article{Lucas2014,
   author = {Gale M. Lucas and Jonathan Gratch and Aisha King and Louis Philippe Morency},
   doi = {10.1016/j.chb.2014.04.043},
   issn = {07475632},
   journal = {Computers in Human Behavior},
   pages = {94 -- 100},
   publisher = {Elsevier Ltd},
   title = {It's only a computer: Virtual humans increase willingness to disclose},
   volume = {37},
   year = {2014},
}

@article{Philip2020,
   author = {Pierre Philip and Lucile Dupuy and Marc Auriacombe and Fushia Serre and Etienne de Sevin and Alain Sauteraud and Jean Arthur Micoulaud-Franchi},
   doi = {10.1038/s41746-019-0213-y},
   issn = {23986352},
   issue = {1},
   journal = {npj Digital Medicine},
   month = {12},
   publisher = {Nature Research},
   title = {Trust and acceptance of a virtual psychiatric interview between embodied conversational agents and outpatients},
   volume = {3},
   year = {2020},
}

@article{Lattie2022,
   author = {Emily G. Lattie and Colleen Stiles-Shields and Andrea K. Graham},
   doi = {10.1038/s44159-021-00003-1},
   issn = {27310574},
   issue = {2},
   journal = {Nature Reviews Psychology},
   month = {2},
   pages = {87 -- 100},
   publisher = {Nature Publishing Group},
   title = {An overview of and recommendations for more accessible digital mental health services},
   volume = {1},
   year = {2022},
}

@article{Nebeker2021,
   author = {Camille Nebeker and Rebecca J. Bartlett Ellis and John Torous},
   doi = {10.1093/tbm/ibz074},
   issn = {16139860},
   issue = {4},
   journal = {Translational Behavioral Medicine},
   month = {8},
   pages = {1004 -- 1015},
   pmid = {31120511},
   publisher = {Oxford University Press},
   title = {Development of a decision-making checklist tool to support technology selection in digital health research},
   volume = {10},
   year = {2021},
}

@article{Kalam2024,
   author = {Khondoker Tashya Kalam and Jannatul Mabia Rahman and Md Rabiul Islam and Syed Masudur Rahman Dewan},
   doi = {10.1002/hsr2.1912},
   issn = {23988835},
   issue = {2},
   journal = {Health Science Reports},
   month = {2},
   publisher = {John Wiley and Sons Inc},
   title = {ChatGPT and mental health: Friends or foes?},
   volume = {7},
   year = {2024},
}

@article{Prince2007,
   author = {Martin Prince and Vikram Patel and Shekhar Saxena and Mario Maj and Joanna Maselko and Michael R Phillips and Atif Rahman},
   doi = {10.1016/S0140},
   journal = {859 Global Mental Health},
   title = {No health without mental health},
   volume = {8},
   year = {2007},
}

@article{Manwell2014,
   author = {Laurie A. Manwell and Skye P Barbic and Karen Roberts and Zachary Durisko and Cheolsoon Lee and Emma Ware and Kwame McKenzie},
   doi = {10.1136/bmjopen-2014},
   issn = {20446055},
   issue = {11},
   journal = {BMJ Open},
   pages = {1 -- 11},
   pmid = {25377009},
   publisher = {BMJ Publishing Group},
   title = {What is mental health? Evidence towards a new definition from a mixed methods multidisciplinary international survey},
   volume = {4},
   year = {2014},
}

@article{Fusar-Poli2020,
   author = {Paolo Fusar-Poli and Gonzalo Salazar de Pablo and Andrea De Micheli and Dorien H. Nieman and Christoph U. Correll and Lars Vedel Kessing and Andrea Pfennig and Andreas Bechdolf and Stefan Borgwardt and Celso Arango and Therese van Amelsvoort},
   doi = {10.1016/j.euroneuro.2019.12.105},
   issn = {18737862},
   journal = {European Neuropsychopharmacology},
   month = {2},
   pages = {33 -- 46},
   pmid = {31901337},
   publisher = {Elsevier B.V.},
   title = {What is good mental health? A scoping review},
   volume = {31},
   year = {2020},
}

@book{WHO2005,
   author = {World Health Organization and others},
   isbn = {928901377X},
   publisher = {World Health Organization, Europe},
   title = {Mental health: facing the challenges, building solutions: report from the WHO European Ministerial Conference},
   year = {2005},
}

@article{Brunt2023,
   author = {Thomas J. Brunt and Oliver Gale-Grant},
   doi = {10.1192/bja.2022.42},
   issn = {2056-4678},
   issue = {4},
   journal = {BJPsych Advances},
   month = {7},
   pages = {230 -- 238},
   publisher = {Royal College of Psychiatrists},
   title = {Telepsychiatry: what clinicians need to know about digital mental healthcare},
   volume = {29},
   year = {2023},
}

@article{Fitzpatrick2017,
   author = {Kathleen Kara Fitzpatrick and Alison Darcy and Molly Vierhile},
   doi = {10.2196/mental.7785},
   issn = {23687959},
   issue = {2},
   journal = {JMIR Mental Health},
   month = {4},
   pmid = {28588005},
   publisher = {JMIR Publications Inc.},
   title = {Delivering cognitive behavior therapy to young adults with symptoms of depression and anxiety using a fully automated conversational agent (Woebot): A randomized controlled trial},
   volume = {4},
   year = {2017},
}

@article{Schmitz2022,
   author = {Anne Schmitz and Ana M. Díaz-Martín and M. Jesús Yagüe Guillén},
   doi = {10.1016/j.chb.2022.107183},
   issn = {07475632},
   journal = {Computers in Human Behavior},
   month = {5},
   publisher = {Elsevier Ltd},
   title = {Modifying UTAUT2 for a cross-country comparison of telemedicine adoption},
   volume = {130},
   year = {2022},
}

@article{Meininger2022,
   author = {Lea Meininger and Julia Adam and Elena von Wirth and Paula Viefhaus and Katrin Woitecki and Daniel Walter and Manfred Döpfner},
   doi = {10.1186/s13034-022-00494-7},
   issn = {17532000},
   issue = {1},
   journal = {Child and Adolescent Psychiatry and Mental Health},
   month = {12},
   publisher = {BioMed Central Ltd},
   title = {Cognitive-behavioral teletherapy for children and adolescents with mental disorders and their families during the COVID-19 pandemic: a survey on acceptance and satisfaction},
   volume = {16},
   year = {2022},
}

@online{DemografiePortal,
   author = "{Bundesinstitut für Bevölkerungsforschung}",
   year = "2022",
   title = {Altersstruktur der Bevölkerung},
   url = {https://www.demografie-portal.de/DE/Fakten/bevoelkerung-altersstruktur.html},
   note = "[online; in German; accessed 08.09.2024]"
}

@online{BÄK2023,
   author = {Bundesärztekammer Arbeitsgemeinschaft der deutschen Ärztekammern},
   title = {Ärztestatistik zum 31. Dezember 2023 Bundesgebiet gesamt},
   year = {2023},
   url = {https://www.bundesaerztekammer.de/fileadmin/user_
upload/BAEK/Ueber_uns/Statistik/AErztestatistik_2023_18.04.2024.pdf},
   note = "[online; German only; accessed 08.09.2024]"
}

@article{Hennemann2016,
   author = {Severin Hennemann and Manfred E. Beutel and Rüdiger Zwerenz},
   doi = {10.2196/jmir.6003},
   issn = {14388871},
   issue = {12},
   journal = {Journal of Medical Internet Research},
   month = {12},
   pmid = {28011445},
   publisher = {JMIR Publications Inc.},
   title = {Drivers and barriers to acceptance of web-based aftercare of patients in inpatient routine care: A cross-sectional survey},
   volume = {18},
   year = {2016},
}

@online{PTK,
author = "Psychotherapeutenkammer Bayern (PTK Bayern)",
title = {Psychotherapeutenkammer Bayern (PTK Bayern)},
url = {https://www.ptk-bayern.de/},
year = {2024},
note = "[online; German only; accessed 19.10.2024]"
}

@book{wilcox2011introduction,
  title={Introduction to robust estimation and hypothesis testing},
  author={Wilcox, Rand R},
  year={2011},
  publisher={Academic press}
}

@article{erhardt2023,
  title={Generalised anxiety and panic symptoms in the German National Cohort (NAKO)},
  author={Erhardt, Angelika and Gelbrich, G{\"o}tz and Klinger-K{\"o}nig, Johanna and Streit, Fabian and Kleineidam, Luca and Riedel-Heller, Steffi G and NAKO Investigators and Schmidt, B{\"o}rge and Schmiedek, Florian and Wagner, Michael and others},
  journal={The World Journal of Biological Psychiatry},
  volume={24},
  number={10},
  pages={881 -- 896},
  year={2023},
  publisher={Taylor \& Francis}
}

@online{Clifton2021,
author = "Jim Clifton",
title = {The Next Global Pandemic: Mental Health},
url = {https://www.gallup.com/workplace/357710/next-global-pandemic-mental-health.aspx},
year = {2021},
note = "[online; accessed 27.10.2024]"
}

@misc{Hausärzteverband2024,
   author = {Hausärztinnen- und Hausärzteverband e. V.},
   pages = {1 -- 6},
   title = {Künstliche Intelligenz (KI) in der Hausarztpraxis},
   url = {https://www.haev.de/fileadmin/user_upload/News_Dateien/2024/2024_07_04_HAEV_Positionspapier_KI.pdf},
   year = {2024},
    note = "[online; in German; accessed 27.10.2024]"
}

@inproceedings{Paikens2020,
   author = {Pēteris Paikens and Artūrs Znotiņš and Guntis Bārzdiņš},
   doi = {10.1007/978-3-030-51310-8_25},
   isbn = {9783030513092},
   issn = {16113349},
   booktitle = {Lecture Notes in Computer Science (including subseries Lecture Notes in Artificial Intelligence and Lecture Notes in Bioinformatics)},
   pages = {277 -- 284},
   publisher = {Springer},
   title = {Human-in-the-loop conversation agent for customer service},
   volume = {12089 LNCS},
   year = {2020},
}

@inproceedings{Kero2023,
   author = {Sandra Kero and Sarah Yasemin Akyürek and Florian Golo Flaßhoff},
   doi = {10.1145/3442188.3445922},
   isbn = {9781450383097},
   month = {3},
   pages = {1 -- 4},
   publisher = {MeMo:KI - Meinungsmonitor Künstliche Intelligenz},
   title = {Bekanntheit und Akzeptanz von ChatGPT in Deutschland},
   year = {2023},
}

@online{dpa2023,
   author = {dpa},
   title = {Ein Jahr ChatGPT: So nutzen Menschen in Deutschland KI},
   url = {https://t3n.de/news/chatgpt-menschen-deutschland-ki-nutzung-1592840/},
   note = "[online; German only; accessed 08.11.2024]",
   year = {2023},
}

@article{Gulliver2010,
   author = {Amelia Gulliver and Kathleen M. Griffiths and Helen Christensen},
   doi = {10.1186/1471-244X-10-113},
   issn = {1471244X},
   journal = {BMC Psychiatry},
   month = {12},
   pmid = {21192795},
   title = {Perceived barriers and facilitators to mental health help-seeking in young people: A systematic review},
   volume = {10},
   year = {2010},
}

@online{Benecke2024,
   author = {Andrea Benecke},
   month = {6},
   title = {Digitalisierung und Künstliche Intelligenz in der Psychotherapie},
   year = {2024},
   url = {https://www.bptk.de/newsletter/2-2024/digitalisierung-und-kuenstliche-intelligenz-in-der-psychotherapie/},
   note = "[online; German only; accessed 08.11.2024]",
}

@article{Wong2024,
   author = {Rebecca Shin Yee Wong},
   doi = {10.1186/s41983-024-00791-2},
   issn = {16878329},
   issue = {1},
   journal = {Egyptian Journal of Neurology, Psychiatry and Neurosurgery},
   month = {12},
   publisher = {Springer Science and Business Media Deutschland GmbH},
   title = {ChatGPT in psychiatry: promises and pitfalls},
   volume = {60},
   year = {2024},
}

@online{Walker2023,
   author = {Lauren Walker},
   month = {3},
   title = {Belgian man dies by suicide following exchanges with chatbot},
   url = {https://www.brusselstimes.com/430098/belgian-man-commits-suicide-following-exchanges-with-chatgpt},
   year = {2023},
   note = "[online; accessed 08.11.2024]",
}

@article{Staples2019,
   author = {Lauren G. Staples and Blake F. Dear and Milena Gandy and Vincent Fogliati and Rhiannon Fogliati and Eyal Karin and Olav Nielssen and Nickolai Titov},
   doi = {10.1016/j.genhosppsych.2018.11.003},
   issn = {18737714},
   journal = {General Hospital Psychiatry},
   month = {1},
   pages = {13 -- 18},
   pmid = {30508772},
   publisher = {Elsevier Inc.},
   title = {Psychometric properties and clinical utility of brief measures of depression, anxiety, and general distress: The PHQ-2, GAD-2, and K-6},
   volume = {56},
   year = {2019},
}

@article{DeclarationOfHelsinki2024,
   author = {World Medical Association and 13A chemin du Levant and 01210 Ferney-Voltaire},
   doi = {10.1001/jama.2024.21972},
   issn = {0098-7484},
   journal = {JAMA},
   month = {10},
   title = {World Medical Association Declaration of Helsinki},
   url = {https://jamanetwork.com/journals/jama/fullarticle/2825290},
   year = {2024},
   note = "[online; accessed 08.11.2024]",
}

@online{surveymonkeyCalculator,
   author = {SurveyMonkey Europe UC},
   title = {Sample size calculator},
   url = {https://www.surveymonkey.com/mp/sample-size-calculator/},
   note = "[online; accessed 08.11.2024]",
}

@online{ClickWorker,
   author = {clickworker GmbH},
   title = {Umfrageteilnehmer für Online-Umfragen},
   url = {https://www.clickworker.de/umfragen},
   note = "[online; accessed 08.11.2024]",
}

@online{UmfrageOnline,
   author = {enuvo GmbH},
   title = {Umfrage erstellen, einfach professionell},
   url = {https://www.umfrageonline.com/},
   note = "[online; accessed 08.11.2024]",
}

@online{Deepl,
   author = {DeepL SE},
   title = {Übersetzer},
   url = {https://www.deepl.com/},
   note = "[online; accessed 08.11.2024]",
}
